\documentclass[aps,prb,twocolumn,superscriptaddress]{revtex4-2}

\usepackage{amsmath}    

\usepackage{amsfonts}
\usepackage{amssymb}
\usepackage[colorlinks=true,citecolor=blue,linkcolor=red]{hyperref}
\usepackage{graphicx}
\usepackage{bbold}					
\usepackage[dvipsnames]{xcolor}

\usepackage{bm} 
\usepackage{dcolumn}
\usepackage{mathtools}

\usepackage{array,booktabs,ragged2e}	
\usepackage{soul}						
\usepackage{cancel}						
\usepackage{tabularx}
\usepackage{multirow}
\usepackage{hhline}
\usepackage{adjustbox}
\usepackage{sidecap}
\usepackage{threeparttable}
\usepackage{colortbl} 
\usepackage{tabularx}
\usepackage[utf8]{inputenc}
\usepackage[T1]{fontenc}

\def\bra#1{\left<{#1}\right|}					
\def\ket#1{\left|{#1}\right>}					
\def\braket#1#2{\left<{#1}|{#2}\right>}			

\newcolumntype{P}[1]{>{\raggedleft\arraybackslash}p{#1}}
\newcolumntype{R}[1]{>{\centering\arraybackslash}p{#1}}

\usepackage{bm,color,amsmath,amssymb,mathrsfs,latexsym,graphicx,psfrag,mathtools}
\usepackage{color}
\usepackage[dvipsnames]{xcolor}

\usepackage{empheq}

\newcommand{\bsl}[1]{\boldsymbol{#1}}

\newcommand{\h}{\mathfrak{H}}
\newcommand{\ii}{\mathrm{i}}

\newcommand{\dsZ}{\mathbb{Z}}
\newcommand{\dsN}{\mathbb{N}}
\newcommand{\dsR}{\mathbb{R}}

\newcommand{\dsC}{\mathbb{C}}
\newcommand{\Tr}{\mathop{\mathrm{Tr}}}

\newcommand{\eqnref}[1]{Eq.\,\eqref{#1}}

\newcommand{\refcite}[1]{Ref.\,\cite{#1}}

\newcommand{\mat}[1]{\left(\begin{matrix}#1\end{matrix}\right)}

\newcommand{\eq}[1]{\begin{equation} #1 \end{equation}}

\newcommand{\eqa}[1]{\begin{align}\begin{split} #1 \end{split}\end{align}}

\let\oldAA\AA
\renewcommand{\AA}{\text{\normalfont\oldAA}}

\newcommand{\ie}{{\emph{i.e.}}}
\newcommand{\eg}{{\emph{e.g.}}}
\newcommand{\TR}{\mathcal{T}}

\newcommand{\W}{\mathcal{W}}
\newcommand{\G}{\mathcal{G}}

\renewcommand{\P}{\mathcal{P}}

\newcommand{\N}{\mathcal{N}}

\newcommand{\A}{\mathcal{A}}

\newcommand{\diag}{\text{diag}}
\newcommand{\F}{\mathcal{F}}

\newcommand{\Ch}{\text{Ch}}

\newcommand{\K}{\text{K}}

\newcommand{\BZ}{\text{1BZ}}

\newtheorem{proposition}{Proposition}

\newcommand{\proof}[1]{\noindent\textbf{Proof:} #1 \\ \noindent\textbf{End of Proof.}}

\usepackage{comment}

\usepackage{cleveref}
\crefname{appendix}{App.\,}{Apps.\,}
\crefname{equation}{Eq.\,}{Eqs.\,}
\crefname{figure}{Fig.\,}{Figs.\,}
\crefname{table}{Tab.\,}{Tabs.\,}
\crefname{section}{Sec.\,}{Secs.\,}
\creflabelformat{appendix}{[#2#1#3]}

\newcommand{\appDirichlet}{App.\,[\text{\color{blue}A}]} 

\newcommand{\appWLbound}{App.\,[\text{\color{blue}B}]} 
 
\newcommand{\appidealZtwoband}{App.\,[\text{\color{blue}C2-3}]} 
\newcommand{\appmonotonicflow}{App.\,[\text{\color{blue}D}]} 
\newcommand{\apptMoTe}{App.\,[\text{\color{blue}E1}]} 
\newcommand{\appmoirerashbaandinversion}{App.\,[\text{\color{blue}E2-3}]}

\begin{document}

\title{Wilson-Loop-Ideal Bands and General Idealization}

\author{Awwab A. Azam}
\affiliation{Department of Electrical and Computer Engineering, University of Florida, Gainesville, FL 32611, USA}

\author{Biao Lian}
\affiliation{Department of Physics, Princeton University, Princeton, New Jersey 08544, USA}

\author{Shinsei Ryu}
\affiliation{Department of Physics, Princeton University, Princeton, New Jersey 08544, USA}

\author{Jiabin Yu}
\email{yujiabin@ufl.edu}
\affiliation{Department of Physics and Quantum Theory Project, University of Florida, Gainesville, FL 32611, USA}

\begin{abstract}
Quantum geometry is universally bounded from below by Wilson-loop windings.
In this work, we define an isolated set of bands to be Wilson-loop-ideal, if their quantum metric saturates the Wilson-loop lower bound.
The definition naturally incorporates the known Chern-ideal and Euler-ideal bands, and allows us to define other types of ideal bands, such as Kane-Mele $\dsZ_2$-ideal and inversion-fragile-ideal bands.
In particular, we find that in the case of zero total Chern number, an isolated WL-ideal set of two bands with non-singular nonabelian Berry curvature and nontrivial normal Wilson-loop winding always admits a Chern-ideal gauge, without the need of a global good quantum number (such as spin).
This enables the direct construction of new topologically ordered states, such as fractional topological insulator wavefunctions.
We further propose a general framework of constructing monotonic flows that achieve Wilson-loop-ideal states starting from non-ideal bands through band mixing, where Wilson-loop-ideal states are not energy eigenstates but have smooth projectors similar to isolated bands.
We apply the constructed flows to the realistic model of $3.89^\circ$ twisted bilayer MoTe$_2$, a moir\'e Rashba model and another moiré time-reversal-breaking models, and numerically find Chern-ideal, $\dsZ_2$-ideal and inversion-fragile states, respectively, with relative error in the integrated quantum metric below $5\times 10^{-3}$.
Our exact-diagonalization calculations on the numerically ideal states demonstrate the potential of our general definition of Wilson-loop-ideal bands and general procedure of constructing Wilson-loop-ideal states for future study of novel correlated physics.
\end{abstract}

\maketitle

\section{Introduction}

Quantum geometry provides a crucial foundation for the study of strongly correlated physics.
Besides contributing significantly to a variety of correlated quantities, such as superfluid weight~\cite{Torma2015SWBoundChern,Torma2016SuperfluidWeightLieb,Liang2017SWBandGeo,Hu2019MATBGSW,Xie2020TopologyBoundSCTBG,Torma2020SFWTBG,Rossi2021CurrentOpinion,Yu2022EOCPTBG,Torma2023WhereCanQuantumGeometryLeadUs,Tian2023QuantumGeoSC,Han2024QG_Nesting,Millis_2025_SuperfluidWeight,BJY2025GeometricSuperfluidWeightQuasicrystals,Huang2025QGSWMultiband}, electron-phonon coupling~\cite{Yu05032023GeometryEPC,Alexandradinata2024ShiftCurrent,Wang2025Optical_QG_EPC,Wang2025_QG_EPC_Ferroelectric}, and correlated charge fluctuations~\cite{Yu2024_QG_Charge_Fluctuation,Wu2024QGCornerChargeFluctuationsManyBody}, quantum geometry can offer a direct route to constructing many-body wavefunctions for correlated phases, which is the focus of this work.
(See \cite{Yu2025QGReview,Nagaosa_08012025_QG_Review} for recent reviews.)
%
%
One known example is the Chern-ideal band~\cite{Jie2021IdealBands,Parker2023IdealBands,Valentin2023IdealBands,Dong_2023_ideal_Higher_Chern,liu2024theorygeneralizedlandaulevels,Roy2014QGChernBound, claassen2015position, northe2022interplay,Queiroz_2024_TMD_Chiral,ji2024quantum,Roy2024FCINonLL,Sun2025IdealChern,Oreg2025UnifyingFrameworkFCI,Liu_2025_idealTHF}, which has a quantum metric that saturates the topological lower bound determined by the Chern number ($\Ch$)~\cite{Roy2014QGChernBound,Bellissard1994QGChernBound}. Wavefunctions for fractional Chern insulators (FCIs)~\cite{neupert, sheng, regnault,Sun2011,Tang11} can be constructed by inserting vortices into such a band (\ie, acting $z_i - z_j$ on the state with $z_i = x_i - \ii y_i$ the complex 2D coordinate for the $i$th particle).
As a natural generalization of the ideal-Chern band, Euler-ideal bands, which saturate the lower bound~\cite{Xie2020TopologyBoundSCTBG,Yu2022EOCPTBG,BJY2024EulerBoundQG,Slager2024EulerOptical} of the integrated quantum metric due to the Euler number~\cite{Ahn2018MonopoleNLSM,Ahn2019TBGFragile,Song2019TBGFragile}, have also been discussed~\cite{BJY2024EulerBoundQG}.

However, the topological lower bound of quantum geometry is known to be far more general than the Chern/Euler bounds~\cite{Yu2025Z2bound,Torma2016QGChiralWinding,Herzog-Arbeitman2021QGOAI,Slager2025QG_bound_spin_chern}.
In particular, \refcite{Yu2025Z2bound} identified a general Wilson-loop (WL) lower bound of quantum geometry, derived from WL winding.
This naturally raises a fundamental question: can bands that saturate a WL lower bound (different from the Chern/Euler bound) provide new avenues for constructing many-body wavefunctions for correlated phases beyond FCIs—such as fractional topological insulators (FTIs)~\cite{Zhang2006QSH,Levin_Stern,Neupert2011FTI,Stern2015review,Neupert_2015}?

Beyond this conceptual generalization, another critical question is how to apply those ideal-band constructions of many-body wavefunctions in real materials.
In practice, achieving ideal bands is extremely challenging in experiments---we currently have only one confirmed example: the lowest Landau level.
In other experimentally realized systems, ideal bands may be closely approximated, but never exactly realized. 
For instance, in twisted bilayer MoTe$_2$, which has been extensively studied both experimentally~\cite{cai2023signatures,zeng2023integer,park2023observation,Xu2023FCItMoTe2,Ji2024LocalProbetMoTe2,Young2024MagtMoTe2,Kang2024_tMoTe2_2.13,xu2024interplaytopologycorrelationssecond,An_2025_FM_tMoTe2,Jia_2025_SC_tMoTe2,Xu_2025_Interplay_SecondMoireBand_tMoTe2,Kang_2025_TRbreaking_FQSHE,Park_2025_tMoTe2_gap,Xu_2025_SC_tMoTe2,Mak_2025_SC_tWSe2} and theoretically~\cite{wu_topological_2019,yu2020giant,pan_band_2020,zhang_electronic_2021,li2021spontaneous,devakul_magic_2021,morales2023pressure,wang_fractional_2024,reddy_fractional_2023,qiu_interaction-driven_2023,dong2023composite,wang_topological_2023,goldman2023zero,morales2024magic,liu2024gate,xu_maximally_2024,reddy_toward_2023,yu_fractional_2024,abouelkomsan_band_2024,li2024electrically,jia_moire_2024,mao2024transfer,zhang_polarization-driven_2024,wang_topology_2023,li2024contrasting,sheng2024quantum,reddy_NonAbelian_2024,ahn_NonAbelian_2024,wang_Higher_2024,shen2024stabilizing,wang2024phase,xu_Multiple_2024,song2024phase,wu2024time,Kwan_2024_FTI,PhysRevResearch.6.L032063,zaklama2025structure,Zhang2024UniversalMoireModel,Wu_Fengcheng_2025_Topological_magnons_tMoTe2,Goncalves_2025_Quasiparticle_tMoTe2,Wu_Fengcheng_2025_tMoTe2_quasi_particles}, 
the top electron band in one valley at a twist angle of $2.13^\circ$ closely resembles a Chern ideal band, yet its integrated quantum metric still deviates by over 10\% from the lower bound~\cite{wang_Higher_2024,xu_Multiple_2024,reddy_NonAbelian_2024,ahn_NonAbelian_2024,Zhang2024UniversalMoireModel}.
If we can develop a general strategy to achieve ideal bands/states in realistic materials, it would significantly enhance the applicability of corresponding many-body wavefunction constructions for capturing strongly correlated phenomena.

In this work, we address both of these questions.
We first define an isolated set of bands to be WL-ideal if their quantum metric saturates the WL lower bound.
This definition naturally incorporates the known Chern/Euler-ideal bands, and allows us to define new types of ideal bands, such as Kane-Mele $\dsZ_2$-ideal and inversion-fragile-ideal bands, which saturate the lower bound~\cite{Yu2025Z2bound} of quantum metric determined by the Kane-Mele $\dsZ_2$ index~\cite{Kane2005Z2,Zhang2006QSH,Kane2005QSH,Bernevig2006BHZ} protected by spinful time-reversal (TR) symmetry and the inversion-protected fragile topology~\cite{song2020}. 
In particular, we find that in the case of zero total Chern number, an isolated WL-ideal set of two bands always has a Chern-ideal gauge and allows a direct construction of topologically-ordered states, as long as the bands
have non-singular nonabelian Berry curvature and nontrivial normal WL winding.
Specifically, a Chern-ideal gauge means that the two bands can be mixed into two states such that each of them is Chern-ideal, and the topologically-ordered states can be constructed by inserting vortices into those Chern-ideal states in a complex-conjugate way due to their opposite Chern numbers, similar to the construction of FCI wavefunctions by inserting vortices in Chern-ideal bands~\cite{Jie2021IdealBands,Parker2023IdealBands,Dong_2023_ideal_Higher_Chern}.

We further propose a general framework for constructing a monotonic flow of the projector of the periodic parts of the Bloch states, which can be used to construct flows that lead to WL-ideal states.
Specifically, we use the framework to construct three different flows that can start from non-ideal topological bands and drive them to ideal ``bands'' through band mixing.
The final WL-ideal ``bands'' are not necessarily an isolated set of energy bands, and thus we refer to them as WL-ideal states---nevertheless, they still have a smooth projector in momentum space.
As an illustration of the three flows, we apply them to the realistic model of $3.89^\circ$ twisted bilayer MoTe$_2$ in \refcite{Zhang2024UniversalMoireModel} a moir\'e Rashba model adapted from \refcite{Liu_2025_Moire_Rashba} and a moir\'e TR-breaking model that we construct, and we find various numerically Chern-ideal states, numerically $\dsZ_2$-ideal states and numerically inversion-fragile-ideal state, respectively, with relative error in the integrated quantum metric below $5\times 10^{-3}$.
By performing exact diagonalization (ED) calculations on the numerically obtained Chern-ideal states for $t$MoTe$_2$, we find that its FCI state has a dispersion of the anyonic excitations is very similar to that of the FCI state in the top electron band, implying that the analytical construction of many-body states from the ideal states obtained through our flow may faithfully capture the physical behavior of the experimental system in the thermodynamic limit.

\section{Wilson-loop Ideal Bands}

The WL-ideal bands are defined via the WL bound of the integrated quantum metric in \refcite{Yu2025Z2bound}.
Consider an isolated set of $N$ bands in a 2D lattice system, and we label the periodic part of Bloch states as $\ket{u_{\bsl{k}}}=(\ket{u_{\bsl{k},1}}, \ket{u_{\bsl{k},2}}, \ldots, \ket{u_{\bsl{k},N}})$, where $\bsl{k}$ is the Bloch momentum.
The quantum metric of this set of bands is defined as $
[g_{\bsl{k}}]_{ij} = \frac{1}{2}\Tr[\partial_{k_i} P_{\bsl{k}} \partial_{k_j} P_{\bsl{k}}]
$, where $i,j$ ranges over all the  Cartesian directions, and $P_{\bsl{k}} = \ket{u_{\bsl{k}}} \bra{u_{\bsl{k}}} = \sum_{n} \ket{u_{\bsl{k},n}} \bra{u_{\bsl{k},n}}$. 
The Wilson line is generally defined for any continuous path $\gamma$ in the \BZ:
$
W(\gamma) = \lim_{L\rightarrow\infty}\bra{u_{\bsl{k}_1}} P_{\bsl{k}_1} P_{\bsl{k}_2} \cdots P_{\bsl{k}_{L-1}} P_{\bsl{k}_{L}}\ket{u_{\bsl{k}_L}}\ ,
$
where $\bsl{k}_1,\bsl{k}_2,\ldots,\bsl{k}_{L}$ are aligned sequentially along the path $\gamma$. 
The Wilson line becomes a WL when $\gamma$ is closed.
We can continuously deform the loop $\gamma$ to cover certain region $D$ of the $\BZ$, and the WL winding $w$ is then defined for the WL eigenvalues along that deformation.
When $m$ symmetry-related copies of $D$ covers $\BZ$ without overlap, we have the WL bound of $\sqrt{\det(g_{\bsl{k}})}$ and thus $\Tr(g_{\bsl{k}})$, which reads 
\eqa{
\label{main_eq:WL_bound}
\Tr \G  & \geq 2 \text{vol}_g \geq \int_{\BZ}d^2 k\ \rho(F_{\bsl{k}}) \geq 2\pi m |w| \ ,
}
where $\Tr \G = \int_{\BZ}d^2 k\ \Tr(g_{\bsl{k}})$, $\text{vol}_g = \int_{\BZ}d^2 k\ \sqrt{\det(g_{\bsl{k}})}$ is the quantum volume, $\rho(A) = \Tr[\sqrt{A A^\dagger}]$ is the Schatten 1-norm, and $F_{\bsl{k}}$ is the nonabelian Berry curvature.
The bound for a general region (instead of just $\BZ$) is reviewed in  {\appWLbound}.
We note that mathematically the integration of $\Tr(g_{\bsl{k}})$ on a domain $D$ is the Dirichlet functional/energy of the map from $D$ to the space of the projectors of rank $N$, which is called Grassmannian $\text{Gr}(N,\dsC^M)$ with $M$ the number components of eigenvector, as discussed in {\appDirichlet}.

We define an isolated set of bands to be WL-ideal if all inequalities in \cref{main_eq:WL_bound} are saturated.
This recovers the known Chern-~\cite{Jie2021IdealBands,Parker2023IdealBands} and Euler-ideal~\cite{BJY2024EulerBoundQG} cases, when $w$ is the nonzero Chern number~\cite{Roy2014QGChernBound,Bellissard1994QGChernBound} and twice Euler number~\cite{Xie2020TopologyBoundSCTBG,Yu2022EOCPTBG,BJY2024EulerBoundQG,Slager2024EulerOptical} , respectively, for which $m=1$.
Importantly, the definition allows us to go beyond the known Chern/Euler-ideal bands to define ideal bands, such as for Kane-Mele $\dsZ_2$ and inversion fragile topology.
Specifically, for bands with nonzero Kane-Mele $\dsZ_2$ index, we have $m=2$ as the TR symmetry divides the 1BZ into two halves and $w=1$, and the $\dsZ_2$-ideal band saturates \cref{main_eq:WL_bound}, which is simply the $\dsZ_2$ bound proposed in \refcite{Yu2025Z2bound}.
The definition also applies to isolated two-band sets in 2D with inversion-protected fragile topology, characterized by a pair of parity-even (parity-odd) states at only one of the four time-reversal-invariant momenta, with parity-odd (parity-even) states at the other three.
In this case, the inversion symmetry gives $m=2$ and the fragile topology gives $w=1$, and the inversion-fragile-ideal bands saturates \cref{main_eq:WL_bound} which becomes the inversion-fragile bound and has never discussed before. 
The properties of them are studied in the next section.

\section{WL-Ideal Bands with Zero Total Chern number}
\label{sec:ideal_zero_Chern}

In this section, we study the properties of an isolated set of two bands ($\ket{u_{\bsl{k}}}=(\ket{u_{\bsl{k},1}},\ket{u_{\bsl{k},2}})$) that have zero total Chern number.
We focus on the common case where the nonabelian Berry curvature is everywhere nonsingular, \ie, $\det(F_{\bsl{k}})\neq 0$ for all $\bsl{k}$, which implies $\det(g_{\bsl{k}})\neq 0$ as well.
We show that if such bands are WL-ideal with nonzero normal WL winding $w$ in \cref{main_eq:WL_bound}, then they always admit Chern-ideal gauges and hence allow direct construction of many-body topological-order wavefunctions.
Here ``normal'' means that the WL is taken along a primitive reciprocal lattice vector.

Owing to the non-singular nature of the nonabelian Berry curvature and the zero total Chern number, one eigenvalue $f_1(\bsl{k})$ of $F_{\bsl{k}}$ is strictly positive and the other, $f_2(\bsl{k})$, is strictly negative over the entire 1BZ.
Otherwise, at least one eigenvalue would cross zero and render $\det(F_{\bsl{k}})=0$  due to $\int_{\BZ} d^2 k [ f_1(\bsl{k}) + f_2(\bsl{k}) ] = 0$.
The ideal condition of saturating \cref{main_eq:WL_bound} further implies $\int_{\BZ} d^2 k [ f_1(\bsl{k}) - f_2(\bsl{k}) ] = 2\pi m |w|$, which, combined with the zero total Chern number, leads to
\eq{
\label{main_eq:f_l_int}
\frac{1}{2\pi}\int_{\BZ} d^2 k f_l(\bsl{k}) = -(-1)^{l} \frac{ m|w| }{2}\ ,
}
which means $m|w|$ has to be even a priori.

We have so far discussed the eigenvalues of $F_{\bsl{k}}$. 
When does $F_{\bsl{k}}$ become diagonal?
As discussed in {\appidealZtwoband}, we find that we can always find a parallel transport gauge of $\ket{u_{\bsl{k}}}$ to make $F_{\bsl{k}}$ diagonal.
One of such parallel transport gauges is 
\eqa{
\label{main_eq:u_tilde_parallel_transport_expression}
\ket{\widetilde{u}_{\bsl{k}}} & = 
P_{\bsl{k}\leftarrow \frac{k_1}{2\pi}\bsl{b}_1}P_{\frac{k_1}{2\pi}\bsl{b}_1\leftarrow \bsl{0}}\ket{u_{\bsl{0}}} \widetilde{R} \ ,
}
where $\widetilde{R}$ is an $N\times N$ unitary matrix that diagonalizes $F_{\bsl{0}}$, and
$
P_{\bsl{k}'\leftarrow \bsl{k}} = \lim_{L\rightarrow\infty}  P_{\bsl{k}_{L}=\bsl{k}'}  P_{\bsl{k}_{L-1}} \cdots P_{\bsl{k}_2} P_{\bsl{k}_1=\bsl{k}} 
$
with $\bsl{k}_1,\bsl{k}_2,\ldots,\bsl{k}_{L}$  aligned sequentially along the straight line from $\bsl{k}$ to $\bsl{k}'$.
By choosing $\widetilde{R}$ to diagonalize $F_{\bsl{0}}$, we find that the nonabelian Berry curvature calculated for $\ket{\widetilde{u}_{\bsl{k}}}$, labeled as $\widetilde{F}_{\bsl{k}}$, is simply $\diag(f_1(\bsl{k}),f_2(\bsl{k}))$.

Crucially, the ideal condition always leads to the following properties of $\ket{\widetilde{u}_{\bsl{k}}}$:
$
\braket{\widetilde{u}_{2,\bsl{k}}}{\nabla_{\bsl{k}}\widetilde{u}_{1,\bsl{k}}} = 0
$.
As a result, the abelian Berry curvature of $\ket{\widetilde{u}_{l,\bsl{k}}}$ is simply $f_{l}(\bsl{k})$, meaning that the Chern number of $\ket{\widetilde{u}_{l,\bsl{k}}}$, labeled as $\widetilde{\Ch}_l$, takes the value $\widetilde{\Ch}_l = (-1)^{l-1} |w|$ according to \cref{main_eq:f_l_int}.
Furthermore, the total quantum metric of the two bands is simply the sum of the quantum metrics of $\ket{\widetilde{u}_{l,\bsl{k}}}$, \ie, $g_{\bsl{k}} = \sum_{l} g_{l,\bsl{k}}$ with $g_{l,\bsl{k}}$ the quantum metric of $\ket{\widetilde{u}_{l,\bsl{k}}}$.
Combined with the ideal condition of saturating \cref{main_eq:WL_bound}, we arrive at $
\frac{1}{2\pi}\int_{\BZ} d^2 k \Tr[g_{l,\bsl{k}}] = |\widetilde{\Ch}_l| =  m|w| /2
$, meaning that both $\ket{\widetilde{u}_{1,\bsl{k}}}$ and $\ket{\widetilde{u}_{2,\bsl{k}}}$ are Chern-ideal states.
Here we choose not to refer to $\ket{\widetilde{u}_{1,\bsl{k}}}$ and $\ket{\widetilde{u}_{2,\bsl{k}}}$ as Chern-ideal bands, because they are not actual eigenstates of the Hamiltonian, though they still have smooth projectors in the momentum space just like those for isolated bands.
We can also refer to $\ket{\widetilde{u}_{\bsl{k}}}$ as the ideal-Chern gauge of $\ket{u_{\bsl{k}}}$.

Because both $\ket{\widetilde{u}_{1,\bsl{k}}}$ and $\ket{\widetilde{u}_{2,\bsl{k}}}$ are Chern-ideal, we can construct two FCI wavefunctions by vortexing~\cite{Jie2021IdealBands,Parker2023IdealBands,Dong_2023_ideal_Higher_Chern} them by acting $x-\text{sgn}(\widetilde{\Ch}_l) \ii y$.
This produces a pair of FCIs with opposite fractional Hall conductances, which remain topologically ordered so long as charge conservation is preserved, due to the noninvertibility of FCIs~\cite{Cheng_2026_Ordering_TopoOrder}.
The $\dsZ_2$-ideal and inversion-fragile-ideal bands are two examples of such WL-ideal states.
Vortexing the former yields FTIs, while vortexing the latter yields inversion-symmetric topologically ordered states.

\section{Monotonic Flows for Idealization}
\label{sec:monotonic_flow}

With WL-ideal bands defined by saturation of \cref{main_eq:WL_bound}, we now ask how to obtain such ideal bands.
Our approach is to introduce a flow equation for the projector $P_{t,\bsl{k}}$, where $P_{0,\bsl{k}}$ is the projector onto a chosen isolated set of bands, and increasing $t$ gradually mixes in other bands until the projector becomes ideal, where $t$ is the parameter that governs the flow.
As the final ideal $P_{t,\bsl{k}}$ is not necessarily the projector of energy bands, we refer to it as ideal states.

To realize this idealization procedure, we construct flow equations of $P_{t,\bsl{k}}$ by choosing a functional $S_t$ and then taking the functional derivative:
\eq{
\label{main_eq:P_flow_monotonic}
\partial_t P_{t,\bsl{k}} =  \alpha\ P_{t,\bsl{k}} \left[ \sum_{ij}  \partial_{k_i} \left(\frac{\delta S_t}{\delta \left[g_{t,\bsl{k}}\right]_{ij} }\partial_{k_j} P_{t,\bsl{k}} \right) \right] \bar{Q}_{t,\bsl{k}} + h.c.\ ,
}
where  $\alpha>0$, $\bar{Q}_{t,\bsl{k}}$ is the projector to certain chosen states orthogonal to $P_{t,\bsl{k}}$---if we include all states that are orthogonal to $P_{t,\bsl{k}}$, we have $\bar{Q}_{t,\bsl{k}} = Q_{t,\bsl{k}} = 1 - P_{t,\bsl{k}}$.
In particular, we require that the $t$ dependence of $S_t$ only comes from $\left[g_{t,\bsl{k}}\right]_{ij} = \frac{1}{2}\Tr\left[\partial_{k_i} P_{t,\bsl{k}} \partial_{k_j} P_{t,\bsl{k}} \right]$, and the flow maintains the transformation rule of $P_{t,\bsl{k}}$ under shifting reciprocal lattice vectors.
With this requirement, the flow in \cref{main_eq:P_flow_monotonic} $P_{t,\bsl{k}}$ will monotonically decrease $S_t$ as $t$ increases, as the right-hand side is equivalent to projecting the functional derivative of $S_t$ with respect to $P_{t,\bsl{k}}$ to the space of $\partial_t P_{t,\bsl{k}}$.
As long as $\frac{\delta S_t}{\delta \left[g_{t,\bsl{k}}\right]_{ij}}$ is continuous in $\bsl{k}$, the topology of $P_{t,\bsl{k}}$ is unchanged.
(See the more details in {\appmonotonicflow}.)

\cref{main_eq:P_flow_monotonic} provides a general framework for constructing a monotonic flow.
The remaining task is to specify $S_t$ so that, when WL-ideal states exist in the relevant Hilbert space, the flow can reach them.
Our first choice is $
S_t = \Tr\G_t\equiv \int_{1BZ} d^d k\ \Tr[g_{t,\bsl{k}}]
$.
With this choice, the flow in \cref{main_eq:P_flow_monotonic} monotonically decreases the integrated trace of the quantum metric.
This is the continuous version of the disentangling step in the Wannierization procedure of Souza, Marzari, and Vanderbilt~\refcite{Vanderbilt2001MLWFMultibands}, so we call it the SMV flow.
Unlike standard Wannierization, which starts from trial atomic states, here the flow starts from the projector of topological bands.
A useful feature of the SMV flow is its monotonic decrease of $\Tr\G$, but it has no flexibility: for a given initial projector, the final projector is uniquely fixed.

To improve the flexibility, we propose a new flow, dubbed static-target flow by choosing 
$
S_t = \int d^d k  \sum_{ij} \left( \left[\bar{g}_{\bsl{k}}\right]_{ij} - \left[g_{t,\bsl{k}}\right]_{ij}\right)^2
$, 
and another new flow, dubbed dynamical-targe flow, by choosing 
$
S_t = \int d^d k  \sum_{ij} \left( \left[\bar{g}_{t,\bsl{k}}\right]_{ij} - \left[g_{t,\bsl{k}}\right]_{ij}\right)^2 
$,
where the $\bar{g}_{\bsl{k}}$ and $\bar{g}_{t,\bsl{k}}$ serve as the static and dynamic targets for the quantum metric.
Different flows can lead to different final ideal states even if the starting states are the same.

\begin{figure}[t]
    \centering
    \includegraphics[width=\linewidth]{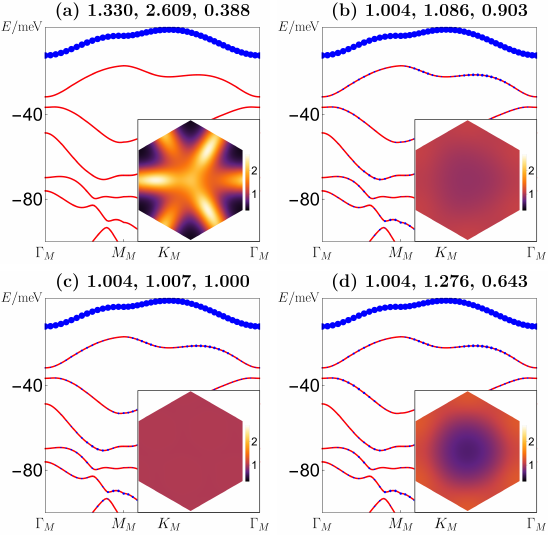}
    \caption{The Chern-ideal states obtained from different geometric flows in the one-valley model of twisted bilayer MoTe$_2$ at twist angle $\theta = 3.89^\circ$. 
    In all plots, the red lines are the band structure of the model.
    The insets show the distributions of the trace of quantum metrics $g_{\bsl{k}}$ of (a) the top electron band, (b) the Chern-ideal states obtained from the SMV flow, (c) the Chern-ideal states obtained from the static-target flow with target $\left[\bar{g}_{\bsl{k}}\right]_{ij} = 2 \pi m|w|/(\Omega^*) \delta_{ij}$, and (d) the Chern-ideal states obtained from the dynamical-target flow with target $\left[\bar{g}_{t,\bsl{k}}\right]_{ij} = 2 m |w| \Tr[g_{t,\bsl{k}}]/(2 \Tr\G_t) \delta_{ij}$.
    The three numbers in each caption of (a-d) are (from left to right) the integral, the maximum and the minimum of the trace of quantum metric over the 1BZ, where the integral is divided by $2\pi$.
    $g_{\bsl{k}}$ is always measured in unit of $\Omega/(2\pi)$ with $\Omega$ the moir\'e unit cell area.
    The area of each blue dot on the band structure is proportional to the amplitude (square root of probability) of the inset states at that momentum and energy.
    }
    \label{fig:tMoTe2}
\end{figure}

\begin{figure}[t]
    \centering
    \includegraphics[width=1.0\linewidth]{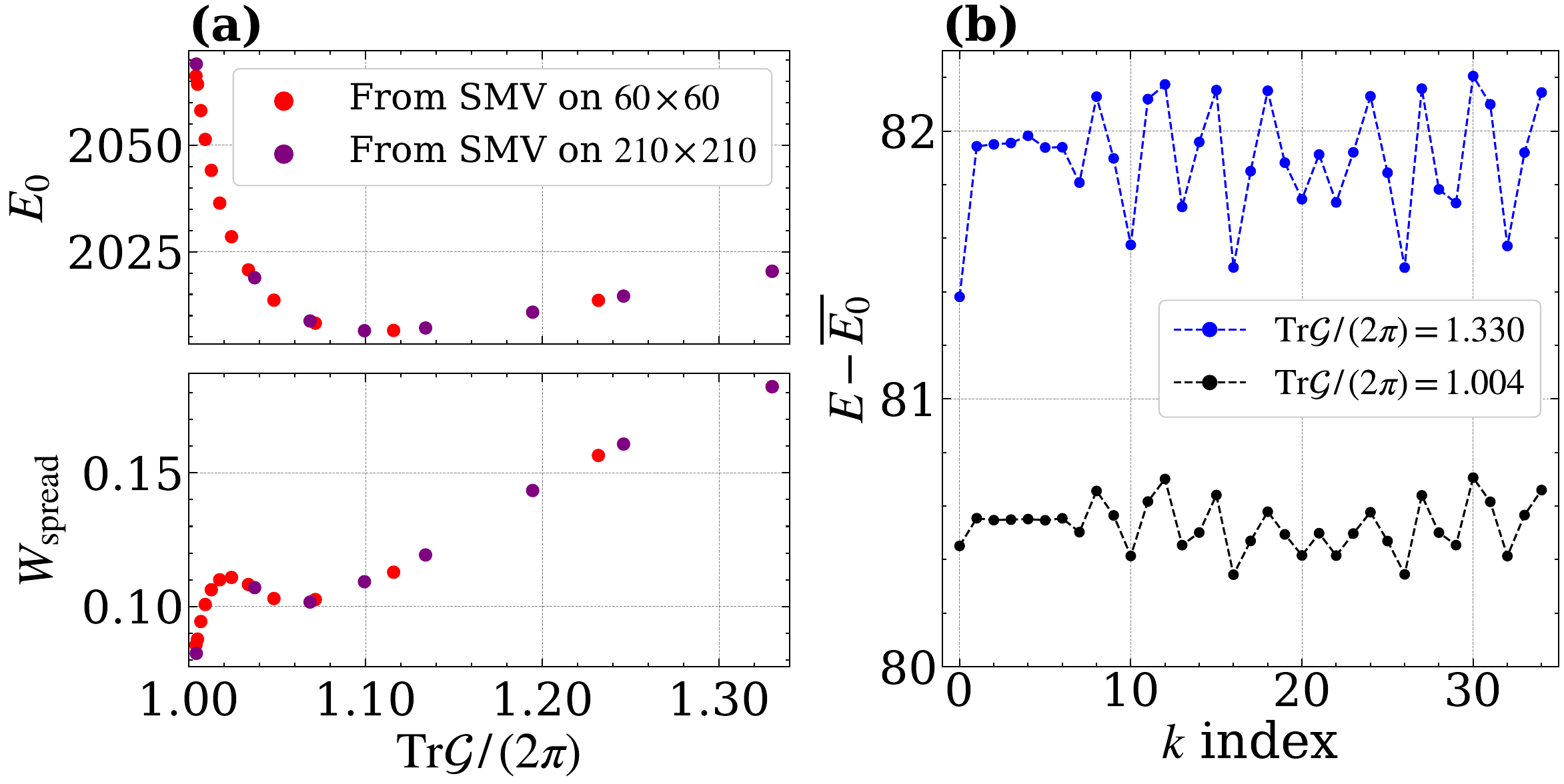}
    \caption{We show the results of the ED calculation for the twisted bilayer MoTe$_2$ by projecting the many-body Hamiltonian to the subspace determined by the single-particle projector $P_{\bsl{k}}$ in the SMV flow.
    In (a), ED calculation is performed with 24 holes on the $6\times 6$ mesh, with the subspace determined by $P_{\bsl{k}}$ in the $60\times 60$ (red) and $210\times 210$ (blue) SMV flows;.
    $\Tr\G/(2\pi)$ of $P_{\bsl{k}}$ is shown on the horizontal axis. 
    $E_0$ is the ground state energy, and $W_{spread}$ is the spread of the three topologically nearly degenerate states.
    In (b), we plot the charge $2e/3$ spectrum given by the 24-hole $5\times 7$ ED calculation in the subspace determined by the initial (blue) and numerically-ideal (black) $P_{\bsl{k}}$ from the $210\times 210$ SMV flow, where $-e$ is the charge of electron, and $\overline{E_0}$ is the average energy of the three topologically (nearly) degenerate states in the 24-hole $6\times 6$ ED following the method in \refcite{Goncalves_2025_Quasiparticle_tMoTe2}.
    $k$ index is the linearized index for many-body momentum.
    }
    \label{fig:SMV_ED}
\end{figure}

\section{Model Studies}

We now apply the three different flows in three moir\'e models.

The first model is the faithful model for twist angle $\theta=3.89^\circ$ provided in \refcite{Zhang2024UniversalMoireModel}, which is constructed directly from density functional theory (DFT) calculations without any continuous parameter fitting.
Twisted bilayer MoTe$_2$ consists of two valleys, $\K$ and $\K'$, related by TR symmetry, and we only focus on $\K$ valley here.
The form of the model and the details for the calculation are shown in {\apptMoTe}, and here we summarize the key results.
The band structure of the model is shown in \cref{fig:tMoTe2}.
The top electron band in \cref{fig:tMoTe2} has $\Ch=1$ and $\Tr\G/(2\pi) = 1.330$ (\cref{fig:tMoTe2}(a)), which is not Chern-ideal.

Now we apply the three flows---the SMV flow, the static-target flow, and the dynamical-target flow---discussed in \cref{sec:monotonic_flow} to the top electron band in \cref{fig:tMoTe2}(a), \ie, choosing the starting $P_{t=0,\bsl{k}}$ in \cref{main_eq:P_flow_monotonic} to be the projector of the top electron band.
Note that for all flows, we allow band mixing between all bands, \ie, choosing $\bar{Q}_{t,\bsl{k}} = Q_{t,\bsl{k}} = 1 - P_{\bsl{k}}$ in \cref{main_eq:P_flow_monotonic}.
The results are shown in \cref{fig:tMoTe2}(b-d).
All three flows achieve numerically Chern-ideal states with relative error $(\Tr\G - 2\pi|\Ch| )/(2\pi |\Ch|) < 5\times 10^{-3}$.
Nevertheless, the static-target flow achieves a quantum metric that is most uniform, which makes sense since the target is the uniform ideal quantum metric, while the dynamical-target flow has the most fluctuations.
Interestingly, the SMV flow achieves the ideal states with least band mixing (or largest remaining probability in the top electron band): the final projectors has probability of 0.942, 0.939 and 0.926 remaining in the top electron bands for the SMV flow, the static-target flow and the dynamical-target flow, respectively.
We also perform the SMV flow on a $210\times 210$ mesh, and converge to a very similar numerical ideal states, demonstrating the numerical convergence.

To demonstrate the potential usefulness of our flow for the study of many-body physics, we add double-gate screened Coulomb interaction (with gate distance $20$nm and relative dielectric constant $10$) and perform 
exact-diagonalization (ED) 
on $6\times 6$ and $5\times 7$ clusters
for the projected many-body hole Hamiltonian in the subspace determined by single-particle projectors given by the SMV flow on $60\times 60$ and/or $210\times 210$ meshes.
As shown in \cref{fig:SMV_ED}(a), we find that the flow can help us find FCI states with lower many-body energy and better topological degeneracy than that obtained from the top electron band.
The lowest many-body energy occurs at smaller $\Tr\G$ when interaction becomes larger (\apptMoTe).
The fact that the band mixing can lower many-body energy  is because the interaction strength is often larger than the band gap and bandwidth in moir\'e systems~\cite{yu_fractional_2024,xu_maximally_2024,abouelkomsan_band_2024}.
In particular, \cref{fig:SMV_ED}(b) shows that the dispersion of the anyon excitations is quite similar for the FCI state in the numerically ideal state and that in the top electron band, indicating that the analytical construction of FCI from the ideal state obtained from our flow can faithfully capture the physical behaviors of the experimental system in the thermodynamic limit.
The same conclusion holds for the neutral excitations, and interestingly, the anyon dispersion has less bandwidth for smaller $\Tr\G$ even if the single-particle dispersion is neglected.
(See more details in \apptMoTe.)

We also consider an adapted moiré Rashba model whose lowest two bands have $\nu_{\dsZ_2} = 1$, and another moiré TR-breaking model whose two bands closest to zero energy have inversion-protected fragile topology. 
We apply the flows discussed in \cref{sec:monotonic_flow} to the topologically nontrivial bandsm and achieve numerically $\dsZ_2$-ideal states and inversion-fragile-ideal states, respectively, with relative error $(\Tr\G - 2\pi |w|  )/(2\pi |w|) < 5 \times 10^{-3}$.
(See details in {\appmoirerashbaandinversion}.)

\section{Discussion}

In sum, we have provided a general definition of WL-ideal bands and a general framework of idealization.
Here we apply our projector flows only to moiré models, which in principle have an infinite number of bands.
One future direction is to apply our flows to models with a finite number of bands, \eg, tight-binding models.
Recently, the ideal topological heavy fermion models~\cite{Liu_2025_idealTHF} are constructed, and they host Chern-ideal bands.
It would be interesting to see whether the topological heavy fermion models can be constructed to realize other types of WL-ideal bands/states. 
Furthermore, one can choose functional $S_t$ to be minimized at certain target quantum metrics that are not WL-ideal, and then drive the projector to those non-ideal states.
It is worth studying whether such a procedure can realize states that are analogous to higher Landau levels.

\section{Acknowledgment}

A.A.A.'s work is supported by the AI Scholars program at the University of Florida and by the UF CCMS Undergraduate Fellowship. B. L. is supported by the National Science Foundation through Princeton University’s Materials Research Science and Engineering Center DMR-2011750, and the National Science Foundation under award DMR-2141966.
S.R. is supported by a Simons Investigator Grant from the
Simons Foundation (Award No. 566116).
This work is supported by the Gordon and Betty Moore 
Foundation EPiQS initiative, Grant GBMF8685.01.
J. Y.'s work is supported by startup funds at University of Florida.

\bibliography{bibfile_references}

\appendix
\onecolumngrid
\clearpage

\tableofcontents

\section{Integrated Traced Quantum Metric as Dirichlet Energy/Functional}

In this section, we show that the integrated traced quantum metric
\eq{
\Tr\G = \int_{\BZ} d^2k \Tr[g_{\bsl{k}}] 
}
is a simplified expression of the Dirichlet energy/function for the map $P_{\bsl{k}}$ from the $\BZ$ to the space of the projectors.

We first define the Dirichlet energy/function between two Riemannian manifolds $A$ and $B$ with well defined metric $d$ and $h$, respectively.
Explicitly, given a local coordinate $x^i$ for a neighbor hood in $A$, we have $ds^2 = \sum_{ij} d_{ij}(x) dx^i dx^j$; similarly, given a local coordinate $y^\mu$ for a neighbor hood in $B$, we have $ds^2 = \sum_{\mu \nu} h_{\mu \nu}(y) dy^\mu dy^\nu$.
For any smooth map $f:A\rightarrow B$, the Dirichlet energy/functional on an open neighborhood $M\subset A$ is defined as
\eq{
\label{eq:Dirichlet}
E(f) = \int_M dvol_x  \sum_{ij}\sum_{\mu\nu}d^{ij}(x) h_{\mu \nu}(\widetilde{f}^{\mu}(x)) \frac{\partial \widetilde{f}^{\mu}(x)}{\partial x^i} \frac{\partial  \widetilde{f}^{\nu}(x)}{\partial x^j}\ ,
}
where $vol_x$ is the volume element, $d^{ij}$ is the inverse of $d_{ij}$, $\widetilde{f}$ is the map between local coordinates induced by $f$, \ie,
\eq{
\widetilde{f}(x(a)) = y(f(a))
}
for all $a\in M$, $x$ the local coordinate on $M$ and $y$ the local coordinate on $f(M)$.
Clearly, \cref{eq:Dirichlet} is coordinate invariant since the indices are contracted properly.

The $\BZ$ is a Riemannian manifold: we can choose flat coordinate $\bsl{k}$ with flat metric $d_{ij}(\bsl{k}) = \delta_{ij}$.
The space of the projectors for $n$ bands of a $M\times M$ Hamiltonian (at each $\bsl{k}$) is the Grassmannian $\text{Gr}(n,C^{M})$, for which we can inherit the metric from the ambient Hermitian matrix space.
The coordinate of $\text{Gr}(n,C^{M})$ can be expressed as $y(P)$ that satisfies $P = \frac{n}{M} \mathbb{1}_{M\times M} + \sum_\mu y^\mu(P) \Lambda_mu $ with $\Lambda_\mu$ $(\mu=1,...,M^2-1)$ the traceless basis for the $M\times M$ Hermitian matrix.
Then, the metric in this coordinate system reads
\eq{
h_{\mu \nu} = \frac{1}{2}\Tr[\Lambda_\mu \Lambda_\nu ]\ ,
}
which means
\eq{
ds^2 = \sum_{\mu\nu} h_{\mu \nu} dy^{\mu} dy^{\nu} = \frac{1}{2}\Tr[\Lambda_\mu \Lambda_\nu ] dy^{\mu} dy^{\nu} = \frac{1}{2}\Tr[dP dP]\ .
}

For the map $f: \BZ \rightarrow \text{Gr}(n,C^{M})$ with $\widetilde{f}(\bsl{k})$ satisfying $P_{\bsl{k}} = \frac{n}{M} \mathbb{1}_{M\times M} + \sum_\mu  \widetilde{f}^{\mu}(\bsl{k}) \Lambda_\mu $.
Then, the corresponding Dirichlet energy/functional reads
\eq{
\label{eq:TrG_as_Dirichlet}
E(f) = \int_{\BZ} d^2 k  \sum_{ij}\sum_{\mu\nu}\delta^{ij} \frac{1}{2}\Tr[\Lambda_\mu \Lambda_\nu ]  \frac{\partial \widetilde{f}^{\mu}(\bsl{k})}{\partial k^i} \frac{\partial  \widetilde{f}^{\nu}(\bsl{k})}{\partial k^j}= \int_{\BZ} d^2 k  \sum_{ij}\delta^{ij} \frac{1}{2}\Tr[\frac{\partial P_{\bsl{k}}}{\partial k^i} \frac{\partial  P_{\bsl{k}}}{\partial k^j}] = \int_{\BZ} d^2 k  \Tr[g_{\bsl{k}}] = \Tr\G  \ .
}
Therefore, $\Tr\G $ is simply the Dirichlet energy/functional under the flat coordinates for $\BZ$.

\section{Wilson-Loop Bound of Integrated Quantum Metric}
\label{app:WL_bound}

In this appendix, we briefly review the proof for the Wilson-loop (WL) bound of the integrated quantum metric in \refcite{Yu2025Z2bound}, review the known examples of Chern, Euler and Kane-Mele $\dsZ_2$ bounds, and propose a new inversion-fragile bound.

Consider an isolated set of $N$ bands in a 2D system, and we label the periodic part of Bloch states as $\ket{u_{\bsl{k}}}=(\ket{u_{\bsl{k},1}}, \ket{u_{\bsl{k},2}}, \ldots, \ket{u_{\bsl{k},N}})$.
Here $\bsl{k}$ is the Bloch momentum taking the form
\eq{
\bsl{k} = \frac{k_1}{2\pi}\bsl{b}_1 + \frac{k_2}{2\pi}\bsl{b}_2\ ,
}
where $\bsl{b}_1$ and $\bsl{b}_2$ are the primitive reciprocal lattice vectors.
The Wilson line is defined as 
\eq{
W(\gamma) = \lim_{L\rightarrow\infty}\bra{u_{\bsl{k}_i}} P_{\bsl{k}_1} P_{\bsl{k}_2} \cdots P_{\bsl{k}_{L-1}} P_{\bsl{k}_{L}}\ket{u_{\bsl{k}_f}}\ ,
}
where $P_{\bsl{k}} = \ket{u_{\bsl{k}}}\bra{u_{\bsl{k}}} $, $\gamma$ is a continuous path of $\bsl{k}$, $\bsl{k}_i$ is at the start of $\gamma$, $\bsl{k}_f$ is at the end of $\gamma$, and $\bsl{k}_1,\bsl{k}_2,\ldots,\bsl{k}_{L}$ are aligned sequentially along $\gamma$.

To derive the WL bound, we need to first consider a proper deformation of a simply connected region.
Specifically, consider a simply connected region that depends smoothly on a continuous parameter $s\in[0,s_f]$, labeled as $D_s$, with $D_0$ having zero area.
Here, being proper means that $D_s\subset D_{s'}$ for any $s\leq s'$.

At a specific value of $s$, $W(\partial D_s)$ is the WL along the boundary of the $D_s$ counter-clockwise.
We choose the initial point of $W(\partial D_s)$  to be $\bsl{k}_s \in \partial D_s$, and we choose $\bsl{k}_s$ to be a smooth function of $s \in [0,s_f]$, and label the path of $\bsl{k}_s$ as a function of $s$: 
\eq{
\gamma_{ref} = \{ \bsl{k}_s | s \in [0,s_f] \}\ .
}
The WL is related to the nonabelian Berry curvature by the nonabelian Stokes' theorem~\cite{Arefeva1980NonAbelianstokes}.
To fully exploit the nonabelian Stokes' theorem, we need to choose a global reference point, which we choose to be $\bsl{k}_0$.
Then, we define $\overline{\partial D_s}$ to be the following path
\eq{
\label{eq:partial_D_s_completion}
\overline{\partial D_s} = \bsl{k}_0 \xrightarrow{\gamma_{ref}} \bsl{k}_s \xrightarrow{\partial D_s} \bsl{k}_s \xrightarrow{\gamma_{ref}} \bsl{k}_0\ ,
}
where $\bsl{k}_s \xrightarrow{\gamma_{ref}} \bsl{k}_{s'}$ means 
the path from $\bsl{k}_s $ to $\bsl{k}_{s'}$ along $\gamma_{ref}$, and $\xrightarrow{\partial D_s} $ means that the path is along $\partial D_s$ counter-clockwise.
Then, $W(\overline{\partial D_s})$ is unitarily related to $W(\partial D_s)$ as 
\eq{
\label{eq:W_bar_partial_Ds}
W(\overline{\partial D_s}) = W(\bsl{k}_0 \xrightarrow{\gamma_{ref}}  \bsl{k}_s) W(\partial D_s) W^{-1}(\bsl{k}_0 \xrightarrow{\gamma_{ref}}  \bsl{k}_s)\ .
}

We can now relate $W(\overline{\partial D_s})$ to non-abelian Berry curvature using the nonabelian Stokes' theorem.
To do so, we should parameterize $\overline{\partial D_s}$ as $\bsl{k}(s,t)$ with $s\in[0,s_f]$ and $t\in[-t_0, 1 + t_0]$,  which satisfies 
\eqa{
\label{eq:k_s_t_parametrization}
& \text{$\bsl{k}(s,-t_0) = \bsl{k}(s,1+t_0) = \bsl{k}_0$ with $t_0\geq 0 $},\\
& \text{$\bsl{k}(s,0) = \bsl{k}(s,1) = \bsl{k}_s$ },\\
& \text{$\bsl{k}(s,t) =\bsl{k}_{\frac{t+t_0}{t_0} s} $ for $t\in [-t_0,0)$, which goes through $\gamma_{ref}$ from $\bsl{k}_0$ to $\bsl{k}_s$ continuously as $t$ increases from $-t_0$ to 0}, \\
& \text{$\bsl{k}(s,t)$ goes through $\partial D_s$ counter-clockwise from $\bsl{k}_s$ to $\bsl{k}_s$ continuously as $t$ increases from 0 to 1}, \\
& \text{$\bsl{k}(s,t) = \bsl{k}_{\frac{1+t_0-t}{t_0} s}$ for $t\in ( 1, 1+t_0 ]$, which goes through $\gamma_{ref}$ from $\bsl{k}_s$ to $\bsl{k}_0$ continuously as $t$ increases from $1$ to $1+t_0$} \ .
}
Then, $W(\overline{\partial D_{s+ds}})W^\dagger(\overline{\partial D_{s}})$ is the WL along the following path
\eq{
\bsl{k}(s,-t_0) \rightarrow  \bsl{k}(s+ds,-t_0) \xrightarrow{\overline{\partial D_{s+ds}}} \bsl{k}(s+ds,1+t_0) 
\rightarrow \bsl{k}(s,1+t_0) \xrightarrow{\overline{\partial D_{s}}} \bsl{k}(s,-t_0) \ ,
}
which in the space of $(s,t)$ is simply the boundary of 
\eq{
\left\{ (s,t) | s\in [ s, s+ds ], t \in [-t_0,1+t_0] \right\}\ .
}
Here $\xrightarrow{\overline{\partial D_s}}$ means that the path is along $\overline{\partial D_s}$ in \cref{eq:partial_D_s_completion}.
Then, according to the nonabelian Stokes' theorem, we have
\eq{
W(\overline{\partial D_{s+ds}})W^\dagger(\overline{\partial D_{s}}) = \prod_{l=0}^{L-1}\h_{s, l dt}e^{-\ii F_{s,l dt} ds dt + O(ds^3, ds^2 dt, ds dt^2, dt^3) }\h_{s,l dt}^{-1} \ ,
}
where $dt = 1/L$ with large positive integer $L$,
\eq{
\h_{s, t} \equiv W(\bsl{k}(s,-t_0) \xrightarrow{\overline{\partial D_s}}\bsl{k}(s,t)) = W(\bsl{k}_0 \xrightarrow{\overline{\partial D_s}}  \bsl{k}(s,t))\ ,
} 
and
\eq{
\F_{s,t} =  \partial_{s}  \A_2(s,t) -  \partial_{t}  \A_1(s,t)  - \ii \left[ \A_1(s,t) , \A_2(s,t)\right]
}
with
\eq{
\bsl{\A}(s,t) = \left( \ii \bra{u_{\bsl{k}}} \partial_{s} \ket{u_{\bsl{k}}} ,  \ii \bra{u_{\bsl{k}}} \partial_{t} \ket{u_{\bsl{k}}} \right)\ .
}
Taking the limit $ds\to 0$ gives
\eq{
[\partial_s W(\overline{\partial D_{s}})]W^\dagger(\overline{\partial D_{s}}) = -\ii\sum_{l=0}^{L-1} \h_{s, l dt} \F_{s,l dt} \h_{s, l dt}^{-1} dt + O(dt) \ ,
}
and then sending $L$ to infinity gives
\eq{
\label{eq:W_to_F_nonabelian_stokes_intermediate}
[\partial_s W(\overline{\partial D_{s}})]W^\dagger(\overline{\partial D_{s}}) = -\ii \int_{-t_0}^{1+t_0} dt\ \widetilde{F}_{s,t} \ ,
}
where
\eq{
\label{eq:F_tilde_s_t}
\widetilde{\F}_{s,t} = \h_{s, t }\F_{s,t} \h_{s, t}^{-1}\ .
}

Before moving forward, we note that the integration over $t$ on the right-hand side of \cref{eq:W_to_F_nonabelian_stokes_intermediate} can be limited to $[0,1]$, while the equation still holds.
It is obviously true for $t_0 = 0$.
For $t_0>0$, we now show that
\eq{
\F_{s,t} = 0 
}
for $t\in[-t_0,0)$ and $t\in(1, 1+t_0]$.
It is because for $t\in[-t_0,0)$, we have
\eq{
\bsl{k}(s,t) = \bsl{k}_{\frac{t+t_0}{t_0} s}\ ,
}
which means
\eqa{
& \partial_{t} \ket{u_{\bsl{k}}} = \frac{s }{t+t_0}\partial_s \ket{u_{\bsl{k}}} \\
& \Rightarrow \left\{ 
\begin{array}{c}
\A_2(s,t)= \frac{s }{t+t_0}  \A_1(s,t) \\
 \braket{\partial_{s}u_{\bsl{k}}}{\partial_{t}u_{\bsl{k}}} = \braket{\partial_{t}u_{\bsl{k}}}{\partial_{s}u_{\bsl{k}}} =  \frac{s }{t+t_0} \braket{\partial_{s}u_{\bsl{k}}}{\partial_{s}u_{\bsl{k}}} \\
\end{array}
\right. \\
& \Rightarrow 
\left\{ 
\begin{array}{c}
  [\A_1(s,t), \A_2(s,t)] =  [\A_1(s,t), \frac{s }{t+t_0} \A_1(s,t)] = 0 \\
  \partial_s \A_2(s,t) - \partial_t \A_1(s,t) = 0 \\
\end{array}
\right.\\
& 
 \Rightarrow  \F_{s,t} = 0 \ .
}
Similarly, for $t\in[1,1+t_0)$, we have
\eq{
\bsl{k}(s,t) = \bsl{k}_{\frac{1+t_0-t}{t_0} s}\ ,
}
which means
\eqa{
& \partial_{t} \ket{u_{\bsl{k}}} = \frac{s }{1+t_0-t}\partial_s \ket{u_{\bsl{k}}} \\
& 
 \Rightarrow  \F_{s,t} = 0 \ .
}
Therefore, we have
\eq{
\label{eq:W_to_F_nonabelian_stokes}
[\partial_s W(\overline{\partial D_{s}})]W^\dagger(\overline{\partial D_{s}}) = -\ii \int_{0}^{1} dt\ \widetilde{F}_{s,t} \ .
}

The next step is relating \eqnref{eq:W_to_F_nonabelian_stokes} to the quantum metric.
To do so, we need to first use the Schatten 1-norm $\rho(A) = \Tr\left[\sqrt{A^\dagger A}\right]$:
\eqa{
\rho\left([\partial_s W(\overline{\partial D_{s}})]W^\dagger(\overline{\partial D_{s}})\right) & = \rho( W^\dagger(\overline{\partial D_{s}})\partial_s W(\overline{\partial D_{s}})) = \rho\left(\int_0^1 dt\ \widetilde{\F}_{s,t} \right) \leq \int_0^1 dt \ \rho\left(\widetilde{\F}_{s,t} \right) = \int_0^1 dt \rho\left(\F_{s,t} \right) \ .
}
Further integrating the expression over $s$ leads to
\eq{
\int_0^{s_f} ds\ \rho( W^\dagger(\overline{\partial D_{s}})\partial_s W(\overline{\partial D_{s}}))  \leq \int_0^{s_f} ds \int_0^1 dt\ \rho\left(\F_{s,t} \right) \ .
}
Now we transform back to $\bsl{k}$.
Note that
\eqa{
\bsl{\A}(s,t) = ( \sum_{i}  A_i(\bsl{k})  \frac{\partial k_i}{\partial s} , \sum_{i}  A_i(\bsl{k})  \frac{\partial k_i }{\partial t} )\ ,
}
and
\eqa{
\F_{s,t} & = \ii \braket{ \partial_s u_{\bsl{k}} }{ \partial_t u_{\bsl{k}} } - \ii \braket{ \partial_t u_{\bsl{k}} }{ \partial_s u_{\bsl{k}} }  - \ii \left[\sum_{i}  A_i(\bsl{k})  \frac{\partial k_i}{\partial s} , \sum_{j}  A_j(\bsl{k})  \frac{\partial k_j }{\partial t}\right] \\
& = \left\{ \ii \braket{ \partial_{k_x} u_{\bsl{k}} }{ \partial_{k_y} u_{\bsl{k}} } - \ii \braket{ \partial_{k_y} u_{\bsl{k}} }{ \partial_{k_x} u_{\bsl{k}} }  - \ii \left[A_x(\bsl{k})  , A_y(\bsl{k}) \right] \right \}(\frac{\partial k_x}{\partial s} \frac{\partial k_y}{\partial t} - \frac{\partial k_x}{\partial t} \frac{\partial k_y}{\partial s}) \\
& = \det\left(\frac{\partial (k_x,k_y)}{\partial (s,t)}\right) F_{\bsl{k}}\ ,
}
where $i=x,y$,
\eq{
A_i(\bsl{k}) =  \ii \bra{u_{\bsl{k}}} \partial_{k_i} \ket{u_{\bsl{k}}} \ ,
}
\eq{
 F_{\bsl{k}} = \ii \braket{ \partial_{k_x} u_{\bsl{k}} }{ \partial_{k_y} u_{\bsl{k}} } - \ii \braket{ \partial_{k_y} u_{\bsl{k}} }{ \partial_{k_x} u_{\bsl{k}} }  - \ii \left[A_x(\bsl{k})  , A_y(\bsl{k}) \right] \ ,
}
and
\eq{
\det\left(\frac{\partial (k_x,k_y)}{\partial (s,t)} \right) = \frac{\partial k_x}{\partial s} \frac{\partial k_y}{\partial t} - \frac{\partial k_x}{\partial t} \frac{\partial k_y}{\partial s}\ .
}
Then,
\eq{
\int_0^{s_f} ds\ \rho( W^\dagger(\overline{\partial D_{s}})\partial_s W(\overline{\partial D_{s}}))  \leq  \int_0^{s_f} ds \int_0^1 dt\ \left|\det\left( \frac{\partial (k_x,k_y)}{\partial (s,t)} \right)\right|\ \rho\left(F_{\bsl{k}} \right) = \int_{D_{s_f}} d^2 k \ \rho\left(F_{\bsl{k}} \right)\ .
}

Owing to \cref{eq:W_bar_partial_Ds} and the triangle inequality of the norm, we know  
\eq{
\rho( W^\dagger(\overline{\partial D_{s}})\partial_s W(\overline{\partial D_{s}})) \geq \rho( W^\dagger({\partial D_{s}})\partial_s W({\partial D_{s}})) \ .
}
Combined with
\eq{
\label{eq:g_lower_bound_by_rho_F}
\Tr[g_{\bsl{k}}] \geq 2\sqrt{\det(g_{\bsl{k}})} \geq \rho(F_{\bsl{k}})\ ,
}
we have
\eq{
\label{eq:WL_bound_intermediate}
\int_0^{s_f} ds\ \rho( W^\dagger(\partial D_{s})\partial_s W(\partial D_{s}))  \leq  \int_{D_{s_f}} d^2 k \ \rho\left(F_{\bsl{k}} \right) \leq \int_{D_{s_f}} d^2 k \ 2\sqrt{\det(g_{\bsl{k}})}  \leq \int_{D_{s_f}} d^2 k \ \Tr[g_{\bsl{k}}]\ ,
}
where
\eq{
\left[g_{\bsl{k}}\right]_{ij} = \frac{1}{2} \Tr\left[ \partial_{k_i} P_{\bsl{k}} \partial_{k_j} P_{\bsl{k}} \right]
}
is the quantum metric.

To further relate to any WL winding, we first specify the degrees of freedom here.
Define a unitary $\W_s$ such that 
\eq{
\label{eq:W_partial_Ds_W_s}
W(\partial D_{s}) = U_s \W_s  U_s^\dagger V\ ,
}
where $U_s$ and $V$ are unitary, $U_s$ smoothly depends on $s$, and $V$ is independent of $s$.
Then, we have
\eq{
\rho( W^\dagger(\partial D_{s})\partial_s W(\partial D_{s})) = \rho( [U_s \W_s  U_s^\dagger]^\dagger \partial_s [U_s \W_s  U_s^\dagger])\ .
}
Define $\phi_{n}(s)$ (with $n=1,2,3,\ldots,N$) to be the continuous phase of the $n$th eigenvalue of $\W_s$, and then we obtain
\eq{
\rho( W^\dagger(\partial D_{s})\partial_s W(\partial D_{s})) = \rho( [U_s \W_s  U_s^\dagger]^\dagger \partial_s [U_s \W_s  U_s^\dagger]) \geq \rho(  \W_s^\dagger \partial_s \W_s ) \geq \sum_n |\partial_s \phi_{n}(s)|\ .
}
With the absolute WL winding defined as
\eq{
\N = \int_{0}^{s_f} ds \sum_n |\partial_s \phi_{n}(s)|\ ,
}
we arrive at 
\eq{
\N \leq \int_0^{s_f} ds\ \rho( W^\dagger(\partial D_{s})\partial_s W(\partial D_{s}))  \leq  \int_{D_{s_f}} d^2 k \ \rho\left(F_{\bsl{k}} \right) \leq \int_{D_{s_f}} d^2 k \ 2\sqrt{\det(g_{\bsl{k}})}  \leq \int_{D_{s_f}} d^2 k \ \Tr[g_{\bsl{k}}]\ .
}
Normally, the WL winding is defined by 
\eq{
\label{eq:w}
w = \frac{1}{2\pi}\int_{0}^{s_f} ds \sum_n  \lambda_n(s) \partial_s \phi_{n}(s)\ ,
}
where $\lambda_n(s) = \pm 1$ depending on the topological invariant of interest, and we choose $\phi_{n}(0)$ to be in $[0,2\pi)$.
Then, we eventually obtain the WL bound of the integrated quantum metric for a generic proper deformation $D_s$ and arbitrary dressing $U_s$ and $V$:
\eq{
\label{eq:WL_bound}
2 \pi |w| \leq \N \leq \int_0^{s_f} ds\ \rho( W^\dagger(\partial D_{s})\partial_s W(\partial D_{s}))  \leq  \int_{D_{s_f}} d^2 k \ \rho\left(F_{\bsl{k}} \right) \leq \int_{D_{s_f}} d^2 k \ 2\sqrt{\det(g_{\bsl{k}})}  \leq \int_{D_{s_f}} d^2 k \ \Tr[g_{\bsl{k}}]
}

\subsection{Review: Chern Bound}
\label{app:Chern_bound}
The simplest example is the Chern bound~\cite{Roy2014QGChernBound,Bellissard1994QGChernBound}.
For Chern number, we can choose $D_s = \{ \bsl{k} = \bsl{k}_0 + \sum_{l=1,2} k_l/(2\pi) \bsl{b}_l | k_1 \in [ 0,s ], k_2 \in [ 0,2\pi ]  \}$, where $\bsl{k}_0 = \sum_{l=1,2} k_{0,l}/(2\pi) \bsl{b}_l $ is the reference point of $W(\partial D_s)$ for all $s$ values, and $\bsl{b}_1,\bsl{b}_2$ are two primitive reciprocal lattice vectors.
Since $\bsl{k}_s = \bsl{k}_0$, we can simply choose $t_0=0$ in \cref{eq:k_s_t_parametrization}.
We choose $s_f=2\pi$, and $D_{2\pi}$ effectively covers the whole 1BZ.
In this case, $W(\partial D_s)$ has the expression
\eq{
\label{eq:W_Ds_Chern}
W(\partial D_s) = W(\bsl{k}_0 \rightarrow \bsl{k}_0 + \frac{s}{2\pi}  \bsl{b}_1) W(\bsl{k}_0 + \frac{s}{2\pi}  \bsl{b}_1 \rightarrow \bsl{k}_0 + \frac{s}{2\pi}  \bsl{b}_1 + \bsl{b}_2) W(\bsl{k}_0 + \frac{s}{2\pi}  \bsl{b}_1 + \bsl{b}_2\rightarrow \bsl{k}_0 + \bsl{b}_2 ) W(\bsl{k}_0 + \bsl{b}_2 \rightarrow \bsl{k}_0  )\ .
}
Owing to the $\ket{u_{\bsl{k}+\bsl{G}}} = e^{-\ii \bsl{G}\cdot \bsl{r}}\ket{u_{\bsl{k}}} $ with $\bsl{r}$ being the position operator, we have
\eq{
W(\bsl{k}_0 + \frac{s}{2\pi}  \bsl{b}_1 + \bsl{b}_2\rightarrow \bsl{k}_0 + \bsl{b}_2 ) = W(\bsl{k}_0 + \frac{s}{2\pi}  \bsl{b}_1 \rightarrow \bsl{k}_0  ) \ .
}
Then, by defining
\eqa{
\label{eq:U_s_V_W_s_Chern}
& U_s = W(\bsl{k}_0 \rightarrow \bsl{k}_0 + \frac{s}{2\pi}  \bsl{b}_1), \\
& V = W(\bsl{k}_0 + \bsl{b}_2 \rightarrow \bsl{k}_0  ),\\
& \W_s = W(\bsl{k}_0 + \frac{s}{2\pi}  \bsl{b}_1 \rightarrow \bsl{k}_0 + \frac{s}{2\pi}  \bsl{b}_1 + \bsl{b}_2)\ ,
}
we can clearly see \cref{eq:W_Ds_Chern} has the form of \cref{eq:W_partial_Ds_W_s}.
We further choose $\lambda_n(s) = 1$ for all $n$ in \cref{eq:w}.
Then, $w$ in \cref{eq:w} is the Chern number, and \cref{eq:WL_bound} tells us 
\eq{
2\pi|\Ch| \leq \int_{\BZ} d^2 k \ 2\sqrt{\det(g_{\bsl{k}})}  \leq \int_{\BZ} d^2 k \ \Tr[g_{\bsl{k}}]\ ,
}
where $\Ch$ is the Chern number.

\subsection{Review: Euler Bound}
A natural generalization of the Chern bound is the Euler bound~\cite{Xie2020TopologyBoundSCTBG,Yu2022EOCPTBG,BJY2024EulerBoundQG,Slager2024EulerOptical}, since there always exists a Chern gauge for bands with nonzero Euler number~\cite{Ahn2018MonopoleNLSM,Ahn2019TBGFragile,Song2019TBGFragile} protected by the combination of time-reversal (TR) and two-fold rotation symmetry.
The choices of $D_s$, $\bsl{k}_s$, $t_0$, and $s_f$ are the same as those of the Chern bound in \cref{app:Chern_bound}.
The expression of $W(\partial D_s)$ is the same as \cref{eq:W_Ds_Chern} and the choice of $U_s$, $V$ and $\W_s$ is the same as \cref{eq:U_s_V_W_s_Chern}.
The only difference is the choice of $\lambda_n(s)$.
For $\lambda_n(s)$, we can choose $\lambda_n(s_*\pm 0^+)=\lambda_{n'}(s_*\mp 0^+)$ across $s_*$ for infinitesimal positive number $0^+$ if $\phi_n(s^*)=\phi_{n'}(s^*)$ is protected by certain symmetries and choose $\lambda_n(s)$ to be continuous otherwise, together with $\lambda_n(0)= (-1)^{n}$.
Then, $w$ in \cref{eq:w} is twice the Euler number $e_2$, and \cref{eq:WL_bound} tells us 
\eq{
4\pi|e_2| \leq \int_{\BZ} d^2 k \ 2\sqrt{\det(g_{\bsl{k}})}  \leq \int_{\BZ} d^2 k \ \Tr[g_{\bsl{k}}]\ ,
}
where $e_2$ is the Chern number.

\subsection{Review: $\dsZ_2$ Bound}
\label{sec:Z_2_bound}

Another example is the quantum geometric bound of the Kane-Mele $\dsZ_2$ index.
In this case, the system has spinful TR symmetry, and the number of bands, $N$, must be even.

The choice of $D_s$ (as well as $\bsl{k}_s$) is the same as the Chern bound case, and we only need to change the value of $s_f$ to $s_f = \pi$.
The expression of $W(\partial D_s)$ is the same as \cref{eq:W_Ds_Chern} and the choice of $U_s$, $V$ and $\W_s$ is the same as \cref{eq:U_s_V_W_s_Chern}.
Then, the $Z_2$ index is $w$ in \cref{eq:w} as long as we choose $\bsl{k}_0$ to be a time-reversal invariant momentum (TRIM), $\phi_{n+1}(s)\geq \phi_{n}(s)$, $\phi_{n}(0)\in [0,2\pi)$, and $\lambda_n(s) = (-1)^n$.
Such a choice is certainly allowed in the derivation, and we arrive at
\eq{
\label{eq:Z2_bound_half_BZ}
2\pi \nu_{\dsZ_2} \leq \int_{\BZ/2} d^2 k \ 2\sqrt{\det(g_{\bsl{k}})}  \leq \int_{\BZ/2} d^2 k \ \Tr[g_{\bsl{k}}]\ ,
}
where $\BZ/2 = \{ \bsl{k} = \bsl{k}_0 +\sum_{l=1,2} \Delta k_l/(2\pi) \bsl{b}_l | \Delta k_1 \in [ 0,\pi ], \Delta k_2 \in [ 0,2\pi ]  \}$.
Owing to the TR symmetry, we have $g_{\bsl{k}} = g_{-\bsl{k}}$, leading to
\eq{
\label{eq:Z2_bound_full_BZ}
4\pi \nu_{\dsZ_2} \leq \int_{\BZ} d^2 k \ 2\sqrt{\det(g_{\bsl{k}})}  \leq \int_{\BZ} d^2 k \ \Tr[g_{\bsl{k}}]\ .
}
We note that \cref{eq:Z2_bound_half_BZ} and \cref{eq:Z2_bound_full_BZ} are equivalent.
In other words, one only needs to bound the quantum metric in half of the Brillouin zone for TR-protected topology.

We note that the choice of $D_s$ for the derivation of \cref{eq:Z2_bound_full_BZ} is not unique.
We can exchange $k_1$ and $k_2$ to reach the same conclusion.
Explicitly, we can choose $s_f=\pi$, and $D_s = \{ \bsl{k} = \bsl{k}_0 + \sum_{l=1,2} \Delta k_l/(2\pi) \bsl{b}_l | \Delta k_1 \in [ 0,2\pi ], \Delta k_2 \in [ 0, s ]  \}$, where $\bsl{k}_0$ is still chosen to be a TRIM.
In this case, $W(\partial D_s)$ has the expression
\eq{
\label{eq:W_Ds_Z_2_k_1_k_2_swapped}
W(\partial D_s) = W(\bsl{k}_0 \rightarrow \bsl{k}_0 + \bsl{b}_1) W(\bsl{k}_0 + \bsl{b}_1 \rightarrow \bsl{k}_0 + \bsl{b}_1 + \frac{s}{2\pi}  \bsl{b}_2) W(\bsl{k}_0 + \bsl{b}_1 + \frac{s}{2\pi}  \bsl{b}_2\rightarrow \bsl{k}_0  + \frac{s}{2\pi}  \bsl{b}_2 ) W(\bsl{k}_0  + \frac{s}{2\pi}  \bsl{b}_2  \rightarrow \bsl{k}_0  )\ .
}
Owing to the $\ket{u_{\bsl{k}+\bsl{G}}} = e^{-\ii \bsl{G}\cdot \bsl{r}}\ket{u_{\bsl{k}}} $ with $\bsl{r}$ the position operator, we have
\eq{
 W(\bsl{k}_0 + \bsl{b}_1 \rightarrow \bsl{k}_0 + \bsl{b}_1 + \frac{s}{2\pi}  \bsl{b}_2) =  W(\bsl{k}_0  \rightarrow \bsl{k}_0 + \frac{s}{2\pi}  \bsl{b}_2) \ .
}
Then, by defining
\eqa{
\label{eq:U_s_V_W_s_Z_2_k_1_k_2_swapped}
& U_s = W(\bsl{k}_0 \rightarrow \bsl{k}_0 + \bsl{b}_1) W(\bsl{k}_0 \rightarrow \bsl{k}_0 + \frac{s}{2\pi}  \bsl{b}_1) \\
& V = W(\bsl{k}_0 \rightarrow \bsl{k}_0 + \bsl{b}_1)\\
& \W_s = W(\bsl{k}_0 + \bsl{b}_1 + \frac{s}{2\pi}  \bsl{b}_2\rightarrow \bsl{k}_0  + \frac{s}{2\pi}  \bsl{b}_2 )\ ,
}
we again see \cref{eq:W_Ds_Z_2_k_1_k_2_swapped} has the form of \cref{eq:W_partial_Ds_W_s}.
As a result, $w$ in \cref{eq:w} becomes the $\dsZ_2$ index as long as we choose $\lambda_n(s)$ to be the same as that in the Euler class, leading to \cref{eq:Z2_bound_half_BZ} with $\BZ/2 = \{ \bsl{k} = \bsl{k}_0 +\sum_{l=1,2} \Delta k_l/(2\pi) \bsl{b}_l | \Delta k_1 \in [ 0,2\pi ], \Delta k_2 \in [ 0,\pi ]  \}$, which is again equivalent to \cref{eq:Z2_bound_half_BZ}.

\subsection{Inversion-Fragile Bound}
\label{eq:2D_inversion_fragile_bound}
The inversion symmetry (or spinless two-fold rotatoin symmetry) in 2D system can preserves fragile topology for an isolated set of two bands.
Specifically, an isolated set of two bands have inversion-protected fragile topology if (i) it has total Chern number to be zero, and (ii) it has a pair of parity-even (parity-odd) states at one of the four time-reversal invariant momenta while all states are parity-odd (parity-even) the other three~\cite{Song2020TwistedFragile}.
The fragile topology is characterized by a connected Wilson loop spectrum.
To see this, let us just consider $W(k_1\bsl{b}_1/(2\pi)\rightarrow k_1\bsl{b}_1/(2\pi) + \bsl{b}_2)$ with $k_1\in [-\pi,\pi].$
Consider
\eq{
W(-\bsl{b}_2/2\rightarrow \bsl{b}_2/2 ) =W(-\bsl{b}_2/2\rightarrow \bsl{0} )W(\bsl{0}\rightarrow \bsl{b}_2/2 )\ .
}
Under inversion, we have
\eq{
\P \ket{\psi_{\bsl{\kappa},n}} = \ket{\psi_{\bsl{\kappa},n}}\xi_{\bsl{\kappa}} \Rightarrow \P \ket{u_{\bsl{\kappa},n}} = e^{\ii 2\bsl{\kappa}\cdot\bsl{r}}\ket{u_{\bsl{\kappa},n}}\xi_{\bsl{\kappa}}  = \ket{u_{-\bsl{\kappa},n}}\xi_{\bsl{\kappa}}
}
with $\P$ the inversion operation, $\ket{\psi_{\bsl{k},n}}$ ($n=1,2$) the Bloch states for the isolated set, $\ket{\psi_{\bsl{k}+\bsl{G},n}}=\ket{\psi_{\bsl{k},n}}$ for any reciprocal lattice vector $\bsl{G}$, $\bsl{\kappa}$ a time-reversal invariant momentum and $\xi_{\bsl{\kappa}}=\pm 1$.
Then, under inversion symmetry, we have
\eq{
W(-\bsl{b}_2/2\rightarrow \bsl{0} ) = \xi_{\bsl{b}_2/2} W(\bsl{b}_2/2\rightarrow \bsl{0} ) \xi_{\bsl{0}}\ ,
}
which means
\eq{
W(-\bsl{b}_2/2\rightarrow \bsl{b}_2/2 ) =\xi_{\bsl{b}_2/2} \xi_{\bsl{0}} W(\bsl{b}_2/2\rightarrow \bsl{0} ) W(\bsl{0}\rightarrow \bsl{b}_2/2 ) = \xi_{\bsl{b}_2/2} \xi_{\bsl{0}}\mathrm{1}_{2\times 2} \ .
}
Similarly, we have
\eq{
W(\bsl{b}_1/2-\bsl{b}_2/2\rightarrow \bsl{b}_1/2+\bsl{b}_2/2 ) =W(\bsl{b}_1/2-\bsl{b}_2/2\rightarrow \bsl{b}_1/2 )W(\bsl{b}_1/2\rightarrow \bsl{b}_1/2 +\bsl{b}_2/2 )\ ,
}
and
\eq{
W(\bsl{b}_1/2-\bsl{b}_2/2\rightarrow \bsl{b}_1/2 ) = \xi_{\bsl{b}_1/2-\bsl{b}_2/2} W(-\bsl{b}_1/2+\bsl{b}_2/2\rightarrow -\bsl{b}_1/2 ) \xi_{\bsl{b}_1/2}= \xi_{\bsl{b}_1/2-\bsl{b}_2/2} W(\bsl{b}_1/2+\bsl{b}_2/2\rightarrow \bsl{b}_1/2 ) \xi_{\bsl{b}_1/2}\ ,
}
leading to
\eq{
W(\bsl{b}_1/2-\bsl{b}_2/2\rightarrow \bsl{b}_1/2+\bsl{b}_2/2 ) =\xi_{\bsl{b}_1/2-\bsl{b}_2/2}\xi_{\bsl{b}_1/2} W(\bsl{b}_1/2+\bsl{b}_2/2\rightarrow \bsl{b}_1/2 )W(\bsl{b}_1/2\rightarrow \bsl{b}_1/2 +\bsl{b}_2/2 ) = \xi_{\bsl{b}_1/2-\bsl{b}_2/2}\xi_{\bsl{b}_1/2}  \mathrm{1}_{2\times 2}\ .
}
As $\xi_{\bsl{b}_1/2-\bsl{b}_2/2}\xi_{\bsl{b}_1/2}$ and $\xi_{\bsl{b}_1/2-\bsl{b}_2/2}\xi_{\bsl{b}_1/2}$ have opposite signs, one of $W(-\bsl{b}_2/2\rightarrow \bsl{b}_2/2 )$ and $W(\bsl{b}_1/2-\bsl{b}_2/2\rightarrow \bsl{b}_1/2+\bsl{b}_2/2 )$ has two eigenvalues being 1, while the other has $-1$.
Combined with the fact that the total Chern number is zero, we know the Wilson spectrum must be connected in the same way as that of Kane-Mele $\dsZ_2$.
As a result, we have 
\eq{
\label{eq:Inversion_Fragile_bound_full_BZ}
4\pi  \leq \int_{\BZ} d^2 k \ 2\sqrt{\det(g_{\bsl{k}})}  \leq \int_{\BZ} d^2 k \ \Tr[g_{\bsl{k}}]
}
for an isolated set of two bands with the inversion-protected fragile topology.

\section{Wilson-Loop-Ideal Bands}
\label{app:WL_ideal_bands}

\subsection{General Consideration}

Suppose we have an isolated set of $N$ bands with $\G$-protected topology, where $\G$ is a symmetry group that can contain TR symmetry as we do not put any constraint on linearity.
The symmetry group $\G$ would define a fundamental zone, which serves as a ``unit'' and will cover the full Brillouin zone under operations in $\G$.
(More precisely, the fundamental zone contains exactly one element of each set of orbits of $\G$, which a set of orbits is generated by acting all operations in $\G$ on a momentum.
One can always choose the fundamental zone to be simply connected.)
For example, if we only have TR symmetry and identity in $\G$, the fundamental zone is just the half of the Brillouin zone which can cover the whole BZ under $\bsl{k}\rightarrow -\bsl{k}$.

Given a proper deformation $D_s$ with $D_{s_f}$ being either the full 1BZ or a fundamental zone, we have the WL winding $w$ associated with certain dressing $U_s$ and $V$.
Then, the ideal condition for this setup is simply
\eq{
2\pi|w| =  \int_{D_{s_f}} d^2 k \ 2\sqrt{\det(g_{\bsl{k}})}  = \int_{D_{s_f}} d^2 k \ \Tr[g_{\bsl{k}}]\ .
}
Supposing that the area of 1BZ is $m$ times the area of $D_{s_f}$ ($m$ is always an integer), the ideal condition can be equivalently written as
\eq{
\label{eq:WL_ideal}
2\pi m |w| =  \int_{\BZ} d^2 k \ 2\sqrt{\det(g_{\bsl{k}})}  = \int_{\BZ} d^2 k \ \Tr[g_{\bsl{k}}]\ .
}
It is because a generic operation $g$ in $\G$ always changes $\bsl{k}$ to $\bsl{k}_g = R\bsl{k}$ with $R$ orthogonal real matrix, and then we have $\Tr[g_{\bsl{k}}] = \Tr[g_{\bsl{k}_g}] $ and $\det[g_{\bsl{k}}] = \det[g_{\bsl{k}_g}] $.
The set of bands is defined to be WL-ideal if \eqnref{eq:WL_ideal} is satisfied.

There is one important consequence (among others) for a set of bands to be WL-ideal.
With the parametrization $\bsl{k}(s,t)$ in \cref{eq:k_s_t_parametrization} for $\partial D_s$, we have
\eq{
\rho\left(\int_0^1 dt\ \widetilde{\F}_{s,t} \right) \leq \int_0^1 dt \ \rho\left(\widetilde{\F}_{s,t} \right) 
}
as an intermediate step of the WL bound, where $\widetilde{\F}_{s,t}$ is the dressed nonabelian Berry curvature defined in \cref{eq:F_tilde_s_t}.
The WL idealness immediately infers
\eq{
\label{eq:F_tilde_s_t_rho_bound_saturation}
\rho\left(\int_0^1 dt\ \widetilde{\F}_{s,t} \right) = \int_0^1 dt \ \rho\left(\widetilde{\F}_{s,t} \right) 
}
for all $s\in[0,s_f]$.
Regardless of the value of $N$, \cref{eq:F_tilde_s_t_rho_bound_saturation} tells us
\eq{
\label{eq:F_tilde_s_t_rho_bound_saturation_pairwise}
\rho\left(\widetilde{\F}_{s,t_1} + \widetilde{\F}_{s,t_2} \right)  = \rho\left(\widetilde{\F}_{s,t_1} \right) + \rho\left(\widetilde{\F}_{s,t_2} \right) 
}
for any $t_1,t_2\in[0,1]$.
To see that, let us first note that the dressed nonabelian Berry curvature $\widetilde{F}_{s,t}$ is Hermitian.
Then, we can use the following proposition.
\begin{proposition}
\label{prop:rho_inequality_saturation}
    Given a set of matrices of the same dimension $\{ M_i | i \in I \}$, then 
    \eq{
    \rho\left(\sum_{i\in I} M_i \right) = \sum_{i\in I} \rho(M_i) \Rightarrow \rho(M_{i_1}+M_{i_2}) = \rho(M_{i_1}) + \rho(M_{i_2})\ \forall\ i_1,i_2\in I 
    \ . }
\end{proposition}
\proof{$\forall\ i_1,i_2\in I$, we always have
\eqa{
\label{eq:proof_intermediate_1}
\rho\left(\sum_{i\in I} M_i \right) &  =  \rho\left(M_{i_1} + M_{i_2} + \sum_{i\in I-\{ i_1 ,i_2\}} M_i \right)  \leq \rho\left(M_{i_1} + M_{i_2}\right)   + \rho\left(\sum_{i\in I-\{ i_1 ,i_2\}} M_i \right)  \\
& \leq \rho(M_{i_1}) + \rho(M_{i_2}) + \rho\left(\sum_{i\in I-\{ i_1 ,i_2\}} M_i \right) \\
& \leq \rho(M_{i_1}) + \rho(M_{i_2}) + \sum_{i\in I-\{ i_1 ,i_2\}} \rho\left(M_i \right) = \sum_{i\in I} \rho(M_i)\ .
}
As $\rho\left(\sum_{i\in I} M_i \right) = \sum_{i\in I} \rho(M_i)$, it means that all the inequalities in \cref{eq:proof_intermediate_1} are saturated, which gives
\eqa{
& \rho\left(M_{i_1} + M_{i_2}\right)   + \rho\left(\sum_{i\in I-\{ i_1 ,i_2\}} M_i \right)  = \rho(M_{i_1}) + \rho(M_{i_2}) + \rho\left(\sum_{i\in I-\{ i_1 ,i_2\}} M_i \right) \\
& \Rightarrow \rho\left(M_{i_1} + M_{i_2}\right) = \rho(M_{i_1}) + \rho(M_{i_2}) \ .
}
}

From \cref{prop:rho_inequality_saturation} to \cref{eq:F_tilde_s_t_rho_bound_saturation_pairwise}, we would need to be careful about the boundary, \ie, $t=0,1$.
In the case of $\widetilde{\F}_{s,t}$, we can safely include $t=0,1$ in \cref{eq:F_tilde_s_t_rho_bound_saturation_pairwise} owing to the fact that $\widetilde{F}_{s,t}$ is a continuous function of $t$.  
\cref{eq:F_tilde_s_t_rho_bound_saturation_pairwise} is a strong condition.
For $N=1$, we immediately know all Berry curvature $\F_{s,t}$ have the same sign for a fixed $s$, as $\widetilde{\F}_{s,t} = \F_{s,t}$.
For $N=2$, we have the following proposition.
\begin{proposition}
\label{prop:N_2_F_tilde_ideal_relation}
For $N=2$, \cref{eq:F_tilde_s_t_rho_bound_saturation_pairwise} means either (i) $\widetilde{\F}_{s,t_1}$ and $ \widetilde{\F}_{s,t_2}$ commute and their corresponding eigenvalues have the same sign, or (ii) $\widetilde{\F}_{s,t_1}$ and $ \widetilde{\F}_{s,t_2}$ do not commute but all of their eigenvalues have the same sign.
\end{proposition}
This is validated by the following proposition.
\begin{proposition}
    Given a two $2\times 2$ Hermitian matrices, $A$ and $B$, then $\rho(A+B) = \rho(A) + \rho (B) $ infers either of the two following statements
    \begin{itemize}
    \item $A$ and $B$ commute, and the corresponding eigenvalues have the same sign, where ``corresponding'' means coming from the same eigenvector. 
    \item $A$ and $B$ do not commute, and all eigenvalues of $A$ and $B$ have the same sign. 
    \end{itemize}
    Here we consider $0$ to have the same sign as all numbers.
\end{proposition}
\proof{
Suppose $v_m$'s are orthonormal eigenvectors of $A+B$ with eigenvalues $\lambda_m$, $x_m$'s are orthonormal eigenvectors of $A$ with eigenvalues $a_m$, and $y_m$'s are orthonormal eigenvectors of $B$ with eigenvalues $b_m$.

If $A$ and $B$ commute, we can choose $x_m=y_m=v_m$. 
In the basis of $v_m$, we have
\eq{
A+B = \mat{ a_1 + b_1 & \\ & a_2 + b_2}\ ,
}
which means
\eq{
\rho(A+B) = |a_1 + b_1| + |a_2+b_2| = |a_1| + |a_2| + |b_1| + |b_2| \ , 
}
where the second equality uses $\rho(A+B) = \rho(A) + \rho(B)$.
As a result, we know $a_1 b_1 \geq 0$ and $a_2 b_2 \geq 0$, \ie, the corresponding eigenvalues have the same sign.

If $A$ and $B$ do not commute, then we immediately know that $v_m^\dagger x_n$ cannot be zero for any $m$ and $n$.
It can be seen from contradiction. Suppose there exist $m_0$ and $n_0$ such that $v_{m_0}^\dagger x_{n_0} = 0 $.
It means $x_{m_0} \propto v_{3-n_0} $ and $x_{3-m_0} \propto v_{n_0} $, \ie, $x_1$ and $x_2$ diagonalizes $A+B$.
Combined with $x_m$ diagonalizes $A$, we obtain $[A+B,A]=0 \Rightarrow [B,A] = 0$, which contradicts to that $A$ and $B$ do not commute.
The same logic holds for $v_m^\dagger y_n$: $v_m^\dagger y_n$ cannot be zero for any $m$ and $n$.
Combined with
\eqa{
\rho(A+B) & = \left| v_1^\dagger (A+B) v_1\right| + \left| v_2^\dagger (A+B) v_2\right| \\
& = \left| v_1^\dagger (x_1 x_1^\dagger a_1 + x_2 x_2^\dagger a_2 + y_1 y_1^\dagger b_1 + y_2 y_2^\dagger b_2) v_1\right| + \left| v_2^\dagger (x_1 x_1^\dagger a_1 + x_2 x_2^\dagger a_2 + y_1 y_1^\dagger b_1 + y_2 y_2^\dagger b_2)  v_2\right| \\
& = \left| |v_1^\dagger x_1|^2 a_1 + |v_1^\dagger x_2|^2 a_2 + |v_1^\dagger y_1|^2 b_1 + |v_1^\dagger y_2|^2 b_2 \right| + \left| |v_2^\dagger x_1|^2 a_1 + |v_2^\dagger x_2|^2 a_2 + |v_2^\dagger y_1|^2 b_1 + |v_2^\dagger y_2|^2 b_2 \right| \\
& \leq |v_1^\dagger x_1|^2 |a_1| + |v_1^\dagger x_2|^2 |a_2| + |v_1^\dagger y_1|^2 |b_1| + |v_1^\dagger y_2|^2 |b_2| +   |v_2^\dagger x_1|^2 |a_1| + |v_2^\dagger x_2|^2 |a_2| + |v_2^\dagger y_1|^2 |b_1| + |v_2^\dagger y_2|^2 |b_2|\\
& = \rho (A)+\rho(B)\ ,
}
we know $\rho(A+B) = \rho (A)+\rho(B)$ infers
\eqa{
& \left| |v_1^\dagger x_1|^2 a_1 + |v_1^\dagger x_2|^2 a_2 + |v_1^\dagger y_1|^2 b_1 + |v_1^\dagger y_2|^2 b_2 \right| + \left| |v_2^\dagger x_1|^2 a_1 + |v_2^\dagger x_2|^2 a_2 + |v_2^\dagger y_1|^2 b_1 + |v_2^\dagger y_2|^2 b_2 \right| \\
& = |v_1^\dagger x_1|^2 |a_1| + |v_1^\dagger x_2|^2 |a_2| + |v_1^\dagger y_1|^2 |b_1| + |v_1^\dagger y_2|^2 |b_2| +   |v_2^\dagger x_1|^2 |a_1| + |v_2^\dagger x_2|^2 |a_2| + |v_2^\dagger y_1|^2 |b_1| + |v_2^\dagger y_2|^2 |b_2|\ ,
}
resulting in $a_1$, $a_2$, $b_1$ and $b_2$ all have the same sign.
}

As a direct consequence, if $\widetilde{\F}_{s,t_1}$ have two nonzero eigenvalues of opposite signs for $N=2$, \cref{eq:F_tilde_s_t_rho_bound_saturation_pairwise} can directly infer $[\widetilde{\F}_{s,t_1},\widetilde{\F}_{s,t_2}]=0$.
This is particularly important for the case of $\dsZ_2$-ideal band, as discussed in \cref{app:ideal_Z2_band}.

\subsection{ $\dsZ_2$-Ideal Bands}
\label{app:ideal_Z2_band}

As a notable case of the WL-ideal band, let us consider the case where the WL winding is the KM $\dsZ_2$ index~\cite{Kane2005Z2,Zhang2006QSH,Kane2005QSH,Bernevig2006BHZ}.
We particularly focus on an isolated set of two bands ($\ket{u_{\bsl{k}}}=(\ket{u_{\bsl{k},1}},\ket{u_{\bsl{k},2}})$) with $\nu_{\dsZ_2} = 1$, and we consider the case where $g_{\bsl{k}}$ is everywhere non-singular, \ie, $\det(g_{\bsl{k}})\neq 0$ for all $\bsl{k}$.
As we are considering the $\dsZ_2$-ideal bands, we require
\eq{
\label{eq:WL_ideal_Z2}
4\pi = \int_{\BZ} d^2 k \ \rho(F_{\bsl{k}})  = \int_{\BZ} d^2 k \ 2\sqrt{\det(g_{\bsl{k}})}  = \int_{\BZ} d^2 k \ \Tr[g_{\bsl{k}}]\ .
}
The non-singular $g_{\bsl{k}}$ means $F_{\bsl{k}}$ must have nonzero rank everywhere---it cannot be a zero matrix at any $\bsl{k}$.
Therefore, the eigenvalues of $F_{\bsl{k}}$ cannot be both zero simultaneously.

\subsubsection{Gapped Spectrum of nonabelian Berry curvature}

The first key property of $F_{\bsl{k}}$ for this set of ideal bands is that $F_{\bsl{k}}$ cannot have two nonzero eigenvalues of the same sign, as we show now.
As discussed in \cref{sec:Z_2_bound}, the derivation of the $\dsZ_2$ bound can be done by choosing
\eq{
\label{eq:D_s_choice_1}
D_s = \{ \bsl{k} = \bsl{k}_0 + \sum_{l=1,2} \Delta k_l/(2\pi) \bsl{b}_l |\Delta  k_1 \in [ 0,s ],\Delta  k_2 \in [ 0,\pi ]  \}
}
or 
\eq{
\label{eq:D_s_choice_2}
D_s = \{ \bsl{k} = \bsl{k}_0 + \sum_{l=1,2}\Delta  k_l/(2\pi) \bsl{b}_l | \Delta k_2 \in [ 0,s ],\Delta  k_1 \in [ 0,\pi ]  \}
}
as long as $\bsl{k}_s = \bsl{k}_0$ is a 
time-reversal-invariant momentum (TRIM).
For the first choice \cref{eq:D_s_choice_1}, $\bsl{k}(s,t)$ can take the form
\eq{
\label{eq:k_s_t_choice_1}
\bsl{k}(s,t) = \bsl{k}_0+ \left\{\begin{array}{ll}
   t \frac{4 s}{2\pi} \bsl{b}_1  &,\ t\in[0,1/4) \\
    \frac{ s}{2\pi} \bsl{b}_1 + 4 (t-1/4) \bsl{b}_2  & ,\ t\in[1/4,2/4] \\
   (3/4 - t) \frac{4 s}{2\pi} \bsl{b}_1 + \bsl{b}_2  & ,\ t\in(2/4,3/4] \\
   4(1 - t)  \bsl{b}_2  & ,\ t\in(3/4,1] 
\end{array}\right.\ .
}
We again focus on $t\in[1/4,2/4]$, where $F_{s,t}$ is non-vanishing.
First we note that for $t_1,t_2\in[1/4,2/4]$, we have $k_1 = k_{0,1} + s$ and $k_2 =k_{0,2} + 8\pi (t-1/4) $, which gives
\eq{
\det\left(\frac{\partial (k_x,k_y)}{\partial (s,t)}\right) = 8\pi\det\left(\frac{\partial (k_x,k_y)}{\partial (k_1,k_2)}\right) = 8\pi/\Omega
}
and thus
\eq{
\F_{s,t} = \frac{8\pi}{\Omega} F_{\bsl{k}(s,t)}\ ,
}
where $\Omega$ is the unit cell area.
Combined with
\eq{
\h_{s,t} = W(\bsl{k}_0 \rightarrow \bsl{k}_0 + \frac{s}{2\pi}\bsl{b}_1) W(\bsl{k}_0 + \frac{s}{2\pi}\bsl{b}_1 \rightarrow \bsl{k}(s,t))
}
for $t\in[1/4,2/4]$, we obtain the form of  $\widetilde{\F}_{s,t}$ in \cref{prop:N_2_F_tilde_ideal_relation}
\eq{
\widetilde{\F}_{s,t} = \frac{8\pi}{\Omega} \widetilde{F}_{\bsl{k}_0,\bsl{k}(s,t)}\ ,
}
where
\eq{
\label{eq:F_tilde_k}
\widetilde{F}_{\bsl{k}_0,\bsl{k}} \equiv h_{\bsl{k}_0,\bsl{k}} F_{\bsl{k}} h_{\bsl{k}_0,\bsl{k}}^{-1}
}
with
\eq{
h_{\bsl{k}_0,\bsl{k}}  = W(\bsl{k}_0 \rightarrow \frac{k_1}{2\pi}\bsl{b}_1 + \frac{k_{0,2}}{2\pi}\bsl{b}_2) W(\frac{k_1}{2\pi}\bsl{b}_1 + \frac{k_{0,2}}{2\pi}\bsl{b}_2 \rightarrow \bsl{k})\ .
}
Then, \cref{prop:N_2_F_tilde_ideal_relation} infers that for any $t_1,t_2\in[1/4,2/4]$ and for any $s\in[0,\pi]$, either $\widetilde{F}_{\bsl{k}_0,\bsl{k}(s,t_1)}$ and $\widetilde{F}_{\bsl{k}_0,\bsl{k}(s,t_2)}$  commute and their corresponding eigenvalues have the same sign or $\widetilde{F}_{\bsl{k}_0,\bsl{k}(s,t_1)}$ and $\widetilde{F}_{\bsl{k}_0,\bsl{k}(s,t_2)}$ do not commute but all their eigenvalues have the same sign.
As a result, if $F_{\bsl{k}(s,t)}$ (with $t\in [1/4,2/4]$) has the two nonzero eigenvalues of the same sign, $F_{\bsl{k}(s,t')}$ must have eigenvalues of the same sign as those of $F_{\bsl{k}(s,t)}$, for any $t'\in [1/4,2/4]$ and any $s\in[0,\pi]$.
Note that $t\in[1/4,2/4]$ always cover $k_2\in [0,2\pi]$ regardless of $\bsl{k}_0$ owing to the periodicity of $F_{\bsl{k}}$;.
Furthermore, $\bsl{k}_0 = \bsl{0}$ would let $s$ cover $k_1\in[0,\pi]$, and $\bsl{k}_0 = -\bsl{b}_1/2$ would let $s$ cover $k_1\in[-\pi,0]$.
Thus, combining the two cases, we arrive at this statement: if $F_{\bsl{k}}$ has the two nonzero eigenvalues of the same sign, $F_{\bsl{k}'}$ must have eigenvalues of the same sign as those of $F_{\bsl{k}}$, as long as $k_1' = k_1$.

For the choice of $D_s$ in \cref{eq:D_s_choice_2}, $\bsl{k}(s,t)$ can take the form
\eq{
\label{eq:k_s_t_choice_2}
\bsl{k}(s,t) = \bsl{k}_0+ \left\{\begin{array}{ll}
   4t  \bsl{b}_1  &,\ t\in[0,1/4) \\
    \bsl{b}_1 + 4 (t-1/4)\frac{ s}{2\pi} \bsl{b}_2  & ,\ t\in[1/4,2/4) \\
   4 (3/4 - t)  \bsl{b}_1 + \frac{ s}{2\pi} \bsl{b}_2  & ,\ t\in [2/4,3/4] \\
   4(1 - t)  \frac{4 s}{2\pi}\bsl{b}_2  & ,\ t\in(3/4,1] 
\end{array}\right.\ .
}
We now focus on $t_1,t_2\in[2/4,3/4]$, where $F_{s,t}$ is non-vanishing.
First we note that for $t_1,t_2\in[2/4,3/4]$, we have $k_1 = k_{0,1} + 8\pi (3/4 - t)$ and $k_2 =k_{0,2} + s $, which gives
\eq{
\det\left(\frac{\partial (k_x,k_y)}{\partial (s,t)}\right) = 8\pi\det\left(\frac{\partial (k_x,k_y)}{\partial (k_2,-k_1)}\right) = 8\pi/\Omega
}
and thus
\eq{
\F_{s,t} = \frac{8\pi}{\Omega}F_{\bsl{k}(s,t)}
}
still holds.
Combined with
\eq{
\h_{s,t} = W(\bsl{k}_0 \rightarrow \bsl{k}_0 +\bsl{b}_1) W(\bsl{k}_0 +\bsl{b}_1 \rightarrow \bsl{k}_0 +\bsl{b}_1  + \frac{s}{2\pi}\bsl{b}_2)W(\bsl{k}_0 +\bsl{b}_1  + \frac{s}{2\pi}\bsl{b}_2 \rightarrow \bsl{k}(s,t))
}
for $t\in[1/4,2/4]$, we obtain the form of  $\widetilde{\F}_{s,t}$ in \cref{prop:N_2_F_tilde_ideal_relation}:
\eq{
\widetilde{\F}_{s,t} = \frac{8\pi}{\Omega} \bar{F}_{\bsl{k}_0,\bsl{k}(s,t)}\ ,
}
where
\eq{
\label{eq:F_bar_k}
\bar{F}_{\bsl{k}_0,\bsl{k}} \equiv \bar{h}_{\bsl{k}_0,\bsl{k}} F_{\bsl{k}} \bar{h}_{\bsl{k}_0,\bsl{k}}^{-1}
}
with
\eq{
\bar{h}_{\bsl{k}_0,\bsl{k}}  = W(\bsl{k}_0 \rightarrow \bsl{k}_0 +\bsl{b}_1) W(\bsl{k}_0 +\bsl{b}_1 \rightarrow \bsl{b}_1  + \frac{k_{0,1}}{2\pi} \bsl{b}_1 + \frac{k_2}{2\pi}\bsl{b}_2)W(\bsl{b}_1  + \frac{k_{0,1}}{2\pi} \bsl{b}_1 + \frac{k_2}{2\pi}\bsl{b}_2 \rightarrow \bsl{k})\ .
}
Then, \cref{prop:N_2_F_tilde_ideal_relation} infers that for any $t_1,t_2\in[2/4,3/4]$ and for any $s\in[0,\pi]$, either (i) $\bar{F}_{\bsl{k}_0,\bsl{k}(s,t_1)}$ and $\bar{F}_{\bsl{k}_0,\bsl{k}(s,t_2)}$ commute and their corresponding eigenvalues have the same sign or (ii) $\bar{F}_{\bsl{k}_0,\bsl{k}(s,t_1)}$ and $\bar{F}_{\bsl{k}_0,\bsl{k}(s,t_2)}$ do not commute but all their eigenvalues have the same sign.
As a result, if $F_{\bsl{k}(s,t)}$ (with $t\in [2/4,3/4]$) has two nonzero eigenvalues of the same sign, the eigenvalues $F_{\bsl{k}(s,t')}$ must have the same sign as those of $F_{\bsl{k}(s,t)}$, for any $t'\in [2/4,3/4]$ and any $s\in[0,\pi]$.
Note that $t\in[2/4,3/4]$ always covers $k_1\in [0,2\pi]$ regardless of $\bsl{k}_0$ owing to the periodicity of $F_{\bsl{k}}$.
Furthermore, $\bsl{k}_0 = \bsl{0}$ would let $s$ cover $k_2\in[0,\pi]$, and $\bsl{k}_0 = -\bsl{b}_2/2$ would let $s$ cover $k_2\in[-\pi,0]$.
Thus, combining the two cases, we arrive at this statement: if $F_{\bsl{k}}$ has the two nonzero eigenvalues of the same sign, $F_{\bsl{k}'}$ must have eigenvalues of the same sign as those of $F_{\bsl{k}}$, as long as $k_2' = k_2$.

Now suppose $F_{\bsl{k}}$ has two positive eigenvalues, which implies $F_{k_1 \bsl{b}_1 - k_2 \bsl{b}_2}$ must have either two positive eigenvalues or one positive eigenvalue and one zero eigenvalue.
Furthermore, owing to the TR symmetry, we know $F_{-\bsl{k}}$ has two negative eigenvalues, which means $F_{k_1 \bsl{b}_1 - k_2 \bsl{b}_2}$ must have either two negative eigenvalues or one negative eigenvalue and one zero eigenvalue.
This is a contradiction---$F_{\bsl{k}}$ cannot have two positive eigenvalues.
The same logic implies that $F_{\bsl{k}}$ cannot have two negative eigenvalues, leading to $F_{\bsl{k}}$ cannot have two nonzero eigenvalues of the same sign.

The fact that $F_{\bsl{k}}$ cannot have two nonzero eigenvalues of the same sign tells us $F_{\bsl{k}}$ must have two different eigenvalues.
It is because if the two eigenvalues of $F_{\bsl{k}}$ are the same, 
they are either both positive, both negative, or both zero; none of them can happen.
Furthermore, the two bands of $F_{\bsl{k}}$ are very special: one has to be nonnegative and the other has to be nonpositive, since they cannot be both negative or both positive.

\subsubsection{Non-Singular nonabelian Berry Curvature and Chern-Ideal Gauge}

We now further restrict to the case where $F_{\bsl{k}}$ is non-signular---$\det(F_{\bsl{k}})$ is non-vanishing throughout the 1BZ.
We will show that in this case, the gapped spectrum of $F_{\bsl{k}}$ directly indicates the existence of a Chern-ideal gauge.

To see that, let us first note that the fact that $F_{\bsl{k}}$ is nonsingular means $F_{\bsl{k}}$ must have one strictly positive eigenvalue and one strictly negative eigenvalue, throughout the 1BZ.
The same also holds for $\widetilde{F}_{\bsl{k}_0,\bsl{k}}$ in \cref{eq:F_tilde_k} and $\bar{F}_{\bsl{k}_0,\bsl{k}}$ in \cref{eq:F_bar_k}.

For $\widetilde{F}_{\bsl{k}_0,\bsl{k}}$ in \cref{eq:F_tilde_k}, we immediately know that  $\widetilde{F}_{\bsl{k}_0,\bsl{k}}$ and $\widetilde{F}_{\bsl{k}_0,\bsl{k}'}$ can never have the same sign for all their eigenvalues.
Then, \cref{prop:N_2_F_tilde_ideal_relation} implies that $\widetilde{F}_{\bsl{k}_0,\bsl{k}}$  and $\widetilde{F}_{\bsl{k}_0,\bsl{k}'}$ must commute and their corresponding eigenvalues have the same sign, as long as $k_1 ' = k_1 \in [k_{0,1}, k_{0,1} + \pi]$ and $k_2,k_2'\in[0,2\pi]$.
If we choose $\bsl{k}_0 = \bsl{0}$, the statement holds for $k_1 ' = k_1 \in [0,  \pi]$ and  $k_2,k_2'\in[0,2\pi]$; if we choose $\bsl{k}_0 = \bsl{b}_1/2$, the statement holds for $k_1 ' = k_1 \in [\pi,  2\pi]$ and $k_2,k_2'\in[0,2\pi]$.
Combined with
\eq{
\widetilde{F}_{\bsl{b}_1/2,\bsl{k}} = W^{-1}( \bsl{0} \rightarrow\bsl{b}_1/2)\widetilde{F}_{\bsl{0},\bsl{k}} W( \bsl{0} \rightarrow\bsl{b}_1/2) 
}
for $k_1 \in [\pi,2\pi]$, we have $\widetilde{F}_{\bsl{0},\bsl{k}}$ and  $\widetilde{F}_{\bsl{0},\bsl{k}'}$ must commute and their corresponding eigenvalues have the same sign, for $k_1 ' = k_1 \in [0,  2 \pi]$ and  $k_2,k_2'\in[0,2\pi]$.
It means that we can choose a $k_2$-independent matrix $\widetilde{R}_{k_1}$ to diagonalize $\widetilde{F}_{\bsl{0},\bsl{k}}$ for all $k_2 \in [0,2\pi]$, \ie,
\eq{
\widetilde{F}_{\bsl{0},\bsl{k}} = \widetilde{R}_{k_1} \mat{f_1(\bsl{k}) & \\ & f_2(\bsl{k})}  \widetilde{R}_{k_1}^\dagger \ ,
}
where we choose $f_1(\bsl{k})>0$ and $f_2(\bsl{k})<0$ without loss of generality.

Now we consider $\bar{F}_{\bsl{k}_0,\bsl{k}}$ in \cref{eq:F_bar_k}.
Similar to $\widetilde{F}$, $\bar{F}_{\bsl{k}_0,\bsl{k}}$ and $\bar{F}_{\bsl{k}_0,\bsl{k}'}$ can never have the same sign for all their eigenvalues either.
Then, \cref{prop:N_2_F_tilde_ideal_relation} infers that $\bar{F}_{\bsl{k}_0,\bsl{k}}$  and $\bar{F}_{\bsl{k}_0,\bsl{k}'}$ must commute and their corresponding eigenvalues have the same sign, as long as $k_2 ' = k_2 \in [k_{0,1}, k_{0,1} + \pi]$ and $k_1,k_1'\in[0,2\pi]$.
If we choose $\bsl{k}_0 = \bsl{0}$, the statement holds for $k_2 ' = k_2 \in [0,  \pi]$ and  $k_1,k_1'\in[0,2\pi]$; if we choose $\bsl{k}_0 = \bsl{b}_2/2$, the statement holds for $k_2 ' = k_2 \in [\pi,  2\pi]$ and $k_1,k_1'\in[0,2\pi]$.
Combined with
\eqa{
\bar{F}_{\bsl{b}_2/2,\bsl{k}} & = W( \bsl{b}_2/2 \rightarrow \bsl{b}_1 + \bsl{b}_2/2 ) W^{-1}( \bsl{b}_1 \rightarrow \bsl{b}_1 + \bsl{b}_2/2) W^{-1}( \bsl{0} \rightarrow\bsl{b}_1)\bar{F}_{\bsl{0},\bsl{k}}   W( \bsl{0} \rightarrow\bsl{b}_1) W( \bsl{b}_1 \rightarrow \bsl{b}_1 + \bsl{b}_2/2) \\
& \qquad \times W^{-1}( \bsl{b}_2/2 \rightarrow \bsl{b}_1 + \bsl{b}_2/2 )
}
for $k_2 \in [\pi,2\pi]$, we have $\bar{F}_{\bsl{0},\bsl{k}}$ and  $\bar{F}_{\bsl{0},\bsl{k}'}$ must commute and their corresponding eigenvalues have the same sign, for $k_2 ' = k_2 \in [0,  2 \pi]$ and  $k_1,k_1'\in[0,2\pi]$.
It means that we can choose a $k_1$-independent matrix $\bar{R}_{k_2}$ to diagonalize $\widetilde{F}_{\bsl{0},\bsl{k}}$ for all $k_1 \in [0,2\pi]$, \ie,
\eq{
\bar{F}_{\bsl{0},\bsl{k}} = \bar{R}_{k_2} \mat{f_1(\bsl{k}) & \\ & f_2(\bsl{k})}  \bar{R}_{k_2}^\dagger \ .
}

Combining the properties of $\widetilde{F}_{\bsl{0},\bsl{k}}$ and $\bar{F}_{\bsl{0},\bsl{k}}$, we can eliminate the $k_1$-dependence in $\widetilde{R}_{k_1}$ and the $k_2$-dependence in $\bar{R}_{k_2}$.
Specifically, we note that for $k_1 \in [0, 2\pi]$
\eq{
h_{\bsl{0},k_1\bsl{b}_1/(2\pi)} = W( \bsl{0} \rightarrow \frac{k_1}{2\pi}\bsl{b}_1 ) =  W( \bsl{0} \rightarrow \bsl{b}_1)W(\bsl{b}_1  \rightarrow k_1\bsl{b}_1/(2\pi)) = \bar{h}_{\bsl{0},k_1\bsl{b}_1/(2\pi)} \ ,
}
resulting in
\eq{
\widetilde{F}_{\bsl{0},k_1\bsl{b}_1/(2\pi)}   = \bar{F}_{\bsl{0},k_1\bsl{b}_1/(2\pi)}  \ .
}
As we know $\bar{F}_{\bsl{0},k_1\bsl{b}_1/(2\pi)}$ and  $\bar{F}_{\bsl{0},k_1'\bsl{b}_1/(2\pi)}$ must commute and their corresponding eigenvalues have the same sign for $k_1,k_1'\in[0,2\pi]$, we know $\widetilde{F}_{\bsl{0},k_1\bsl{b}_1/(2\pi)}$ and  $\widetilde{F}_{\bsl{0},k_1'\bsl{b}_1/(2\pi)}$ must commute and their corresponding eigenvalues have the same sign for  $k_1,k_1'\in[0,2\pi]$, resulting that $\widetilde{R}_{k_1} = \widetilde{R}$ must be independent of the momentum.
Similarly, for $k_2 \in [0, 2\pi]$
\eq{
h_{\bsl{0}, \bsl{b}_1 + k_2\bsl{b}_2/(2\pi)} = W( \bsl{0} \rightarrow \bsl{b}_1) W(\bsl{b}_1\rightarrow \frac{k_2}{2\pi}\bsl{b}_2 ) = \bar{h}_{\bsl{0},\bsl{b}_1 + k_2\bsl{b}_2/(2\pi)} \ ,
}
resulting in
\eq{
\widetilde{F}_{\bsl{0},\bsl{b}_1 + k_2\bsl{b}_2/(2\pi)}   = \bar{F}_{\bsl{0}\bsl{b}_1 + k_2\bsl{b}_2/(2\pi)}  \ .
}
As we know $\widetilde{F}_{\bsl{0},\bsl{b}_1 + k_2\bsl{b}_2/(2\pi)}$ and  $\widetilde{F}_{\bsl{0},\bsl{b}_1 + k_2'\bsl{b}_2/(2\pi)}$ must commute and their corresponding eigenvalues have the same sign for  $k_2,k_2'\in[0,2\pi]$, we know $\bar{F}_{\bsl{0},\bsl{b}_1 + k_2\bsl{b}_2/(2\pi)}$ and  $\bar{F}_{\bsl{0},\bsl{b}_1 + k_2'\bsl{b}_2/(2\pi)}$ must commute and their corresponding eigenvalues have the same sign for  $k_2,k_2'\in[0,2\pi]$, resulting that $\bar{R}_{k_2} = \bar{R}$ must be independent of the momentum.

So we now have
\eq{
\label{eq:F_tilde_ref_0_diagonal}
\widetilde{F}_{\bsl{0},\bsl{k}} = \widetilde{R} \mat{f_1(\bsl{k}) & \\ & f_2(\bsl{k})}  \widetilde{R}^\dagger \ ,
}
and
\eq{
\label{eq:F_bar_ref_0_diagonal}
\bar{F}_{\bsl{0},\bsl{k}} = \bar{R} \mat{f_1(\bsl{k}) & \\ & f_2(\bsl{k})}  \bar{R}^\dagger \ .
}
To exploit such properties, we can define 
\eq{
\ket{\widetilde{u}_{\bsl{k}}} = \ket{u_{\bsl{k}}} h_{\bsl{0},\bsl{k}}^{-1} \widetilde{R}\ ,
}
and
\eq{
\ket{\bar{u}_{\bsl{k}}} = \ket{u_{\bsl{k}}} \bar{h}_{\bsl{0},\bsl{k}}^{-1} \bar{R}\ .
}
Importantly, the nonabelian Berry curvature of $\ket{\widetilde{u}_{\bsl{k}}}$ is simply $\widetilde{R}^\dagger h_{\bsl{0},\bsl{k}} F_{\bsl{k}} h_{\bsl{0},\bsl{k}}^{-1} \widetilde{R}$, and the nonabelian Berry curvature of $\ket{\bar{u}_{\bsl{k}}}$ is simply $\bar{R}^\dagger \bar{h}_{\bsl{0},\bsl{k}} F_{\bsl{k}} \bar{h}_{\bsl{0},\bsl{k}}^{-1} \bar{R}$.
Combined with \cref{eq:F_tilde_ref_0_diagonal} and \cref{eq:F_bar_ref_0_diagonal}, we know the nonabelian Berry curvatures of $\ket{\widetilde{u}_{\bsl{k}}}$ and $\ket{\bar{u}_{\bsl{k}}}$ are just $\diag( f_1(\bsl{k}) , f_2(\bsl{k}) )$.

Now we show $\ket{\widetilde{u}_{\bsl{k},i}}$ and $\ket{\bar{u}_{\bsl{k},i}}$ both have well-defined Chern numbers and are ideal.
First, we note that $\ket{u_{\bsl{k}}} F_{\bsl{k}} \bra{u_{\bsl{k}}}$ is a smooth hermitian operator for $\bsl{k}\in\dsR^2$, and satisfies 
\eq{
\ket{u_{\bsl{k}+\bsl{G}}} F_{\bsl{k}+\bsl{G}} \bra{u_{\bsl{k}+\bsl{G}}} = e^{-\ii \bsl{r}\cdot\bsl{G}}\ket{u_{\bsl{k}}} F_{\bsl{k}} \bra{u_{\bsl{k}}} e^{\ii \bsl{r}\cdot\bsl{G}}\ .
}
Clearly, $f_1(\bsl{k})$ and $f_2(\bsl{k})$ are isolated bands of $\ket{u_{\bsl{k}}} F_{\bsl{k}} \bra{u_{\bsl{k}}}$, which means that $\ket{\widetilde{u}_{\bsl{k},i}}$ and $\ket{\bar{u}_{\bsl{k},i}}$ must have well-defined Chern numbers, just like the eigenstates of an isolated band of a Hamiltonian having a well-defined Chern number.
Let us use $\widetilde{\Ch}_n$ to label the Chern number of $\ket{\widetilde{u}_{\bsl{k},n}}$, and use $\overline{\Ch}_n$ to label the Chern number of $\ket{\bar{u}_{\bsl{k},n}}$.
We now show that
\eq{
\label{eq:Ch_Chern_gauge}
\widetilde{\Ch}_n = \overline{\Ch}_n = \frac{1}{2\pi} \int_{\BZ} d^2 k\ f_n(\bsl{k}) \ .
}
To show that, we note that $\ket{\widetilde{u}_{\bsl{k}}} $ and $\ket{\bar{u}_{\bsl{k}}}$ are parallel transport gauges.
Explicitly,
\eqa{
\label{eq:u_tilde_u_bar_parallel_transport_expression}
\ket{\widetilde{u}_{\bsl{k}}} & = 
P_{\bsl{k}\leftarrow \frac{k_1}{2\pi}\bsl{b}_1}P_{\frac{k_1}{2\pi}\bsl{b}_1\leftarrow\bsl{0}}\ket{u_{\bsl{0}}} \widetilde{R}, \\
\ket{\bar{u}_{\bsl{k}}} & = 
P_{\bsl{k} \leftarrow \bsl{b}_1  + \frac{k_2}{2\pi}\bsl{b}_2  }P_{\bsl{b}_1  + \frac{k_2}{2\pi}\bsl{b}_2  \leftarrow \bsl{b}_1} P_{ \bsl{b}_1 \leftarrow \bsl{0} }\ket{u_{\bsl{0}}} \bar{R} \ ,
}
where
\eq{
P_{\bsl{k}'\leftarrow \bsl{k}} = \lim_{L\rightarrow\infty}  P_{\bsl{k}_{L}}  P_{\bsl{k}_{L-1}} \cdots P_{\bsl{k}_2} P_{\bsl{k}_1} \ ,
}
and $\bsl{k}_1,\bsl{k}_2,\ldots,\bsl{k}_{L}$ are aligned sequentially along the straight line from $\bsl{k}$ to $\bsl{k}'$ with $\bsl{k}_1 = \bsl{k}$ and $\bsl{k}_L = \bsl{k}'$.
\cref{eq:u_tilde_u_bar_parallel_transport_expression} immediately leads to the following properties
\eqa{
& \bra{\widetilde{u}_{\bsl{k}}} \partial_{k_2}\ket{\widetilde{u}_{\bsl{k}}} = \lim_{\epsilon\rightarrow 0^+} \bra{\widetilde{u}_{\bsl{k}}}\frac{ (P_{\bsl{k}+ \frac{\epsilon}{2\pi}\bsl{b}_2} - P_{\bsl{k}})  }{\epsilon} \ket{\widetilde{u}_{\bsl{k}}}=  \bra{\widetilde{u}_{\bsl{k}}}\partial_{k_2}P_{\bsl{k}}\ket{\widetilde{u}_{\bsl{k}}} =  \bra{\widetilde{u}_{\bsl{k}}}\partial_{k_2}\ket{\widetilde{u}_{\bsl{k}}} + \braket{\partial_{k_2}\widetilde{u}_{\bsl{k}}}{\widetilde{u}_{\bsl{k}}} = 0 \\
& \bra{\bar{u}_{\bsl{k}}} \partial_{k_1}\ket{\bar{u}_{\bsl{k}}} = \lim_{\epsilon\rightarrow 0^+} \bra{\bar{u}_{\bsl{k}}}\frac{ (P_{\bsl{k}- \frac{\epsilon}{2\pi}\bsl{b}_1} - P_{\bsl{k}})  }{-\epsilon} \ket{\bar{u}_{\bsl{k}}} =  \bra{\bar{u}_{\bsl{k}}}\partial_{k_1}P_{\bsl{k}}\ket{\bar{u}_{\bsl{k}}} =  \bra{\bar{u}_{\bsl{k}}}\partial_{k_1}\ket{\bar{u}_{\bsl{k}}} + \braket{\partial_{k_1}\widetilde{u}_{\bsl{k}}}{\bar{u}_{\bsl{k}}} = 0 \ .
}
In particular, we know
\eq{
\bra{\widetilde{u}_{\bsl{k},2}} \partial_{k_2}\ket{\widetilde{u}_{\bsl{k},1}} = \bra{\bar{u}_{\bsl{k},2}} \partial_{k_1}\ket{\bar{u}_{\bsl{k},1}} = 0\ .
}
Owing to the gapped nature of $F_{\bsl{k}}$, we know 
\eq{
\ket{\bar{u}_{\bsl{k},n}} = \ket{\widetilde{u}_{\bsl{k},n}} e^{\ii \theta_n(\bsl{k})}\ ,
}
which leads to
\eq{
\label{eq:decouple_basis}
\braket{\bar{u}_{\bsl{k},1}}{\nabla_{\bsl{k}}\bar{u}_{\bsl{k},2}} = \braket{\widetilde{u}_{\bsl{k},1}}{\nabla_{\bsl{k}}\widetilde{u}_{\bsl{k},2}} = 0\ .
}
\cref{eq:decouple_basis} infers that the abelian Berry curvatures of $\ket{\widetilde{u}_{\bsl{k},n}}$ and $\ket{\bar{u}_{\bsl{k},n}}$ are equal to $f_n(\bsl{k})$, which leads to \cref{eq:Ch_Chern_gauge}.

Finally, we show that $\ket{\bar{u}_{\bsl{k},n}}$ and $\ket{\widetilde{u}_{\bsl{k},n}}$ are Chern-ideal states.
First, since $f_1(\bsl{k})>0$ and $f_2(\bsl{k})<0$, \cref{eq:WL_ideal_Z2} infers
\eq{
\frac{1}{2\pi}\int_{\BZ} d^2 k\ [f_1(\bsl{k}) - f_2(\bsl{k})] = \frac{1}{2\pi}\int_{\BZ} d^2 k\ \rho(F_{\bsl{k}}) = 2\ ,
}
meaning that
\eq{
\widetilde{\Ch}_1 - \widetilde{\Ch}_2 = \overline{\Ch}_1 - \overline{\Ch}_2 = 2\ .
}
Since TR symmetry requires the total Chern number to be zero, we know
\eq{
\widetilde{\Ch}_n = \overline{\Ch}_n = (-1)^{n-1}\ .
}
Furthermore, the quantum metric can be split into
\eq{
\Tr[g_{\bsl{k}}] = \frac{1}{2} \sum_{i=x,y} \Tr[\partial_{k_i} P_{\bsl{k}} \partial_{k_i} P_{\bsl{k}}] = \frac{1}{2} \sum_{i=x,y} \sum_{m,n = 1,2}\Tr[\partial_{k_i} \widetilde{P}_{m,\bsl{k}} \partial_{k_i} \widetilde{P}_{n,\bsl{k}}]\ ,
}
where
\eq{
\widetilde{P}_{n,\bsl{k}} = \ket{\widetilde{u}_{\bsl{k},n}}\bra{\widetilde{u}_{\bsl{k},n}} = \ket{\bar{u}_{\bsl{k},n}}\bra{\bar{u}_{\bsl{k},n}}\ .
}
Combined with
\eq{
\Tr[\partial_{k_i} \widetilde{P}_{1,\bsl{k}} \partial_{k_i} \widetilde{P}_{2,\bsl{k}}] =  \braket{\widetilde{u}_{\bsl{k},2}}{\partial_{k_i}\widetilde{u}_{\bsl{k},1}}\braket{\widetilde{u}_{\bsl{k},1}}{\partial_{k_i} \widetilde{u}_{\bsl{k},2}} + \braket{\partial_{k_i}\widetilde{u}_{\bsl{k},2}}{\widetilde{u}_{\bsl{k},1}}\braket{\partial_{k_i}\widetilde{u}_{\bsl{k},1}}{\widetilde{u}_{\bsl{k},2}} = 0
}
derived from \cref{eq:decouple_basis}, we obtain 
\eq{
\Tr[g_{\bsl{k}}] = \frac{1}{2} \sum_{i=x,y} \Tr[\partial_{k_i} P_{\bsl{k}} \partial_{k_i} P_{\bsl{k}}] = \sum_{n = 1,2}\Tr[\widetilde{g}_{n,\bsl{k}}]\ ,
}
where
\eq{
[\widetilde{g}_{n,\bsl{k}}]_{ij} = \frac{1}{2}\Tr\left[\partial_{k_i} \widetilde{P}_{n,\bsl{k}}\ \partial_{k_j} \widetilde{P}_{n,\bsl{k}} \right]
}
is the quantum metric of $\ket{\widetilde{u}_{\bsl{k},n}}$ and $\ket{\bar{u}_{\bsl{k},n}}$.
Then, \cref{eq:WL_ideal_Z2} infers
\eq{
\label{eq:ideal_chern_intermediate}
\sum_{n = 1,2}\frac{1}{2\pi}\int_{\BZ} d^2k\ \Tr[\widetilde{g}_{n,\bsl{k}}] = 2 \ .
}
As 
\eq{
\frac{1}{2\pi}\int_{\BZ} d^2k\ \Tr[\widetilde{g}_{n,\bsl{k}}] \geq |\widetilde{\Ch}_n|  \ ,
}
\cref{eq:ideal_chern_intermediate} leads to
\eq{
\frac{1}{2\pi}\int_{\BZ} d^2k\ \Tr[\widetilde{g}_{n,\bsl{k}}] = |\widetilde{\Ch}_n| = |\overline{\Ch}_n|\ ,
}
which means both $\ket{\bar{u}_{\bsl{k},n}}$ and $\ket{\widetilde{u}_{\bsl{k},n}}$ are Chern-ideal states for $n=1,2$.
Here we refer to $\ket{\bar{u}_{\bsl{k},n}}$ and $\ket{\widetilde{u}_{\bsl{k},n}}$ as Chern-ideal states instead of Chern-ideal bands because they are not necessarily energy bands, yet their projectors are still smooth and have the embedding relation similar to the eigenstates of an isolated band.
In other words, $\ket{\widetilde{u}_{\bsl{k}}}$ and $\ket{\bar{u}_{\bsl{k}}}$ are Chern-ideal gauges.

\subsection{ Ideal Bands with Zero Chern number and Nonzero Normal Wilson Loop Winding}
\label{app:Chern-Zero-Ideal_band}

We now consider a more general case than \cref{app:ideal_Z2_band}--- an isolated set of two bands ($\ket{u_{\bsl{k}}}=(\ket{u_{\bsl{k},1}},\ket{u_{\bsl{k},2}})$) with (i) zero total Chern number and (ii) $F_{\bsl{k}}$ is everywhere non-singular, \ie, $\det(F_{\bsl{k}})\neq 0$ for all $\bsl{k}$.
We still require
\eq{
\label{eq:WL_ideal}
2\pi |w| = \int_{\BZ} d^2 k \ \rho(F_{\bsl{k}})  = \int_{\BZ} d^2 k \ 2\sqrt{\det(g_{\bsl{k}})}  = \int_{\BZ} d^2 k \ \Tr[g_{\bsl{k}}]\ ,
}
where $w$ is a nonzero winding number for both $W( k_1 \bsl{b}_1/(2\pi) \rightarrow k_1 \bsl{b}_1/(2\pi)  + \bsl{b}_2)$ and $W( k_2 \bsl{b}_2/(2\pi)  + \bsl{b}_1\rightarrow k_2 \bsl{b}_2/(2\pi) )$, which we refer to as the normal Wilson loop. 

Again, the first key property of $F_{\bsl{k}}$ for this set of ideal bands is that $F_{\bsl{k}}$ cannot have two nonzero eigenvalues of the same sign, as we show now.
The reasoning is simple: if $F_{\bsl{k}}$ has two nonzero eigenvalues of the same sign at one specific $\bsl{k}$, it must have two nonzero eigenvalues of the same sign at all $\bsl{k}$ since $F_{\bsl{k}}$ has continuous eigenvalues, which would lead to nonzero total Chern number---contradiction to the zero total Chern number starting point.
As a result, the two eigenvalues of $F_{\bsl{k}}$, $f_1(\bsl{k})$ and $f_{2}(\bsl{k})$ with $f_1(\bsl{k})\geq f_2(\bsl{k})$ without loss of generality, must satisfy $f_1(\bsl{k})>0>f_2(\bsl{k})$.

Now for the two parallel transport gauges $\widetilde{u}_{\bsl{k}}$ and $\ket{\bar{u}_{\bsl{k}}}$, we know that the nonabelian Berry curvature is diagonal for both of them.
Then, we have
\eq{
\widetilde{P}_{n,\bsl{k}} = \ket{\widetilde{u}_{\bsl{k},n}}\bra{\widetilde{u}_{\bsl{k},n}} = \ket{\bar{u}_{\bsl{k},n}}\bra{\bar{u}_{\bsl{k},n}}\ ,
}
and \cref{eq:decouple_basis}, which leads to 
\eq{
\Tr[g_{\bsl{k}}] = \frac{1}{2} \sum_{i=x,y} \Tr[\partial_{k_i} P_{\bsl{k}} \partial_{k_i} P_{\bsl{k}}] = \sum_{n = 1,2}\Tr[\widetilde{g}_{n,\bsl{k}}]\ ,
}
where
\eq{
[\widetilde{g}_{n,\bsl{k}}]_{ij} = \frac{1}{2}\Tr\left[\partial_{k_i} \widetilde{P}_{n,\bsl{k}}\ \partial_{k_j} \widetilde{P}_{n,\bsl{k}} \right]
}
is the quantum metric of $\ket{\widetilde{u}_{\bsl{k},n}}$ and $\ket{\bar{u}_{\bsl{k},n}}$.
As a result, \cref{eq:WL_ideal} infers
\eq{
\label{eq:ideal_chern_intermediate}
\sum_{n = 1,2}\frac{1}{2\pi}\int_{\BZ} d^2k\ \Tr[\widetilde{g}_{n,\bsl{k}}] = |\Ch_n| = \left| \frac{1}{2\pi} \int_{\BZ} d^2k f_n(\bsl{k}) \right| \ ,
}
which means each component $\ket{\widetilde{u}_{\bsl{k},n}}$ is an Chern-ideal state.

\subsubsection{Inversion-Protected Fragile Ideal Bands}

In the case of an isolated set of two bands with inversion-projected fragile topology in \cref{eq:2D_inversion_fragile_bound}, we have $|w|=1$, and the $|\Ch_n|=1$.
Thus, the inversion-fragile ideal states consists of two Chern-ideal states with opposite Chern numbers $\pm 1$, and each Chern-ideal state is inversion symmetric.

 \section{General Framework for Monotonic Flow}
 \label{app:monotonic_flow}

 With the definition of Wilson-loop ideal bands discussed in \cref{app:WL_ideal_bands}, we now address the question of how to have a set of ideal bands.
 In this section, we will provide a general framework that can start from a set of topological bands and make them ideal by mixing them with other bands while maintaining the original topology.
 It is done by setting up the flow equation of of the projector $P_{t,\bsl{k}}$ of a set of periodic parts of $N$ Bloch states, \ie, $\ket{u_{t,\bsl{k}}} = (\ket{u_{t,\bsl{k},1}},\ket{u_{t,\bsl{k},2}},\ldots,\ket{u_{t,\bsl{k},N}})$.
 Here $t$ is the parameter that governs the flow, similar to the time in the Schr\"odinger equation.
 (We note that $t$ is not the physical time---it cannot be flipped by the TR symmetry.)
 Importantly, the projector $P_{t,\bsl{k}}$ must smoothly depend on $\bsl{k}$ and a time variable $t$, and satisfies the following embedding relation
 \eq{
 P_{t,\bsl{k}+\bsl{G}} = V_{\bsl{G}} P_{t,\bsl{k}} V_{\bsl{G}}^\dagger,
 }
 where $\bsl{G}$ ranges over all reciprocal lattice vectors, and $V_{\bsl{G}}$ is unitary and independent of $t$.
 Of course, if $P_{t,\bsl{k}}$ is the projector of an isolated set of $N$ bands, it surely satisfies the requirement. 

 We use $g_{t,\bsl{k}}$ to label the quantum metric of $P_{t,\bsl{k}}$ at each time $t$ and $\bsl{k}$:
 \eq{
 \left[g_{t,\bsl{k}}\right]_{ij} = \frac{1}{2}\Tr\left[\partial_{k_i} P_{t,\bsl{k}} \partial_{k_j} P_{t,\bsl{k}} \right]
 }
 with $i,j$ ranging over all the Cartesian directions.
 The flow equation is set up by choosing a functional $S_t$ of $g_{t,\bsl{k}}$ that satisfies 
 \begin{itemize}
     \item the functional derivative $\frac{\delta S_t}{\delta \left[g_{t,\bsl{k}}\right]_{ij}}$ is invariant under $\bsl{k}\rightarrow \bsl{k}+\bsl{G}$ for any reciprocal lattice vector $\bsl{G}$
     \item the time dependence of $S_t$ only comes from $g_{t,\bsl{k}}$.
 \end{itemize}
 Here, $\delta$ is the functional derivative in the space of functions of $\bsl{k}$, and we treat all components of $g_{t,\bsl{k}}$ to be independent when taking the functional derivative.

 We now provide a procedure to construct a flow of $P_{t,\bsl{k}}$ that monotonically decreases $S_t$ as $t$ increases.
 First, we define a matrix by taking functional derivative on $S_t$:
 \eqa{
 \label{eq:A_general}
  A_{t,\bsl{k}} & =  \sum_{ij} \frac{\delta S_t}{\delta \left[g_{t,\bsl{k}}\right]_{ij} } \partial_{k_i} \partial_{k_j} P_{t,\bsl{k}} + \sum_{ij}  \left(\partial_{k_i} \frac{\delta S_t}{\delta \left[g_{t,\bsl{k}}\right]_{ij} } \right) \partial_{k_j} P_{t,\bsl{k}} \\
  & =  \sum_{ij}  \partial_{k_i} \left(\frac{\delta S_t}{\delta \left[g_{t,\bsl{k}}\right]_{ij} }\partial_{k_j} P_{t,\bsl{k}} \right) 
 }
 Second, we require $P_{t,\bsl{k}}$ to satisfy
 \eq{
 \label{eq:P_flow_monotonic}
 \partial_t P_{t,\bsl{k}} =  \alpha  (P_{t,\bsl{k}} A_{t,\bsl{k}} \bar{Q}_{t,\bsl{k}} + \bar{Q}_{t,\bsl{k}} A_{t,\bsl{k}} P_{t,\bsl{k}})\ ,
 }
 where $\bar{Q}_{t,\bsl{k}}$ is the projector to the part of the states orthogonal to $P_{t,\bsl{k}}$, \ie,
 \eq{
 \bar{Q}_{t,\bsl{k}}^2 = \bar{Q}_{t,\bsl{k}}\ ,\ Q_{t,\bsl{k}} \bar{Q}_{t,\bsl{k}} Q_{t,\bsl{k}} =  \bar{Q}_{t,\bsl{k}}\ , \ \Tr[\bar{Q}_{t,\bsl{k}}] \leq \Tr[Q_{t,\bsl{k}}]\ ,
 }
 and $Q_{t,\bsl{k}} = 1 - P_{t,\bsl{k}}$.

 By further choosing $\alpha > 0$, we arrive at
 \eqa{
 \frac{d}{dt} S_t & =  \int_{\BZ} d^d k\   \sum_{ij} \frac{\delta S_t}{\delta \left[g_{t,\bsl{k}}\right]_{ij} } \partial_t \left[g_{t,\bsl{k}}\right]_{ij} =  \int_{\BZ} d^d k\   \sum_{ij} \frac{\delta S_t}{\delta \left[g_{t,\bsl{k}}\right]_{ij} } \Tr[ \partial_{k_i} \partial_t P_{t,\bsl{k}}  \partial_{k_j} P_{t,\bsl{k}} ] \\
 & =  - \int_{\BZ} d^d k\   \sum_{ij} \frac{\delta S_t}{\delta \left[g_{t,\bsl{k}}\right]_{ij} } \Tr[  \partial_t P_{t,\bsl{k}}  \partial_{k_i} \partial_{k_j} P_{t,\bsl{k}} ] - \int_{\BZ} d^d k\   \sum_{ij} \left(\partial_{k_i} \frac{\delta S_t}{\delta \left[g_{t,\bsl{k}}\right]_{ij}  }\right) \Tr[  \partial_t P_{t,\bsl{k}}   \partial_{k_j} P_{t,\bsl{k}} ] \\
 & = - \int_{\BZ} d^d k  \Tr[  \partial_t P_{t,\bsl{k}} A_{t,\bsl{k}} ] = - \alpha   \int_{\BZ} d^d k  \Tr[ (P_{t,\bsl{k}} A_{t,\bsl{k}} \bar{Q}_{t,\bsl{k}} + \bar{Q}_{t,\bsl{k}} A_{t,\bsl{k}} P_{t,\bsl{k}})A_{t,\bsl{k}} ] \\
 & = - 2 \alpha   \int_{\BZ} d^d k  \Tr[ P_{t,\bsl{k}} A_{t,\bsl{k}} \bar{Q}_{t,\bsl{k}}  \bar{Q}_{t,\bsl{k}} A_{t,\bsl{k}}P_{t,\bsl{k}} ]\ ,
 }
 which means 
 \eq{
 \frac{d}{dt} S_t \leq 0\ ,
 }
 \ie, the flow equation of $P_{t,\bsl{k}}$ in \cref{eq:P_flow_monotonic} monotonically decreases $S_t$.

 In the following, we will use this framework to construct three flows that will be used to find ideal bands from a generic set of topological bands.

 \subsection{Souza-Marzari-Vanderbilt Flow}
 \label{app:SMV_flow}

 For the first flow, we choose
 \eq{
 S_t = \int_{1BZ} d^d k\ \Tr[g_{t,\bsl{k}}]\ ,
 }
 which is the integral of the trace of the quantum metric.
 In this case, the functional derivative of $S_t$ reads
 \eq{
 \frac{\delta S_t}{\delta \left[g_{t,\bsl{k}}\right]_{ij} } = \sum_{i'} \frac{\partial \left[g_{t,\bsl{k}}\right]_{i'i'}}{\partial\left[g_{t,\bsl{k}}\right]_{ij} } = \sum_{i'}  \delta_{ii'}\delta_{ji'} = \delta_{ij}\ .
 }
 Combined with \cref{eq:A_general}, $A_{t,\bsl{k}}$ in \cref{eq:P_flow_monotonic} reads
 \eq{
 A_{t,\bsl{k}}= \nabla_{\bsl{k}}^2 P_{t,\bsl{k}}\ ,
 }
 leading to the following explicit flow equation
 \eq{
 \label{eq:Vanderbilt_flow}
 \partial_t P_{\bsl{k}} =  \alpha  (P_{t,\bsl{k}}\nabla_{\bsl{k}}^2 P_{t,\bsl{k}} \bar{Q}_{t,\bsl{k}} + \bar{Q}_{t,\bsl{k}} \nabla_{\bsl{k}}^2 P_{t,\bsl{k}} P_{t,\bsl{k}})\ .
 }
 \cref{eq:Vanderbilt_flow} is simply the continuous version of the disentangling part of the Wannierization algorithm proposed by Souza, Marzari and Vanderbilt in \refcite{Vanderbilt2001MLWFMultibands}, which is designed to find the subspace with minimal integrated trace of quantum metric. 
 The key difference between the usage of \cref{eq:Vanderbilt_flow} and the procedure in \refcite{Vanderbilt2001MLWFMultibands} is that the flow here will start from topological states while that in \refcite{Vanderbilt2001MLWFMultibands} starts from trial atomic states.
 Nevertheless, we refer to \cref{eq:Vanderbilt_flow} as the Souza-Marzari-Vanderbilt (SMV) flow.

 To demonstrate the connection between \cref{eq:Vanderbilt_flow} and the discrete disentangling flow in \refcite{Vanderbilt2001MLWFMultibands}, let us consider a discrete mesh for the Bloch momentum, and the disentangling algorithm in \refcite{Vanderbilt2001MLWFMultibands} can be equivalently 
 re-written as follows.
 Given a specific $\bsl{k}$ point in the mesh, $\bsl{k}$ has neighbors labeled by $\bsl{k}+\bsl{q}$, where the number of different values of $\bsl{q}$ is equal to the number of neighbors considered.
 If we include one specific $\bsl{q}_0$, we should include all other $\bsl{q}$ that have the same magnitude as $\bsl{q}_0$, and then we can further define a weight $w_{|\bsl{q}|}$ that satisfies
 \eq{
 \sum_{\bsl{q}} w_{|\bsl{q}|}  q_i q_j = \delta_{ij} \ .
 }
 For example, for the square lattice with lattice constant 1, we can choose the mesh as $(2\pi l_1 /L, 2\pi l_2 /L )$ with $l_1,l_2=0,1,2,\ldots, L-1$, which means $\bsl{q}=(\pm 2\pi/L,0), (0,\pm 2\pi/L)$ and $w_{|\bsl{q}|} = L^2/(8\pi^2)$.
 Then, for the iteration step $i$ (for which the projector is labeled as $P^{(i)}_{\bsl{k}}$), we can generate the projector $P^{(i+1)}_{\bsl{k}}$ for the step $i+1$ by
 diagonalizing
 \eq{
 \label{eq:Y_matrix}
 Y^{(i)}_{\bsl{k}} = \frac{1}{\sum_{\bsl{q}} w_{|\bsl{q}|} }\sum_{\bsl{q}} w_{|\bsl{q}|}  P_{\bsl{k}+\bsl{q}}^{(i)}
 }
 in the subspace specified by $P_{\bsl{k}}^{(i)}$ and chosen $\bar{Q}_{\bsl{k}}^{(i)}$.
 Then, $P^{(i+1)}_{\bsl{k}}$ is the projector given by the eigenvectors for the $N$ largest eigenvalues, where $N=\Tr[P^{(i)}_{\bsl{k}}]$ is the number of the Bloch states at each $\bsl{k}$.
 Perform the iteration until $P_{\bsl{k}}$ converges. 

 Now let us connect the algorithm to \cref{eq:Vanderbilt_flow}.
 First note that
 \eqa{
 \sum_{\bsl{q}} w_{|\bsl{q}|} P_{\bsl{k}+\bsl{q}} & = \sum_{\bsl{q}} w_{|\bsl{q}|} \left[  P_{\bsl{k}} + \bsl{q}\cdot\nabla_{\bsl{k}} P_{\bsl{k}} + \frac{1}{2}(\bsl{q}\cdot\nabla_{\bsl{k}})^2 P_{\bsl{k}} \right] + O(|\bsl{q}|^3) \\
 & = \sum_{\bsl{q}} w_{|\bsl{q}|}  P_{\bsl{k}} +  \frac{1}{2}\nabla_{\bsl{k}}^2 P_{\bsl{k}}  + O(|\bsl{q}|^1) \ ,
 }
 since $\sum_{\bsl{q}} w_{|\bsl{q}|} = O(|\bsl{q}|^{-2})$.
 Let us define $\alpha = 1/2$ and $dt = (\sum_{\bsl{q}} w_{|\bsl{q}|})^{-1}$, and then the matrix that we diagonalize at the step $i$ is equivalent to
 \eq{
 (P^{(i)}_{\bsl{k}}+\bar{Q}_{\bsl{k}}^{(i)})Y_{\bsl{k}}^{(i)} (P^{(i)}_{\bsl{k}}+\bar{Q}_{\bsl{k}}^{(i)})=  P_{\bsl{k}}^{(i)} + dt\ \alpha (P^{(i)}_{\bsl{k}}+\bar{Q}_{\bsl{k}}^{(i)})\nabla_{\bsl{k}}^2 P_{\bsl{k}}^{(i)} (P^{(i)}_{\bsl{k}}+\bar{Q}_{\bsl{k}}^{(i)}) + O(|\bsl{q}|^3)\ .
 }
 Then, according to perturbation theory, we have
 \eq{
 P_{\bsl{k}}^{(i+1)} = P_{\bsl{k}}^{(i)} + dt\ \alpha \left[ P_{\bsl{k}}^{(i)} \nabla_{\bsl{k}}^2 P_{\bsl{k}}^{(i)} \bar{Q}_{\bsl{k}}^{(i)} + \bar{Q}_{\bsl{k}}^{(i)} \nabla_{\bsl{k}}^2 P_{\bsl{k}}^{(i)} P_{\bsl{k}}^{(i)} \right] + O(|\bsl{q}|^3)\ ,
 }
 which, after taking the continuum limit $|\bsl{q}|\rightarrow 0 \Rightarrow dt\rightarrow 0$,  becomes \cref{eq:Vanderbilt_flow}.

 \subsection{Static-Target Flow}

 Now we device a flow towards a static target $\left[\bar{g}_{\bsl{k}}\right]_{ij}$.
 To do so, we choose the following functional 
 \eq{
 \label{eq:S_static_target_flow}
 S_t = \int d^d k  \sum_{ij} \left( \left[\bar{g}_{\bsl{k}}\right]_{ij} - \left[g_{t,\bsl{k}}\right]_{ij}\right)^2\ ,
 }
 which is the squared $L_2$ distance between $\bar{g}_{\bsl{k}}$ and $g_{t,\bsl{k}}$.
 \cref{eq:S_static_target_flow} leads to
 \eq{
 \label{eq:A_static_target_flow}
 A_{t,\bsl{k}} = \sum_{i j } \partial_{k_i} \left[ - 2 ( \left[\bar{g}_{\bsl{k}}\right]_{ij} - \left[g_{t,\bsl{k}}\right]_{ij}) \partial_{k_j} P_{t,\bsl{k}} \right]\ .
 }
 Then, we refer to \cref{eq:P_flow_monotonic} with \cref{eq:A_static_target_flow} as the static-target flow, since it monotonically decreases $S_t$, which means it drives $g_{t,\bsl{k}}$ towards the static target $\bar{g}_{\bsl{k}}$.

 \subsection{Dynamical-Target Flow}

 For the third flow, we adopt the form of \cref{eq:S_static_target_flow} but choose a dynamical target $\left[\bar{g}_{t,\bsl{k}}\right]_{ij}$ instead of a static one, resulting in 
 \eqa{
 \label{eq:S_dynamical_target}
 S_t & = \int d^d k  \sum_{ij} \left( \left[\bar{g}_{t,\bsl{k}}\right]_{ij} - \left[g_{t,\bsl{k}}\right]_{ij}\right)^2 \ .
 }
 We refer to the flow \cref{eq:P_flow_monotonic} determined by \cref{eq:S_dynamical_target} as the dynamical-target flow.
 As we require the $t$ dependence of $S_t$ to solely come from $g_{t,\bsl{k}}$, $\bar{g}_{t,\bsl{k}}$ has to depend on $g_{t,\bsl{k}}$, and thus the form of the $A_{t,\bsl{k}}$ is not as simple as \cref{eq:A_static_target_flow}.

 As a specific example, let us focus on 2D ($d=2$) the following form of $\bar{g}_{t,\bsl{k}}$:
 \eq{
 \label{eq:dynamical_target_form}
 \left[\bar{g}_{t,\bsl{k}}\right]_{ij} =  \frac{B}{2 \Tr\G_t  } \Tr[g_{t,\bsl{k}}]\delta_{ij}\ ,
 }
 where
 \eq{
 \Tr\G_t = \int d^2 k \Tr[g_{t,\bsl{k}}]\ ,
 }
 and $B$ is a constant.
 With \cref{eq:dynamical_target_form}, the functional now reads
 \eqa{
 \label{eq:S_dynamical_idealization}
 S_t & = \int d^2 k  \sum_{ij} \left( \frac{B}{ 2 \Tr\G_t  } \Tr[g_{t,\bsl{k}}]\delta_{ij} - \left[g_{t,\bsl{k}}\right]_{ij}\right)^2 \\
 & = 2 \left(\frac{B}{2 \Tr\G_t  } \right)^2\int d^2 k  \Tr[g_{t,\bsl{k}}]^2 -  \frac{B}{ \Tr\G_t  } \int d^2 k    \Tr[g_{t,\bsl{k}}]^2 + \int d^2 k  \sum_{ij} \left[g_{t,\bsl{k}}\right]_{ij}^2\\
 }
 where $B$ is a static constant.
 The functional derivative of $S_t$ reads
 \eqa{
 \frac{\delta S_t }{\delta [g_{t,\bsl{k}}]_{ij}} & = \left(\frac{B}{ \Tr\G_t  } \right)^2 \Tr[g_{t,\bsl{k}}] \delta_{ij} - 2 \frac{B}{ \Tr\G_t  } \Tr[g_{t,\bsl{k}}] \delta_{ij}  + 2 \left[g_{t,\bsl{k}}\right]_{ij} \\
 & \qquad - \left(\frac{B}{ \Tr\G_t  } \right)^2 \frac{1}{\Tr\G_t}\int d^2 k  \Tr[g_{t,\bsl{k}}]^2 \delta_{ij} +  \frac{B}{ \Tr\G_t  } \frac{1}{\Tr\G_t}\int d^2 k    \Tr[g_{t,\bsl{k}}]^2 \delta_{ij} \\
 & =  - 2 \left( \frac{B}{ 2 \Tr\G_t  } \Tr[g_{t,\bsl{k}}]\delta_{ij} - \left[g_{t,\bsl{k}}\right]_{ij}\right) + \left[ \frac{B}{ \Tr\G_t  } - 1  \right] \frac{B}{ \Tr\G_t  }\ \delta_{ij}  \left( \Tr[g_{t,\bsl{k}}] -  \frac{1}{\Tr\G_t}\int d^2 k  \Tr[g_{t,\bsl{k}}]^2 \right)\\
 }
 which gives the following expression of $A_{t,\bsl{k}}$:
 \eqa{
 \label{eq:A_dynamical_idealization}
 A_{t,\bsl{k}} & =  \sum_{ij}  \partial_{k_i} \left(- 2 \left( \frac{B}{2 \Tr\G_t  } \Tr[g_{t,\bsl{k}}]\delta_{ij} - \left[g_{t,\bsl{k}}\right]_{ij}\right)  \partial_{k_j} P_{t,\bsl{k}} \right)\\
 & \qquad + \left[ \frac{B}{ \Tr\G_t  } - 1  \right] \frac{B}{ \Tr\G_t  }\ \left( \Tr[g_{t,\bsl{k}}] -  \frac{1}{\Tr\G_t}\int d^2 k  \Tr[g_{t,\bsl{k}}]^2 \right)\nabla_{\bsl{k}}^2  P_{t,\bsl{k}} \\
 & \qquad +  \left[ \frac{B}{ \Tr\G_t  } - 1  \right] \frac{B}{ \Tr\G_t  }\sum_{i}\partial_{k_i}\Tr[g_{t,\bsl{k}}] \partial_{k_i}  P_{t,\bsl{k}}\\
  & =  \sum_{ij}  \partial_{k_i} \left(- 2 \left(\left[\bar{g}_{t,\bsl{k}}\right]_{ij} - \left[g_{t,\bsl{k}}\right]_{ij}\right)  \partial_{k_j} P_{t,\bsl{k}} \right)\\
 & \qquad + \left[ \frac{B}{ \Tr\G_t  } - 1  \right] \frac{B}{ \Tr\G_t  }\ \left( \Tr[g_{t,\bsl{k}}] -  \frac{1}{\Tr\G_t}\int d^2 k  \Tr[g_{t,\bsl{k}}]^2 \right)\nabla_{\bsl{k}}^2  P_{t,\bsl{k}} \\
 & \qquad + 2 \left[ \frac{B}{ \Tr\G_t  } - 1  \right] \sum_{ij}\partial_{k_i}\left[\bar{g}_{t,\bsl{k}}\right]_{ij} \partial_{k_j}  P_{t,\bsl{k}}\ .
 }

 By choosing $B$ to be the WL lower bound of $\Tr\G_t$, $\bar{g}_{t,\bsl{k}}$ in \cref{eq:dynamical_target_form} saturates the WL lower bound.
 Then, \cref{eq:P_flow_monotonic} with \cref{eq:A_dynamical_idealization} is a dynamical-target flow that drives $\bar{g}_{t,\bsl{k}}$ to a dynamical ideal $\bar{g}_{t,\bsl{k}}$, serving as a process of the dynamical idealization.

\section{Model Calculations}

We now use the three flows discussed in \cref{app:monotonic_flow} to find ideal bands in two models: the twisted bilayer MoTe$_2$ in \refcite{Zhang2024UniversalMoireModel}, and the moiré Rashba model adapted from \refcite{Liu_2025_Moire_Rashba}.
In both models, we always choose $\bar{Q}_{t,\bsl{k}} = Q_{t,\bsl{k}}$ in \cref{eq:P_flow_monotonic}, which means we allow band mixing among all bands.
Furthermore, the numerical calculation is always done on a discrete mesh of $\bsl{k}$ and $t$.
Therefore, we should use a discrete version of the flow equation \cref{eq:P_flow_monotonic}, similar to the discretization of the SMV flow discussed in \cref{app:SMV_flow}.
Specifically, given the projector $P_{t_l,\bsl{k}}$ at the iteration step $l$, the projector $P_{t_{l+1},\bsl{k}}$ is projection operator to the eigen-subspace of the $N$ largest eigenvalues of
\eq{
\label{eq:Y_discrete}
 Y_{l,\bsl{k}} =P_{t_{l},\bsl{k}}  +  \alpha\  d t_l\  A_{t_{l},\bsl{k}}\ ,
}
where $d t_l = t_{l+1} - t_l$.
Clearly, \cref{eq:Y_discrete} reduces to \cref{eq:P_flow_monotonic} in the limit $d t_l \rightarrow 0$.

\subsection{Twisted Bilayer MoTe$_2$}

\label{app:tMoTe2}

Twisted bilayer MoTe$_2$ consists of two valleys, $\K$ and $\K'$, related by the TR symmetry, and we only focus on the $\K$ valley here.
We will use the faithful model provided in \refcite{Zhang2024UniversalMoireModel}, which is constructed directly from density functional theory (DFT) calculations without any continuous parameter fitting.
As a brief review, the $\K$ valley model we use takes the following general form
\eq{
\label{eq:H_K_0}
H_{\K,0}  = \sum_{M_x, M_y \in \dsN} \sum_{l,l'=t,b}\int d^2 r \left(\ii^{M_x+M_y} \partial_x^{M_x} \partial_y^{M_y} c^\dagger_{\K,l,\bsl{r}} \right) t^{M_x M_y}_{ll'}(\bsl{r})c_{\K,l',\bsl{r}} \ ,
}
where $c^\dagger_{\K,l,\bsl{r}}$ creates an electron in $\K$ valley on layer $l$ at position $\bsl{r}$, and $\dsN$ is the set of non-negative integers.
In particular, the intralayer and interlayer $t^{M_x M_y}_{ll'}(\bsl{r})$ have the following form
\eqa{
& t^{M_x M_y}_{ll}(\bsl{r}) = \sum_{\bsl{G}_M} V_{l,\bsl{G}_M}^{M_x M_y} e^{-\ii \bsl{G}_M\cdot \bsl{r} }\\
& t^{M_x M_y}_{bt}(\bsl{r}) = \sum_{\bsl{G}_M} w_{bt,\bsl{q}_1+\bsl{G}_M}^{M_x M_y} e^{-\ii (\bsl{q}_1+\bsl{G}_M)\cdot \bsl{r} }\\
& t^{M_x M_y}_{tb}(\bsl{r}) = \sum_{\bsl{G}_M} w_{bt,-\bsl{q}_1+\bsl{G}_M}^{M_x M_y} e^{-\ii (-\bsl{q}_1+\bsl{G}_M)\cdot \bsl{r} }\ ,
}
where $\bsl{G}$ is the moiré reciprocal lattice vector that is the linear combination of the primitive reciprocal lattice vectors $\bsl{b}_{M,1} = \frac{4 \pi}{\sqrt{3} a_{M}} (1,0)$ and $\bsl{b}_{M,2} = \frac{4 \pi}{\sqrt{3} a_{M}} (\frac{1}{2},\frac{\sqrt{3}}{2})$, and $\bsl{q}_1 = \frac{4 \pi}{3 a_{M}} (0,1)$.
Here 
\eq{
a_M = \frac{a_0}{2 \sin\left( \frac{\theta}{2} \right)} 
}
is the moiré lattice constant, $\theta$ is the twist angle, and $a_0$ is the lattice constant of the monolayer MoTe$_2$.
The $V_{l,\bsl{G}_M}^{M_x M_y}$, $w_{bt,\bsl{q}_1+\bsl{G}_M}^{M_x M_y}$ and $w_{bt,-\bsl{q}_1+\bsl{G}_M}^{M_x M_y}$ are model parameters determined from DFT calculations, and \refcite{Zhang2024UniversalMoireModel} provides a faithful way to do so without involving any fitting.
Here we choose $\theta=3.89^\circ$ and use the corresponding parameter values in Table X and Table XI of \refcite{Zhang2024UniversalMoireModel}.
The Hamiltonian is diagonalized in the momentum space with the set of $\bsl{Q}=\pm \bsl{q}_1 +\bsl{G}_M$ points chosen to include all $\bsl{Q}$ that satisfy $|\bsl{Q}|\leq 5|\bsl{q}_1|$---there are 60 $\bsl{Q}$ points in total.

We apply all three flows in \cref{app:monotonic_flow} to this model.
The starting projector $P_{t=0,\bsl{k}}$ is always that of the top energy band, which has the Chern number $1$ and $\Tr\G/(2\pi) = 1.330$.
All the flows are calculated on a $60\times 60$ momentum mesh, \ie, $\bsl{k} = (l_1/L) \bsl{b}_{M,1} + (l_2/L) \bsl{b}_{M,2}$ with $l_1,l_2 = 0,1,2,...,L-1$ and $L=60$, unless specified otherwise.

For the SMV flow, we choose $\alpha d t_l$ in \cref{eq:Y_discrete} to be $\alpha d t_l = |\bsl{q}_1|^2/40000$ for all steps, and we stop at iteration step $12800$, which already achieves $\Tr\G/(2\pi) = 1.004$.
Besides the SMV flow on $60\times 60$ mesh, we also perform the flow on $210\times 210$ mesh for the exact-diagonalization (ED) calculations in \cref{sec:ED_tMoTe2}.
For the flow on $210\times 210$ mesh, we choose $\alpha d t_l$ in \cref{eq:Y_discrete} to be $\alpha d t_l = |\bsl{q}_1|^2/49000$ for all steps, and we stop at iteration step $16400$, which already achieves $\Tr\G/(2\pi) = 1.004$.

For the static-target flow, we choose $ \bar{g}_{\bsl{k}}$ in \cref{eq:S_static_target_flow} as that of the lowest Landau level:
\eq{
\left[\bar{g}_{\bsl{k}}\right]_{ij} = \frac{\Omega}{4\pi} \delta_{ij}\ , 
}
where $\Omega$ is the moiré unit cell area.
For the flow parameters, we choose the value of $\alpha d t_l$ in \cref{eq:Y_discrete} based on the following criterion.
We label the difference between the $N$th and $N+1$th largest eigenvalues of $Y_{l,\bsl{k}}$ as $\Delta y_{l,\bsl{k}}$, and we define $\Delta y_l = \max_{\bsl{k}} y_{l,\bsl{k}}$ to be the maximum value of $y_{l,\bsl{k}}$ at a fixed step $l$.
Then, $d t_{l+1}$ is determined from $d t_l$ via 
\eq{
\label{eq:chi}
(d t_{l+1} / d t_l )^2 = \chi / | 1 - y_{l,\bsl{k}}| \ .
}
We choose $\chi = 0.001$, and $d t_0 \alpha = |\bsl{q}_1|^4/40000$ for the static-target flow.
In addition, in order to avoid instability caused by error accumulation, we need to replace $g_{t,\bsl{k}}$ used in the left-hand side of \cref{eq:P_flow_monotonic} by a smoothened version $g_{t,\bsl{k}}^{smooth}$ every step.
The smooth $g_{t,\bsl{k}}^{smooth}$ is constructed by first Fourier transforming $g_{t,\bsl{k}}$ to the real space
\eq{
g_{t,\bsl{R}_M} = \sum_{\bsl{k}} e^{\ii \bsl{R}_M\cdot \bsl{k}} g_{t,\bsl{k}}\ ,
}
where $\bsl{R}_M$ is the moiré lattice vector that is a linear combination of the basis lattice vectors $\bsl{a}_{M,1} = a_M (-\frac{\sqrt{3}}{2},-\frac{1}{2})$ and $\bsl{a}_{M,2} = C_3 \bsl{a}_{M,1}$ with $C_3$ the three-fold rotation.
Then, we put a cutoff $\Lambda$ on $\bsl{R}_M$, and Fourier transform $g_{t,\bsl{R}_M}$ back to the momentum space, leading to the smoothened $g_{t,\bsl{k}}^{smooth}$:
\eq{
\label{eq:smooth_g_in_flow}
g_{t,\bsl{k}}^{smooth} = \frac{1}{L^2}\sum_{\bsl{R}_M}^{|\bsl{R}_M|\leq \Lambda} e^{-\ii \bsl{R}_M\cdot \bsl{k}} g_{t,\bsl{R}_M} \ .
}
We choose $\Lambda = 2 a_M$ for the static-target flow.

For the dynamical idealization, we dynamically choose $\alpha dt_l$ by specifying $\chi$ in \cref{eq:chi} and choosing $\alpha dt_0 = |\bsl{q}_{1}|^4/400000$.
In addition, in order to avoid instability caused by error accumulation, we again need to replace $g_{t,\bsl{k}}$ used in the left-hand side of \cref{eq:P_flow_monotonic} by a smoothened version $g_{t,\bsl{k}}^{smooth}$ every step.
The $\chi$ in \cref{eq:Y_discrete} and $\Lambda $ in \cref{eq:smooth_g_in_flow} are chosen as follows: 
\begin{itemize}
    \item $\chi = 0.001$ and $\Lambda = 4 |\bsl{a}_{M,1}|$ for the first 7500 steps;
    \item $\chi = 0.0005$ and $\Lambda = 3 |\bsl{a}_{M,1}|$ for the last 3900 steps.
\end{itemize}

The probabilities of the final $P_{t_f, \bsl{k}}$ on the highest 10 energy bands of the $\K$-valley twist bilayer MoTe$_2$ model for all three flows are shown in \cref{tab:prob_tMoTe2}.

\begin{table}[t]
    \centering
    \begin{tabular}{c|cccccccccc}
    \text{band index} & 1 & 2 & 3 & 4 & 5 & 6 & 7 & 8 & 9 & 10 \\
    \hline
    \text{SMV Flow} & 0.942 & 0.016 & 0.012 & 0.008 & 0.005 & 0.005 & 0.002 & 0.004 & 0.002 & 0.001 \\
    \text{static-target Flow} & 0.939 & 0.015 & 0.016 & 0.009 & 0.004 & 0.005 & 0.002 & 0.004 & 0.002 & 0.001 \\
    \text{dynamical-target Flow} & 0.926 & 0.018 & 0.010 & 0.010 & 0.007 & 0.008 & 0.003 & 0.006 & 0.004 & 0.002 \\
    \end{tabular}
    \caption{The probabilities of the final $P_{t_f, \bsl{k}}$ from each flow on the energy bands of the twisted bilayer MoTe$_2$ model.
    The band index $n$ is labeled from high energies to low energies, \ie, $E_{n+1}(\bsl{k})\leq E_{n}(\bsl{k})$, and the highest energy band is band $1$.
    The probability of $P_{t_f, \bsl{k}}$ on band $n$ is calculated as $\sum_{\bsl{k}\in\BZ}\Tr[P_{t_f, \bsl{k}} P_{n, \bsl{k}}]/( \Tr[P_{t_f, \bsl{k}}] L^2)$, where $P_{n, \bsl{k}}$ is the projector of band $n$, and $L^2$ is the number of $\bsl{k}$ points in 1BZ.
    }
    \label{tab:prob_tMoTe2}
\end{table}

\subsubsection{Exact-Diagonalization Results}
\label{sec:ED_tMoTe2}

The many-body Hamiltonian for ED is the projected version of the hole Hamiltonian in the continuum, which reads
\eq{
H_{cont} = H_{\K,0}^h + H_{\K,int}^h\ ,
}
where 
\eq{
H_{\K,0}^h = - \sum_{M_x, M_y \in \dsN} \sum_{l,l'=t,b}\int d^2 r \left(\ii^{M_x+M_y} \partial_x^{M_x} \partial_y^{M_y} \widetilde{c}^\dagger_{\K,l,\bsl{r}} \right) t^{M_x M_y}_{ll'}(\bsl{r})\widetilde{c}_{\K,l',\bsl{r}} 
}
with $t^{M_x M_y}_{ll'}$ in \cref{eq:H_K_0},
\eq{
H_{\K,int}^h = \frac{1}{2}\sum_{l l' } \int d^2 r d^2 r' V(\bsl{r}-\bsl{r}')  \widetilde{c}^\dagger_{\K, l, \bsl{r}} \widetilde{c}^\dagger_{\K, l', \bsl{r}'} \widetilde{c}_{\K, l', \bsl{r}'} \widetilde{c}_{\K, l, \bsl{r}}  \ ,
}
where $\widetilde{c}_{\K, l,\bsl{r}}^\dagger$ creates a hole at position $\bsl{r}$ in the $l$th layer in the $\K$ valley.
Here we choose the double-gated screened Coulomb potential $V(\bsl{r})$ with gate distance  $\xi$:
\eq{
V(\bsl{r}) =\int_{\dsR^2} \frac{d^2 p }{ (2\pi)^2} V(\bsl{p}) e^{\ii \bsl{p}\cdot\bsl{r}}\ ,
}
where
\eq{
\qquad V(\bsl{p}) = \pi \xi^2 \frac{e^2 }{4\pi \epsilon_r \epsilon_0 \xi} \frac{\tanh(\xi |\bsl{p}|/2)}{\xi |\bsl{p}|/2}\ ,
}
$\epsilon_0$ is the vacuum permitivity, and $\epsilon_r$ is the relative dielectric constant.
Note that the hole single-particle part $H_{\K,0}^h$ has the same eigenstates as the electron version in \cref{eq:H_K_0}, and just has opposite eigenvalues.

Given a single-particle rank-1 projector $P_{\bsl{k}}$, the continuum many-body Hamiltonian is projected to the subspace determined by it, and obtain 
\eqa{
    H &= \mathcal{H}_0  + \sum_{\bsl{k}_1', \bsl{k}_2', \bsl{k}_3', \bsl{k}_4'} V(\bsl{k}_1', \bsl{k}_2', \bsl{k}_3', \bsl{k}_4') \widetilde{\gamma}_{\bsl{k}_1'}^\dagger \widetilde{\gamma}_{\bsl{k}_2'}^\dagger \widetilde{\gamma}_{\bsl{k}_3'} \widetilde{\gamma}_{\bsl{k}_4'}\ , \label{eq:ED_hamiltonian}
}
where
\eq{
\mathcal{H}_0 = \sum_{\bsl{k}} \widetilde{\gamma}_{\bsl{k}}^\dagger \widetilde{\gamma}_{\bsl{k}} \epsilon_{\bsl{k}}\ ,
}
\eq{
\widetilde{\gamma}_{\bsl{k}}^{\dagger} = \sum_{\bsl{Q}} \widetilde{c}_{\K, \bsl{k},\bsl{Q}}^{\dagger} U_{\bsl{Q}}(\bsl{k})
}
with $U(\bsl{k}) U^\dagger(\bsl{k}) = P_{\bsl{k}}$ at each $\bsl{k}$ and $\widetilde{c}_{\K, \bsl{k},\bsl{Q}}^\dagger$ the Fourier transformation of $\widetilde{c}^\dagger_{\K,l,\bsl{r}}$.

\begin{figure}[t]
    \centering
    \includegraphics[width=1.0\linewidth]{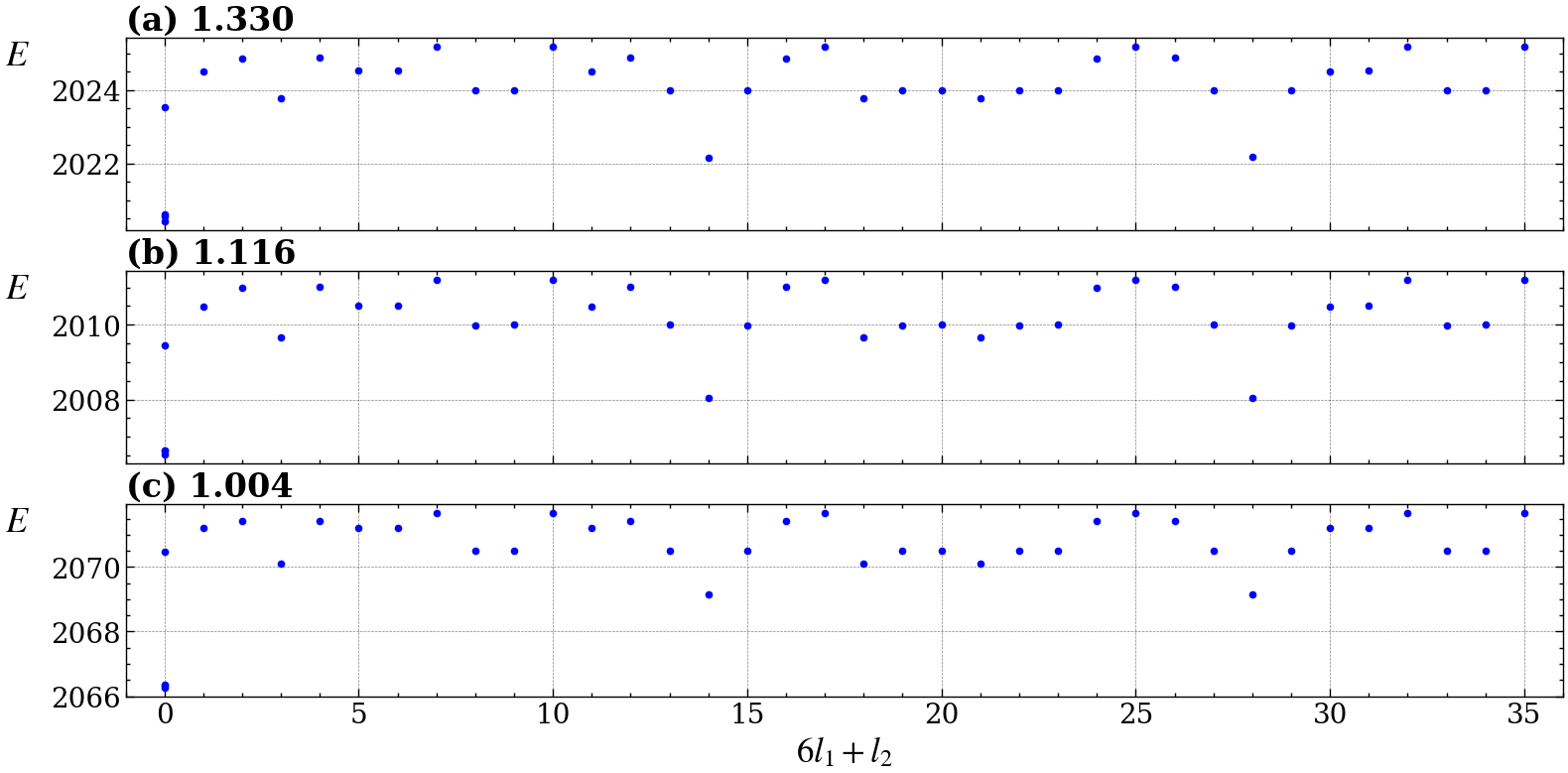}
    \caption{We show the 24-hole $6\times 6$ ED spectrum for the twisted bilayer MoTe$_2$ with $P_{\bsl{k}}$ picked from different steps of the SMV flow on $60\times 60$. 
    The $60\times 60$ $P_{\bsl{k}}$ has $\Tr\G/(2\pi)=1.330$ in (a) (which is the initial step, \ie, the top electron band),  $\Tr\G/(2\pi)=1.116$ in (b), and $\Tr\G/(2\pi)=1.004$ in (c) (which is step 12800, \ie, the numerically ideal state). For visualization, we map the total many-body momentum $l_1/L_1 \bsl{b}_{M,1} + l_1/L_2 \bsl{b}_{M,2}$ to $L_2 l_1 + l_2$. 
    We calculate only the ground state for all momentum sectors except the $(0,0)$ sector, for which we calculate the $4$ lowest-energy eigenstates. Note that near-degeneracy of the three lowest energies is a sign of the topological degeneracy of FCI.}
    \label{fig:SMV_energy_spectra}
\end{figure}

\begin{figure}[t]
    \centering
    \includegraphics[width=0.6\linewidth]{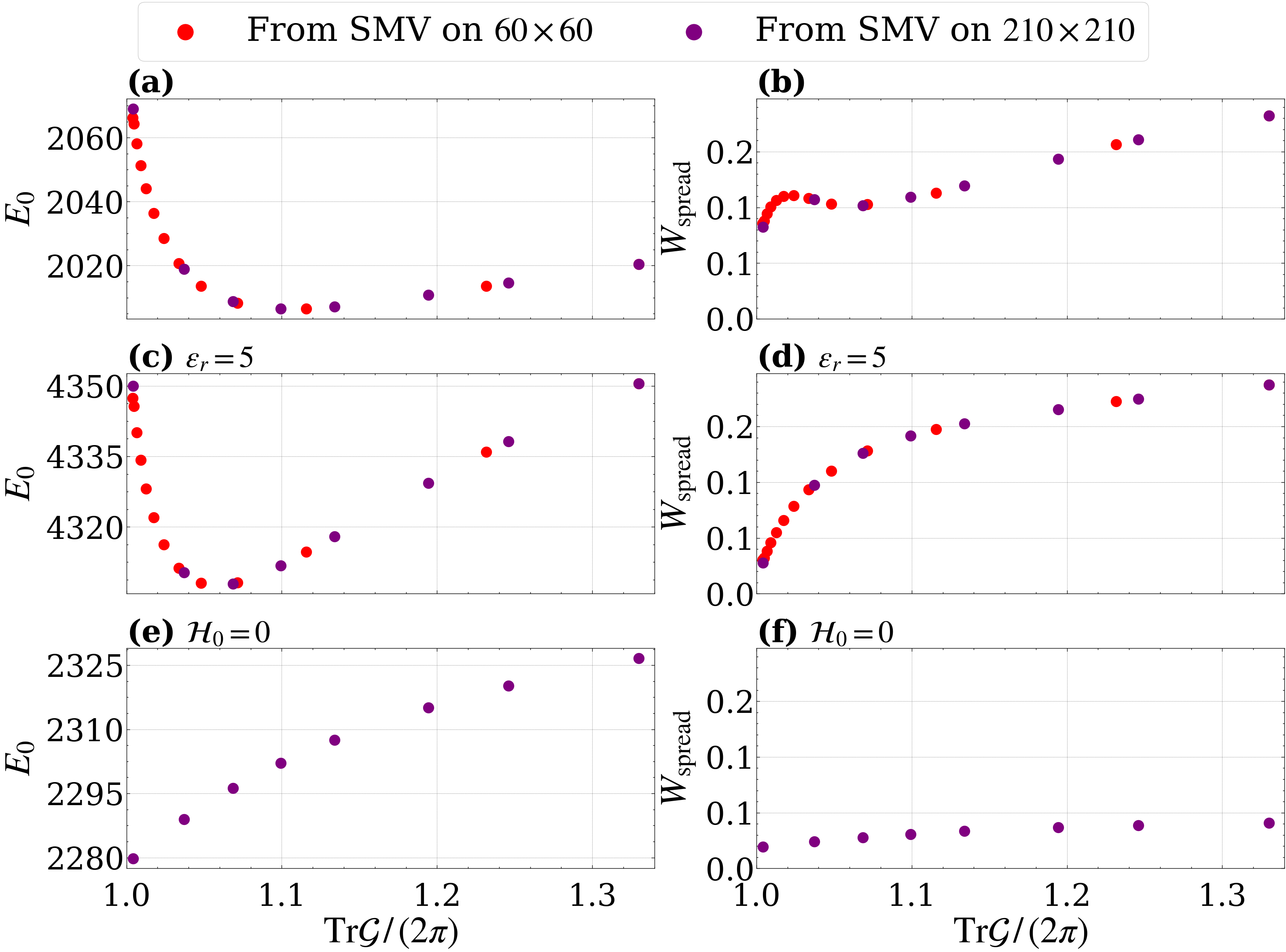}
    \caption{
    Plots showing (a,c,e) the ground-state energy $E_0$ and (b,d,f) the spread $W_{\text{spread}}=E_2-E_0$ (where $E_2$ is the third-lowest energy) as a function of $\Tr\G/(2\pi)$. 
    The energies are calculated via ED on \cref{eq:ED_hamiltonian} with 24 holes on a $6\times 6$ mesh using the $P_{\bsl{k}}$ from the SMV flow on twisted bilayer MoTe$_2$.
    The red and blue dots are from the SMV flow on $60\times 60$ and $210\times210$ meshes, respectively.
    $\Tr\G/(2\pi)$ is calculated for $P_{\bsl{k}}$ on the mesh of the SMV flow.
    $\epsilon_r = 5$ (c-d) means the interaction strength is doubled compared to the $\epsilon_r =10$ (a-b), when the single-particle term is kept.
    $\mathcal{H}_0 = 0 $ for (e-f) means we neglect the single-particle term in \cref{eq:ED_hamiltonian},which corresponds to the infinite interaction limit.
    }
    \label{fig:6x6_ED_plots}
\end{figure}

\begin{figure}[t]
    \centering
    \includegraphics[width=0.8\linewidth]{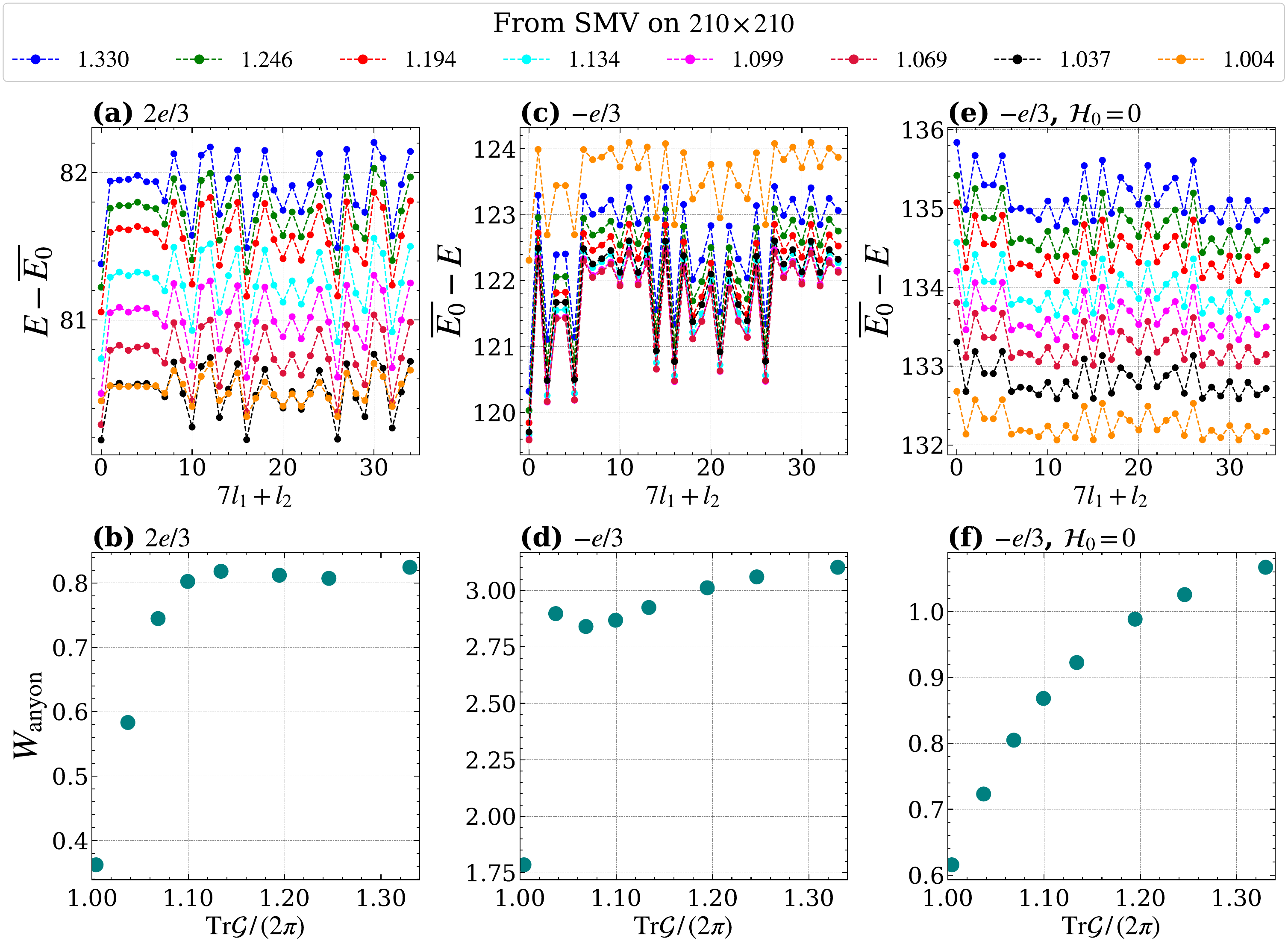}
    \caption{
    In this plot, we show the anyonic excitations.
    For visualization, we map the total many-body momentum $l_1/L_1 \bsl{b}_{M,1} + l_1/L_2 \bsl{b}_{M,2}$ to $L_2 l_1 + l_2$ in (a,c,e), and (b,d,f) are just the bandwidth of the dispersion in (a,c,e), respectively.
    In (a,c,e), the excitations are always plotted for $(L_1, L_2) = (5,7)$ , and $\overline{E_0}$ is the average energy of the three topologically (nearly) degenerate states in the $6\times 6$ ED with 24 holes.
    The subspace is given by $P_{\bsl{k}}$ from the $210\times 210$ SMV flow,  $-e$ is the electron charge, and the numbers in the legend mark $\Tr\G/(2\pi)$ of the $210\times 210$ $P_{\bsl{k}}$.
    We plot in (a) the charge $2e/3$ spectrum, where $E$ is calculated via ED with 24 holes on $5\times 7$ mesh, and $\overline{E_0}$ is the average energy of the three topologically (nearly) degenerate states in the $6\times 6$ ED with 24 holes, and in (c) the charge $-e/3$ spectrum, where $E$ is calculated via ED with 23 holes on $5\times 7$ mesh. 
    (e) is the same as (c) except that the single-particle part of \cref{eq:ED_hamiltonian} is neglected.
    }
    \label{fig:SMV_excitations}
\end{figure}

The $P_{\bsl{k}}$ here is given by a certain step of an SMV flow on the $L\times L$ mesh, and thus the projected many-body Hamiltonian is in principle on the $L\times L$ mesh.
However, the $L\times L$ meshes we use for the SMV flow are too large for the ED calculations.
Thus, for all ED calculations, we limit $\bsl{k}$ in the projected many-body Hamiltonian to a $L_1\times L_2 $ mesh, \ie, $\bsl{k} = (l_1/L_1) \bsl{b}_{M,1} + (l_2/L_2) \bsl{b}_{M,2}$ with $l_i = 0,1,2,...,L_i-1$, with $L_1,L_2 \ll L$.
Clearly, to justify the operation, the $L_1\times L_2 $ mesh has to be a subset of the $L\times L$ mesh.
Equivalently, the process can also be expression in the following way: at a certain step of an SMV flow on $L\times L$ mesh, we have a $P_{\bsl{k}}$ with specific $\Tr\G/(2\pi)$ on the $L\times L$ mesh, and we use the value of $P_{\bsl{k}}$ on the $L_1\times L_2$ mesh, which must be a subset of the $L\times L$ mesh, for the projection of the many-body Hamiltonian.
The ED results are shown in \cref{fig:SMV_energy_spectra,fig:6x6_ED_plots,fig:SMV_excitations}, and the specific configurations for the plot are shown in the captions of the plots.
Here we choose $\xi=20$nm, $\epsilon_r =10$ (unless $\epsilon_r =5$ is specified), and $L=60,210$.

\begin{figure}[t]
    \centering
    \includegraphics[width=0.6\linewidth]{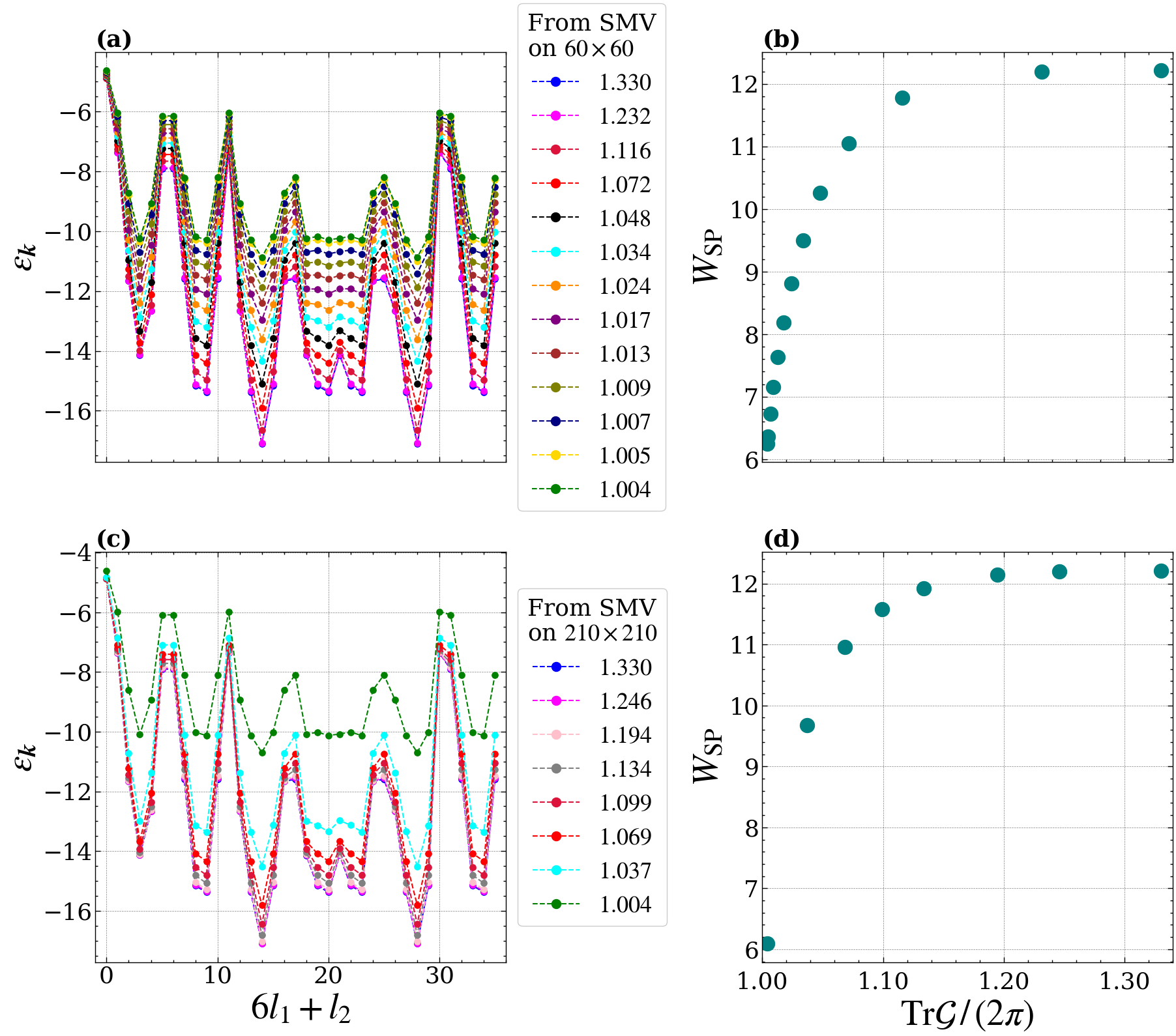}
    \caption{
    We show the single particle dispersion $\epsilon_{\bsl{k}}$ in \cref{eq:ED_hamiltonian}  (a,c) and its bandwidth in (b,d) on the $6\times 6$ mesh.
    (a-b) is from the SMV flow on the $60\times 60$ mesh, and (c-d) is from the SMV flow on the $210\times 210$ mesh.
    }
    \label{fig:SP_plots}
\end{figure}

From \cref{fig:6x6_ED_plots}(a,c,e), it is clear that our flow can help find vortex states that have lower many-body energy. For example, for $\epsilon_r = 10$, we find the lowest energy is around $\Tr\G/(2\pi) = 1.13$. 
As we increase the interaction, the lowest many-body energy happens at a lower $\Tr\G$, and in the infinite interaction limit, the numerically ideal state gives the vortex states with the lowest many-body energy. 
The underlying reasoning comes from band mixing. As the $\Tr\G$ decreases, the single-particle energy increase, as shown in \cref{fig:SP_plots}(a,c). However, the band mixing also changes the form of the projected interaction, which can overcome the increase in single-particle energy and cause the many-body energy to decrease.

As shown in \cref{fig:6x6_ED_plots}(b,d,f), the topological degeneracy improves as $\Tr\G$ decreases. 
However, the change of the projected interaction is now not the only postively contributing factor. 
In fact, as shown in \cref{fig:SP_plots}(b,d), decreasing $\Tr\G$ also reduces the bandwidth of the single-particle dispersion, which also helps improve the topological degeneracy. 
In any case, our results clearly demonstrate that our flow can improve the topological degeneracy in finite-size numerical calculations.

\begin{figure}[t]
    \centering
    \includegraphics[width=0.3\linewidth]{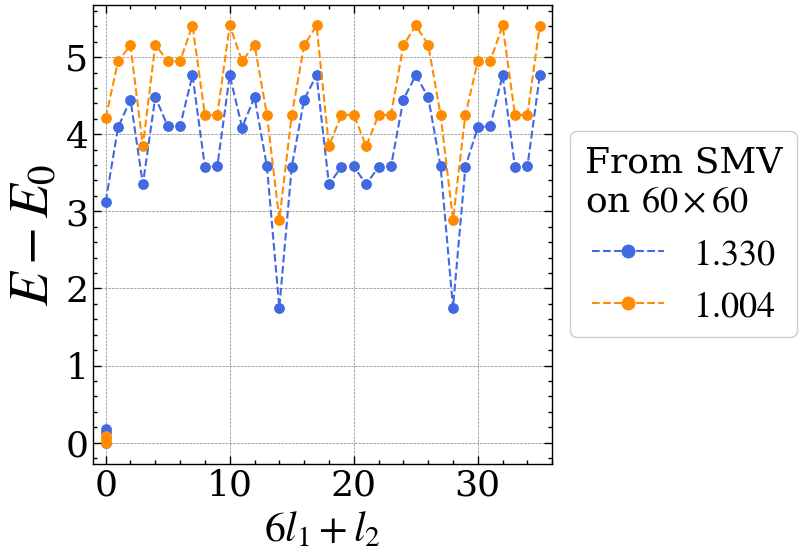}
    \caption{
    We show the 24-hole $6\times 6$ ED spectrum for the twisted bilayer MoTe$_2$ with $P_{\bsl{k}}$ picked from different steps of the SMV flow on $60\times 60$. 
    The $60\times 60$ $P_{\bsl{k}}$ has $\Tr\G/(2\pi)=1.330$ (blue, which is the initial step, \ie, the top electron band), and $\Tr\G/(2\pi)=1.004$ (green, which is step 12800, \ie, the numerically ideal state). 
    We set the zero energy to be the ground-state energy.
    For visualization, we map the total many-body momentum $l_1/L_1 \bsl{b}_{M,1} + l_1/L_2 \bsl{b}_{M,2}$ to $L_2 l_1 + l_2$. 
    The dashed line connects the lowest neutral excitations at all momentum sectors.
    }
    \label{fig:SMV_charge_neutral_excitations}
\end{figure}

In \cref{fig:SMV_excitations,fig:SMV_charge_neutral_excitations}, we find that the numerically ideal state can give the vortex states that have qualitatively the same anyon and neutral excitation spectrum as that from the top electron band.
The bandwidth of the excitation decreases generally as $\Tr\G$ decreases, and change of the projected interaction positively contributes to this decrease, s shown in \cref{fig:SMV_excitations}(f).

\subsection{Moiré Rashba Model}
\label{app:moire_rashba}

\begin{figure}[t]
    \centering
    \includegraphics[width=0.6\linewidth]{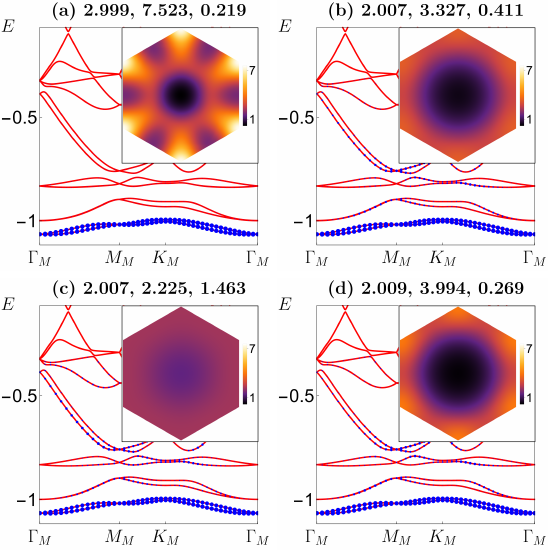}
    \caption{The $\dsZ_2$-ideal states obtained from different geometric flows in the moir\'e Rashba model. 
    In all plots, the red lines are the band structure of the model.
    The insets show the distributions of the trace of quantum metrics $g_{\bsl{k}}$ of (a) the top electron band, (b) the $\dsZ_2$-ideal states obtained from the SMV flow, (c) the $\dsZ_2$-ideal states obtained from the static-target flow with target \cref{eq:ideal_uniform_g_z2}, and (d) the $\dsZ_2$-ideal states obtained from the adapted dynamical-target flow.
    The three numbers in each caption of (a-d) are (from left to right) the integral, the maximum and the minimum of the trace of quantum metric over the 1BZ, where the integral is divided by $2\pi$.
    $g_{\bsl{k}}$ is always measured in unit of $\Omega/(2\pi)$ with $\Omega$ the moiré unit cell area.
    The area of each blue dot on the band structure is proportional to the amplitude (square root of probability) of the inset states at that momentum and energy.
    }
    \label{fig:moire_rashba}
\end{figure}

We now use an adapted moiré Rashba model to obtain ideal $\dsZ_2$ bands.
The original moiré Rashba model in \refcite{Liu_2025_Moire_Rashba} has the following form
\eq{
H_0 = \int d^2 r\ c^\dagger_{\bsl{r}} \left[ - \frac{1}{2 m} \nabla^2 + \lambda (-\ii \partial_y \sigma_x + \ii \partial_x \sigma_y)  + \Delta_1 \sum_{i=1}^{3} \sum_{s=\pm}  e^{ s \ii \bsl{g}_i^{(1)}\cdot\bsl{r}}  \right] c_{\bsl{r}} \ ,
}
where $c^\dagger_{\bsl{r}} = (c^\dagger_{\bsl{r},\uparrow}, c^\dagger_{\bsl{r},\downarrow})$, the primitive reciprocal lattice vectors reads $\bsl{b}_{M,1} = (0,1)$ and $\bsl{b}_{M,2} = C_6 \bsl{b}_{M,1}$, and $\bsl{g}_i^{(1)} = C_3^{i-1}\bsl{b}_{M,1}  $.
The model has spinful TR symmetry, represented as $\TR c^\dagger_{\bsl{r}} \TR^{-1} = c^\dagger_{\bsl{r}} \ii \sigma_y$.
As the Hermiticity requires $\Delta_1$ to be real, the model $H_0$ also has spinful $C_2$ symmetry, represented as $C_2 c^\dagger_{\bsl{r}} C_2^{-1} = c^\dagger_{-\bsl{r}} (-\ii \sigma_z)$.
The combined $C_2\TR$ symmetry will promote the $\dsZ_2$ index of an isolated set of 2 bands to the Euler number, which will simplify the ideal $\dsZ_2$ band to the ideal Euler bands.
To avoid such simplification, we break the $C_2$ symmetry by adding the following term to the Hamiltonian
\eq{
H_1 = \int d^2 r\ c^\dagger_{\bsl{r}} \left[ \ii \Delta_2 \sum_{i=1}^{3} \sum_{s=\pm} s e^{ s \ii \bsl{g}_i^{(1)}\cdot\bsl{r}}   +  \sum_{i=1}^{3} \sum_{s=\pm} (\Delta_3 + s \ii \Delta_4)e^{s \ii \bsl{g}_i^{(2)}\cdot\bsl{r}}  \right] c_{\bsl{r}} \ ,
}
where $\bsl{g}_i^{(2)} = C_3^{i-1}(\bsl{b}_{M,1} + \bsl{b}_{M,2})  $.
$H=H_0+H_1$ is the Hamiltonian that we choose.

For the numerical calculation, we choose
\eq{
2 m = 1,\ \lambda = 1.9,\ \Delta_1 = 0.12,\ \Delta_2 = 0.005,\ \Delta_3 = 0.05,\ \Delta_4 = 0.01\ .
}
The calculation for the flows is again done in the momentum space with the set of $\bsl{G}_M$ vectors including all $\bsl{G}_M$ vectors that satisfy $|\bsl{G}_M\leq 3.5|$.
The band structure is shown in \cref{fig:moire_rashba}, where the lowest two bands are isolated and have $\nu_{\dsZ_2} = 1$.
Again, we choose a $60\times 60$ momentum mesh, \ie, $\bsl{k} = (l_1/L) \bsl{b}_{M,1} + (l_2/L) \bsl{b}_{M,2}$ with $l_1,l_2 = 0,1,2,...,L-1$ and $L=60$.

All flows start from the isolated set of two lowest bands, which has $\nu_{\dsZ_2} = 1$ and $\Tr \G/(2\pi) = 2.999$---far from being ideal.

For the SMV flow, we choose $\alpha dt_l$ in \cref{eq:Y_discrete} to be $\alpha dt_l = |\bsl{b}_{M,1}|^2/120000$, and we stop at step $16000$.
The result is shown in \cref{fig:moire_rashba}(b).

For the static-target flow, we choose $\bar{g}_{\bsl{k}}$ in \cref{eq:S_static_target_flow} as that of a set of ideal $\dsZ_2$ bands with uniform quantum metric:
\eq{
\label{eq:ideal_uniform_g_z2}
\left[\bar{g}_{\bsl{k}}\right]_{ij} = \frac{\Omega}{2\pi} \delta_{ij}\ , 
}
where $\Omega$ is the moiré unit cell area.
We dynamically choose $\alpha dt_l$ by specifying $\chi$ in \cref{eq:chi} and choosing $\alpha dt_0 = |\bsl{b}_{M,1}|^4/1200000$.
In addition, in order to avoid instability caused by error accumulation, we again need to replace $g_{t,\bsl{k}}$ used in the left-hand side of \cref{eq:P_flow_monotonic} by a smoothened version $g_{t,\bsl{k}}^{smooth}$ every step.
The $\chi$ in \cref{eq:Y_discrete} and $\Lambda $ in \cref{eq:smooth_g_in_flow} are chosen as the following.
\begin{itemize}
    \item $\chi = 0.001$ and $\Lambda = 5 |\bsl{a}_{M,1}|$ for the first 4500 steps;
    \item $\chi = 0.002$ and $\Lambda = 4 |\bsl{a}_{M,1}|$ for the next 1200 steps;
    \item $\chi = 0.002$ and $\Lambda = 3 |\bsl{a}_{M,1}|$ for the next 3000 steps;
    \item $\chi = 0.001$ and $\Lambda = 2 |\bsl{a}_{M,1}|$ for the next 1200 steps;
    \item $\chi = 0.0015$ and $\Lambda = 1 |\bsl{a}_{M,1}|$ for the next 1800 steps;
    \item $\chi = 0.0015$ and $\Lambda = 0.5 |\bsl{a}_{M,1}|$ for the last 1200 steps.
\end{itemize}
The result is shown in \cref{fig:moire_rashba}(c).

For the dynamical idealization, we combine the dynamical-target flow with the SMV flow.
Explicitly, we first perform the dynamical-target flow by dynamically choosing $\alpha dt_l$ by specifying $\chi$ in \cref{eq:chi} and choosing $\alpha dt_0 = |\bsl{b}_{M,1}|^4/1.2 \times 10^6$ and by replacing $g_{t,\bsl{k}}$ used in the left-hand side of \cref{eq:P_flow_monotonic} by a smoothened version $g_{t,\bsl{k}}^{smooth}$ every step.
For the dynamical-target part, we choose the $\chi$ in \cref{eq:Y_discrete} and $\Lambda$ in \cref{eq:smooth_g_in_flow} are as follows:
\begin{itemize}
    \item $\chi = 0.002$ and $\Lambda = 6 |\bsl{a}_{M,1}|$ for the first 2100 steps;
    \item $\chi = 0.002$ and $\Lambda = 5 |\bsl{a}_{M,1}|$ for the next 2400 steps;
    \item $\chi = 0.002$ and $\Lambda = 4 |\bsl{a}_{M,1}|$ for the next 4200 steps.
\end{itemize}
The dynamical-target part leads to a projector with $\Tr\G/(2\pi) = 2.061$.
Starting from this project, we then perform the SMV flow for the last $5200$ steps by choosing  $\alpha dt_l$ in \cref{eq:Y_discrete} to be $\alpha dt_l = |\bsl{b}_{M,1}|^2/120000$.
The result is shown in \cref{fig:moire_rashba}(d).

We note that for all three flows, we should constantly keep $P_{t,\bsl{k}}$ TR symmetric---we symmetrize $P_{t,\bsl{k}}$ as each state based on the TR symmetry.

The probabilities of the final $P_{t_f, \bsl{k}}$ on the lowest 10 energy bands of the $C_2$-breaking moiré Rashba model for all three flows are shown in \cref{tab:prob_moire_rashba}.

\begin{table}[t]
    \centering
    \begin{tabular}{c|cccccccccc}
    \text{band index} & 1 & 2 & 3 & 4 & 5 & 6 & 7 & 8 & 9 & 10 \\
    \hline
    \text{SMV Flow} & 0.468 & 0.419 & 0.012 & 0.011 & 0.013 & 0.008 & 0.023 & 0.018 & 0.004 & 0.003 \\
    \text{static-target Flow} & 0.424 & 0.343 & 0.012 & 0.007 & 0.016 & 0.006 & 0.044 & 0.016 & 0.009 & 0.002 \\
    \text{adapted dynamical-target Flow} & 0.450 & 0.412 & 0.013 & 0.016 & 0.015 & 0.011 & 0.018 & 0.024 & 0.005 & 0.006  \\
    \end{tabular}
    \caption{The probabilities of the final $P_{t_f, \bsl{k}}$ from each flow on the energy bands of the $C_2$-breaking Moiré Rashba model.
    The band index $n$ is labeled from low energies to high energies, \ie, $E_{n+1}(\bsl{k})\geq E_{n}(\bsl{k})$, and the lowest energy band is band $1$.
    The probability of $P_{t_f, \bsl{k}}$ on band $n$ is calculated as $\sum_{\bsl{k}\in\BZ}\Tr[P_{t_f, \bsl{k}} P_{n, \bsl{k}}]/( \Tr[P_{t_f, \bsl{k}}] L^2)$, where $P_{n, \bsl{k}}$ is the projector of band $n$, and $L^2$ is the number of $\bsl{k}$ points in 1BZ.
    Here the adapted dynamical-target flow combines the dynamical-target flow with the SMV flow. 
    }
    \label{tab:prob_moire_rashba}
\end{table}

\subsection{Moiré Model for Inversion Fragile Topology}

\begin{figure}[t]
    \centering
    \includegraphics[width=0.6\linewidth]{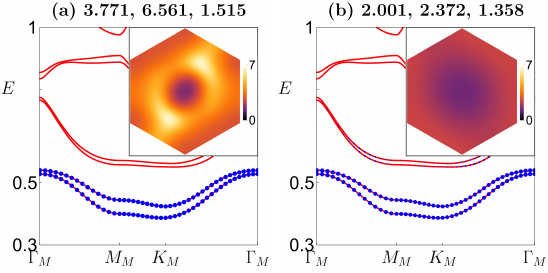}
    \caption{The inversion-fragile-ideal states obtained from SMV flows in the moir\'e model for the inversion fragile topology. 
    In all plots, the red lines are the band structure of the model.
    The insets show the distributions of the trace of quantum metrics $g_{\bsl{k}}$ of (a) the lowest two electron bands in the band plot (not the lowest bands overall) and (b) the inversion-fragile-ideal states obtained from the SMV flow.
    The three numbers in each caption of (a-b) are (from left to right) the integral, the maximum and the minimum of the trace of quantum metric over the 1BZ, where the integral is divided by $2\pi$.
    $g_{\bsl{k}}$ is always measured in unit of $\Omega/(2\pi)$ with $\Omega$ the moiré unit cell area.
    The area of each blue dot on the band structure is proportional to the amplitude (square root of probability) of the inset states at that momentum and energy.
    }
    \label{fig:fragile}
\end{figure}

We construct a continuum moiré Hamiltonian for the inversion fragile topology. 
The Hamiltonian has the form
\begin{equation}
H = H_0 + H_V\ .
\end{equation}
$H_0$ is a BHZ-type parent Hamiltonian~\cite{Bernevig2006BHZ}:
\begin{equation}
H_0
=\int d^2r c^\dagger_{\bsl{r}}\left[ 
\left(m - B(-i\nabla)^2\right)\tau_3\sigma_0
- iA\,\partial_x\,\tau_1\sigma_3
+ iA\,\partial_y\,\tau_2\sigma_0 \right] c_{\bsl{r}}
\end{equation}
where $\boldsymbol{\tau}$ and $\boldsymbol{\sigma}$ are Pauli matrices acting on orbital and spin degrees of freedom, respectively.

$H_V$ is the first-Harmonic moiré potential term:
\eq{
H_V = \int d^2r c^\dagger_{\bsl{r}}\left[ \sum_{j=1}^3 V_j(\bsl{r}) \right] c_{\bsl{r}}
}
\begin{align}
V_j(\mathbf r)
=V_0 \tau_0 \sigma_0 + \sum_{b=0}^{3}
\Big[
&2u_{j0b}\cos(\mathbf g_j\!\cdot\!\mathbf r)\,\tau_0\sigma_b
+2u_{j3b}\cos(\mathbf g_j\!\cdot\!\mathbf r)\,\tau_3\sigma_b \nonumber\\
&-2u_{j1b}\sin(\mathbf g_j\!\cdot\!\mathbf r)\,\tau_1\sigma_b
-2u_{j2b}\sin(\mathbf g_j\!\cdot\!\mathbf r)\,\tau_2\sigma_b
\Big],
\end{align}
where all coefficients $V_0, u_{j,ab}\in\mathbb{R}$, and $\bsl{g}_j = C_3^j (1,0)^T$.

The whole Hamiltonian has inversion symmetry, defined by
\begin{equation}
\P c^\dagger_{\bsl{r}} \P^{-1} = c^\dagger_{-\bsl{r}} \tau_3 \sigma_0\ .
\end{equation}
Without $u_{j,ab}$, the system has additional symmetries such as spinful time-reversal symmetry; adding $u_{j,ab}$ can make sure all symmetries but inversion are broken. 

Numerically, we choose 
\eqa{
&A = 1,\quad B = 1,\quad m = 0.500000,\quad V_1 = 0.200000,\\
&u_{100} = 0.003139,\quad u_{101} = -0.009525,\quad u_{102} = 0.001484,\quad u_{103} = -0.002253,\\
&u_{110} = -0.001829,\quad u_{111} = 0.004652,\quad u_{112} = -0.000967,\quad u_{113} = -0.008295,\\
&u_{120} = 0.003674,\quad u_{121} = -0.006335,\quad u_{122} = -0.002034,\quad u_{123} = 0.006633,\\
&u_{130} = 0.009235,\quad u_{131} = -0.006470,\quad u_{132} = -0.002778,\quad u_{133} = 0.003189,\\
&u_{200} = 0.001100,\quad u_{201} = -0.006258,\quad u_{202} = 0.007501,\quad u_{203} = 0.003793,\\
&u_{210} = -0.005473,\quad u_{211} = 0.007849,\quad u_{212} = -0.005295,\quad u_{213} = 0.008489,\\
&u_{220} = 0.008999,\quad u_{221} = 0.001876,\quad u_{222} = -0.004746,\quad u_{223} = -0.003667,\\
&u_{230} = 0.007456,\quad u_{231} = -0.003916,\quad u_{232} = -0.001605,\quad u_{233} = 0.007108,\\
&u_{300} = 0.002508,\quad u_{301} = 0.000919,\quad u_{302} = -0.001616,\quad u_{303} = -0.001726,\\
&u_{310} = -0.002968,\quad u_{311} = -0.005528,\quad u_{312} = -0.009246,\quad u_{313} = 0.005990,\\
&u_{320} = 0.009098,\quad u_{321} = 0.002407,\quad u_{322} = 0.004135,\quad u_{323} = -0.006816,\\
&u_{330} = -0.005538,\quad u_{331} = 0.006775,\quad u_{332} = 0.003145,\quad u_{333} = -0.007985.
}
With that, we plot the band structure in \cref{fig:fragile}, with the lowest two bands above zero energy having inversion fragile topology with inversion eigenvalues being $++$ at $\Gamma_M$ and $--$ at three other TRIMs.
By choosing a $60\times 60$ momentum mesh, \ie, $\bsl{k} = (l_1/L) \bsl{b}_{M,1} + (l_2/L) \bsl{b}_{M,2}$ with $\bsl{b}_{M,1}=(1,0)$ and $\bsl{b}_{M,2}=C_6 \bsl{b}_{M,1}$, $l_1,l_2 = 0,1,2,...,L-1$ and $L=60$, we find the inversion-fragile bands have $\Tr \G/(2\pi) = 3.771$ far from the lower bound $2$.

The SMV flow is performed starting from that set of bands, while keeping the $60\times 60$ momentum mesh.
The set of $\bsl{G}_M$ vectors including all $\bsl{G}_M$ vectors that satisfy $|\bsl{G}_M|\leq 3.1$.
We  choose $\alpha dt_l$ in \cref{eq:Y_discrete} to be $\alpha dt_l = |\bsl{b}_{M,1}|^2/24000$ for the first 3600 steps and then choose $\alpha dt_l = |\bsl{b}_{M,1}|^2/15000$ for the remaining 1200 steps.
The result is shown in \cref{fig:fragile}(b).

The probabilities of the final $P_{t_f, \bsl{k}}$ on the lowest 10 energy bands of the $C_2$-breaking moiré Rashba model for all three flows are shown in \cref{tab:prob_moire_rashba}.

\begin{table}[t]
    \centering
    \begin{tabular}{c|cccccccccc}
    \text{band index} & -4 & -3 & -2 & -1 & 0 & 1 & 2 & 3 & 4 & 5 \\
    \hline
    \text{SMV Flow} & 0.085 & 0.023 & 0.025 & 0.025 & 0.026 & 0.348 & 0.342 & 0.006 & 0.008 & 0.000 \\
    \end{tabular}
    \caption{The probabilities of the final $P_{t_f, \bsl{k}}$ from the SMV flow on the energy bands of the Moiré model for the inversion-fragile topology.
    The band index $n$ is labeled from low energies to high energies, \ie, $E_{n+1}(\bsl{k})\geq E_{n}(\bsl{k})$, and the lowest energy band above zero energy is band $0$.
    The probability of $P_{t_f, \bsl{k}}$ on band $n$ is calculated as $\sum_{\bsl{k}\in\BZ}\Tr[P_{t_f, \bsl{k}} P_{n, \bsl{k}}]/( \Tr[P_{t_f, \bsl{k}}] L^2)$, where $P_{n, \bsl{k}}$ is the projector of band $n$, and $L^2$ is the number of $\bsl{k}$ points in 1BZ.
    }
    \label{tab:prob_moire_fragile}
\end{table}

\end{document}